\documentclass{emulateapj}
\usepackage{natbib}
\usepackage{epsfig} 
\usepackage{graphicx} 
\usepackage{subfigure}
\usepackage{float} 
\usepackage{amsmath}
\usepackage{color}
\usepackage{amssymb}
\usepackage{amsfonts} 
\usepackage{units} 
\usepackage{mathtools} 
\usepackage{bm}
\usepackage[colorlinks,linkcolor=blue,anchorcolor=green,citecolor=blue]{hyperref}
\def\be{\begin{eqnarray}}
\def\ee{\end{eqnarray}}
 
\shorttitle{Bunching Coherent Curvature Radiation}
\shortauthors{Yang \& Zhang} 
\begin{document} 

\title{Bunching Coherent Curvature Radiation in Three-Dimensional Magnetic Field Geometry: Application to Pulsars and Fast Radio Bursts}

\author{Yuan-Pei Yang\altaffilmark{1,2,3} and Bing Zhang\altaffilmark{4,5,1}}

\affil{$^1$Kavli Institute for Astronomy and Astrophysics, Peking University, Beijing 100871, China; yypspore@gmail.com;\\
$^2$ National Astronomical Observatories, Chinese Academy of Sciences, Beijing 100012, China\\
$^3$ KIAA-CAS Fellow\\
$^4$ Department of Physics and Astronomy, University of Nevada, Las Vegas, NV 89154, USA; zhang@physics.unlv.edu\\
$^5$ Department of Astronomy, School of Physics, Peking University, Beijing 100871, China \\
}

\begin{abstract}
The extremely high brightness temperatures of pulsars and fast radio bursts (FRBs) require their radiation mechanisms to be coherent. Coherent curvature radiation by bunches has been long discussed as the mechanism for radio pulsars and recently for FRBs. Assuming that bunches are already generated in pulsar magnetospheres, we calculate the spectrum of coherent curvature radiation under a three-dimensional magnetic field geometry. Different from previous works assuming parallel trajectories and a mono-energetic energy distribution of electrons, we consider a bunch characterized by its length, curvature radius of the trajectory family, bunch opening angle, and electron energy distribution. We find that the curvature radiation spectra of the bunches are characterized by a multi-segment broken power law, with the break frequencies depending on bunch properties and trajectory configuration. We also emphasize that in a pulsar magnetosphere only the fluctuation of net charges with respect to the background (Goldreich-Julian) outflow can make a contribution to coherent radiation. We apply this model to constrain the observed spectra of pulsars and FRBs. For a typical pulsar ($B_p=10^{12}~\unit{G}$, $P=0.1~\unit{s}$), a small fluctuation of the net charge $\delta n_{\rm GJ}\sim 0.1 n_{\rm GJ}$ can provide the observable flux. For FRBs, the fluctuating net charge may be larger due to its abrupt nature. For $\delta n_{\rm GJ}\sim n_{\rm GJ}$, a neutron star with a strong magnetic field and fast rotation is required to power an FRB in the spindown-powered model. The requirement is less stringent in the cosmic comb model thanks to the larger cross section and compressed charge density of the bunch made by the external astrophysical stream that combs the magnetosphere. 
\end{abstract}

\keywords{radiation mechanisms: non-thermal --- radio continuum: general}

\section{Introduction}\label{sec1}

Both radio pulsars and fast radio bursts (FRBs) show non-thermal radio spectra and a very high brightness temperature, $T_B$, which is much greater than any plausible thermal temperature, $T_e$, of the electrons in the source. 
In general, there are two mechanisms that limit the brightness temperature in a synchrotron source \citep[e.g.][]{mel17}: 1. synchrotron self-absorption (SSA) implies $T_B\lesssim\gamma m_ec^2/k_{\rm B}$, where $\gamma$ is the Lorentz factor of the electrons; 2. inverse Compton scattering implies $T_B\lesssim10^{12}~\unit{K}$ \citep{kel69}.
Observationally, the brightness temperature of radio pulsars can reach $T_B\sim10^{26}~\unit{K}$, and the brightness temperature of FRBs can even reach $T_B\sim10^{37}~\unit{K}$. Therefore, the emission mechanism from pulsars and FRBs must be extremely coherent\footnote{In physics, two waves with the same waveform are perfectly coherent if they have a constant phase difference and the same frequency. Thus, coherence can cause the amplitude of the superposition of two wave enhanced or reduced. In the field of pulsars and FRBs, ``coherent'' is mainly defined as ``coherently enhanced''. We adopt this definition throughout the paper.}, 
which means that the observed emission cannot be explained by the simple summation of the radiation power of individual particles. Rather, the superposition of the electromagnetic waves from each particle should be considered.
Theoretical models usually invoke one of three classes of coherent emission mechanisms \citep[e.g.][]{mel17}: radiation by bunches (related to particle coherence in position space), a reactive instability (related to particle coherence in momentum space), and a maser mechanism (negative absorption).

After decades of studies, the coherent emission mechanism of radio pulsars remains not fully understood \citep{mel17}. The leading mechanism invokes coherent curvature radiation by bunches \citep{gun71,stu71,gin75,rud75,bus76,ben77,pat80,mel00,gil04}, but other mechanisms, such as various maser mechanisms \citep{twi58,mcc66,bla75,mel78,luo92,luo95}, linear acceleration emission \citep{coc73,mel78,kro79}, relativistic plasma emission \citep{wea98,mel99,mel17}, and anomalous Doppler emission \citep{mac79,kaz91,lyu99a,lyu99b} have been also discussed in the literature.
FRBs have even more extreme brightness temperatures \citep[e.g.][]{lor07,tho13,cha17}, which require more stringent coherent conditions than radio pulsars. 
Nonetheless, curvature radiation by bunches \citep{kat14,kat18a,kat18b,kum17,ghi17a,lu17} has been suggested to be the leading mechanism, even though the coherent conditions have been over-simplified in these models. Some maser mechanisms \citep{ghi17b,lyu14,wax17,bel17} have also been proposed to explain FRB emission. However, the maser condition of population inversion is hard to achieve, especially for the extremely high brightness temperature observed in FRBs \citep{lu17}.

In this paper, we mainly focus on the coherent curvature radiation by bunches. 
When a charged particle moves along a curved trajectory, its perpendicular acceleration will result in the so called ``curvature radiation''. For a relativistic electron in the magnetosphere around a neutron star, due to the strong magnetic field, 
the vertical momentum perpendicular to the field line drops to zero in a very short period of time due to synchrotron cooling, e.g., $t_{\rm cool}\sim10^{-18}~\unit{s}(\gamma/1000)(B/10^{12}~\unit{G})$, where $\gamma$ is the Lorentz factor of electron and $B$ is the magnetic field strength. This leads to electrons moving along the field lines. The electron trajectories then essentially track with the field lines\footnote{ Notice that the curved trajectory does not strictly overlap with the field line even in the co-rotating frame. A charged particle moving along a field line must be subjected to a Lorentz force that causes it to follow the curved path, which requires a drift velocity perpendicular to the plane that contains the field lines \citep[e.g.][]{zhe79}.}.
 
Besides the mechanism of forming bunches \citep[e.g.][]{pat80,mel00}, coherent curvature radiation by a three-dimensional bunch has been studied in some earlier papers, e.g. \citet{stu75} and \citet{els76}. In these works, an underlying assumption is that the electron spatial distribution is ``stationary'', which means that the spatial separations among the electrons remain the same as they move out from the magnetosphere, which demands that all the electron trajectories are parallel to each other. Thus the radiation only depends on the initial distribution of electrons. In this case, based on a three-dimensional Fourier transform of the electron distribution, a simple theory of coherent curvature radiation from a three-dimensional bunch could be developed. 
In reality, the magnetic field lines are very likely not parallel to each other. For example, a bunch moving in a dipolar field will expand when it moves away from the dipole field center, so that the electron distribution is not ``stationary''. In order to make effective coherence, the opening angle of a bunch needs to be confined within $1/\gamma$ in the direction of the field line. However, this condition can be hardly maintained due to the curvature and non-parallel nature of the dipole field lines. 

Another simplified assumption in these previous works \citep{stu75,els76} is that the electron distribution is mono-energetic. Theoretical modeling of pulsar magnetosphere and observations suggest that the accelerated electrons should have an energy distribution, the simplest of which is a power law. The calculation of coherent radiation of such power-law-distributed electrons become more complex, since a coherent sum of the amplitudes from electrons with different energy should be considered. \citet{ghi17a} discussed the spectrum of curvature radiation from power-law-distributed electrons, however, they ignored the coherent sum of the amplitudes of electromagnetic waves.

In this paper, we do not discuss the formation of bunches (see detailed discussion in e.g. \citealt{pat80,mel00}), but attempt to calculate the spectrum of the curvature radiation by bunches (assuming that they already exist) under a three-dimensional magnetic field geometry and a power-law distribution of electron energy. We then apply this theory to pulsars and FRBs and investigate the conditions to reproduce their observed brightness temperatures.
We consider that a bunch, consisting of a trajectory family, is characterized by the following parameters: bunch length, curvature radius of the trajectory family, bunch opening angle, and electron energy distribution. 
We find that the radiation spectra of bunches under a three-dimensional magnetic field geometry is a multi-segment broken power law with the break frequencies depending on the above parameters. In particular, we emphasize that the low-frequency index of the spectrum is $2/3$ rather than $1/3$ in most previous works \citep[e.g.][]{kum17,ghi17a}.
Consider that the observed duration of one pulse from pulsars or FRBs, e.g., $T_{\rm obs}\gtrsim1~\unit{ms}$ is much longer than the pulse duration of the curvature radiation, e.g. $T_{p}\sim 1/\nu_{\rm obs}\sim1~\unit{ns}$, there are numerous bunches sweeping cross the line of sight during the observed duration.
We emphasize that not all electrons in the magnetosphere contribute to coherent radiation, and that only the fluctuating net charges with respect to the Goldreich-Julian outflow can make a contribution. Such a fluctuation of charges might originate from the abrupt discharges of the inner gap near the neutron star surface \citep{rud75,zhang96,zhang97,gil00,gil06}, instabilities in the outflow \citep{chen77,ego83,uso87,ged02,kum17,lu17}, or the oscillation of plasma in the acceleration region \citep{lev05,bel07,luo08}.
For all these cases, the fluctuating net charges in one bunch are usually not much larger than the local Goldreich-Julian density for pulsars. We apply the observational data of pulsars and FRBs to constrain these parameters. 

The paper is organized as follows. In Section \ref{sec2}, we briefly review the main properties of the radiation from a single moving charge. 
In Section \ref{sec3}, we discuss the coherent curvature radiation from a point source bunch. In particular, we calculate the coherent radiation spectrum from electrons with a power-law distribution.  
In Section \ref{sec4}, we extend our discussion of coherent curvature radiation to a spatially extended source with: 1. electrons distributed in the same trajectory; 2. electrons distributed in a trajectory family; 3. electrons in the trajectories with different curvature radii. In particular, for the second case, 
different from the previous models for three-dimensional bunches with a stationary distribution in parallel trajectories \citep[e.g.][]{stu75,els76}, we consider a more general case in which the electron trajectories are not parallel to each other. 
In Section \ref{sec5}, the multi-frequency spectra of three-dimensional bunch are derived for different parameter ranges, which are found to be a broken power law.
In Section \ref{sec6} and Section \ref{sec7}, we discuss the applications of this theory to pulsars and FRBs, respectively. For the FRB models, we consider both the traditional polar cap model and the cosmic comb model. The results are summarized in Section \ref{sec8} with some discussions. Some detailed calculations are presented in the Appendix.

\section{Radiation by moving charges}\label{sec2}

Consider an electron that moves along a trajectory $\bm{r}(t)$. The observation point is assumed to be far enough away from the region of space where the acceleration occurs. 
In this case, the energy radiated per unit solid angle per unit frequency interval is given by (see Appendix \ref{seca})
\be
\frac{dI}{d\omega d\Omega}=\frac{e^2\omega^2}{4\pi^2c}\left|\int_{-\infty}^{+\infty}\bm{n}\times(\bm{n}\times\bm{\beta})e^{i\omega(t-\bm{n}\cdot\bm{r}(t)/c)}dt\right|^2.\nonumber\\\label{radiation}
\ee
where $\omega$ is the observed angle frequency, $\bm{\beta}=\dot{\bm{r}}(t)/c$ is the dimensionless velocity, and $\bm{n}$ is the unit vector between the electron and the observation point, which is sensibly constant in time. 
Therefore, one can see that the observed spectrum is determined by the electron trajectory 
over a period of time.

If there is more than one charged particle, a coherent sum of the amplitudes should replace the single amplitude.
In this case, the energy radiated per unit solid angle per unit frequency interval is given by
\be
\frac{dI}{d\omega d\Omega}=\frac{\omega^2}{4\pi^2c}\left|\int_{-\infty}^{+\infty}\sum_j^Nq_j\bm{n}\times(\bm{n}\times\bm{\beta}_j)e^{i\omega(t-\bm{n}\cdot\bm{r}_j(t)/c)}dt\right|^2,\nonumber\\\label{multiemission}
\ee
where $j$ represents the identifier of each charged particle, and $q_j$ is the corresponding charge.

We should note that $dI/d\omega d\Omega$ does not have ``per unit time'' in its dimension. As pointed out by \citet{ryb79}, the coexistence of $dt$ and $d\omega$ would violate the uncertainty relation between $\omega$ and $t$, e.g., $\Delta t\Delta\omega>1$. However, if the pulse repeats on an average time scale $T$, the radiation power may be written as \citep{ryb79}
\be
\frac{dW}{d\omega d\Omega dt}
\equiv\frac{1}{T}\frac{dI}{d\omega d\Omega}.
\label{power}
\ee

For synchrotron radiation, naturally the pulse repeats with the gyration period $T=2\pi/\omega_B$, where $\omega_B=eB/\gamma m_ec$ is the gyration frequency. However, for curvature radiation of a single particle, the charge motion direction only sweeps the line of sight once, so that the definition of radiation power is no longer meaningful. If there is more than one particle sweeping cross the line of sight, $T$ would be the mean time interval between each pair of the particles, and $dI/d\omega d\Omega$ corresponds to the radiation energy of one particle\footnote{Here we have assumed that the emission pulse is incoherent. If the superposition from the electromagnetic wave from each source is coherent,  $T$ would be the mean time interval of each coherent pulse, and $dI/d\omega d\Omega$ corresponds to the radiation energy of one coherent pulse.}.

First, we briefly summarize the curvature radiation of a single electron during instantaneously circular motion (see Appendix \ref{secb0}).
Considering that the instantaneously-circular trajectory has a curvature radius $\rho$ and lies in a trajectory plane, and the angle between the line of sight and the trajectory plane is $\theta$. For an accelerated relativistic electron with Lorentz factor $\gamma$, the radiation is beamed in a narrow cone of $\sim 1/\gamma$ in the direction of the electron's velocity, which can be seen as a short pulse as the beam sweeps cross the observational point. 
Then the energy radiated per unit frequency interval per unit solid angle is given by \citep[e.g.][]{jac62}
\be
\frac{dI}{d\omega d\Omega}&=&\frac{e^2}{3\pi^2c}\left(\frac{\omega\rho}{c}\right)^2\nonumber\\
&\times&\left(\frac{1}{\gamma^2}+\theta^2\right)^2\left[K_{2/3}^2(\xi)+\frac{\theta^2}{(1/\gamma^2)+\theta^2}K_{1/3}^2(\xi)\right],\nonumber\\\label{CRspec}
\ee
where the parameter $\xi$ in the modified Bessel function $K_{\nu}(\xi)$ is defined by $\xi=(\omega\rho/3c)\left(1/\gamma^2+\theta^2\right)^{3/2}$, the first term in the square bracket corresponds to the polarized component in the trajectory plane, and the second term corresponds to the polarized component that is perpendicular to the line of sight and the above polarized component (see Appendix \ref{secb0}). Numerically, $dI/d\omega d\Omega$ is dominated by the first term. 

According to Eq.(\ref{CRspec}), one can find that the typical spread angle depends on frequency (see Appendix \ref{secb0}), e.g.,
\be
\theta_c(\omega)\simeq
\begin{dcases}
\frac{1}{\gamma}\left(\frac{2\omega_c}{\omega}\right)^{1/3}=\left(\frac{3c}{\omega\rho}\right)^{1/3}, &\omega\ll\omega_c\\
\frac{1}{\gamma}\left(\frac{2\omega_c}{3\omega}\right)^{1/2}, &\omega\gg\omega_c
\end{dcases},\label{angle}
\ee
where $\omega_c$ is the critical frequency of the curvature radiation, e.g.,
\be
\omega_c=\frac{3}{2}\gamma^3\left(\frac{c}{\rho}\right),\label{omegac}
\ee
For $\omega\sim\omega_c$, the radiation is confined to angles of the order $\sim 1/\gamma$; for lower frequencies, the spread angle is larger. Note that for frequencies higher than $\omega_c$, one has $\theta_c\propto\omega^{-1/2}$. However, the radiation has become negligible.

On the other hand, the observed spectrum depends on the observation direction. At $\theta=0$, the radiated energy is maximum. Once $\theta\gtrsim\theta_c(\omega)$, the radiated energy will significantly decrease. 
For the case with $\theta=0$, the energy radiated per unit frequency interval per unit solid angle could be approximately given by (see Appendix \ref{secb0})
\be
\frac{dI}{d\omega d\Omega}\simeq\frac{e^2}{c}\left[\frac{\Gamma(2/3)}{\pi}\right]^2\left(\frac{3}{4}\right)^{1/3}\left(\frac{\omega\rho}{c}\right)^{2/3}e^{-\omega/\omega_c}.\label{scr2}
\ee
Thus, one has $dI/d\omega d\Omega\propto\omega^{2/3}$ for $\omega\ll\omega_c$. For the case with $\theta\neq0$, as $\theta$ increases the cut-off frequency of the spectrum will shift to lower frequency, but the spectral index still keeps $2/3$. In this work, in order to analyze the maximum brightness temperature of pulsars or FRBs, we consider that the observed direction is at $\theta=0$.

The spectrum of the total energy emitted by the electron can be found by integrating Eq.(\ref{CRspec}) over angle \citep{wes59}, i.e.
\be
\frac{dI}{d\omega}=\sqrt{3}\frac{e^2}{c}\gamma\frac{\omega}{\omega_c}\int_{\omega/\omega_c}^\infty K_{5/3}(x)dx.\label{syn}
\ee
This equation can give the classical spectrum of synchrotron radiation in astrophysical processes, e.g. $dI/d\omega\propto\omega^{1/3}$ for $\omega\ll\omega_c$. However, one must note that it is the total radiation spectrum in all directions rather than the direction along the line of sight. In most astrophysical sources that invoke synchrotron radiation, since electrons in the magnetic fields have random pitch angles (the angle between the magnetic field direction and the electron velocity direction), or the local magnetic fields where electrons are accelerated have random directions, the incoherent sum of the radiation energy per unit solid angle, e.g., Eq.(\ref{CRspec}), from different electrons will make the classical $\omega^{1/3}$ spectrum as shown in Eq.(\ref{syn}) \citep{yan18}.
However, for curvature radiation, Eq.(\ref{syn}) is not applicable for the following two reasons: 1. the trajectories of charged sources are almost the same at large scales, and the observed spectrum is from the radiation along the line of sight; 2. even for more than one charged sources with different motion directions, a coherent sum of the amplitudes, rather than a simple integration of radiation energy over angles, should be considered.

Finally, we consider the duration of one pulse emitted by such an instantaneously circular motion. For a given frequency with $\omega\ll\omega_c$, the spread in angle is $\theta_c(\omega)\simeq(3c/\omega\rho)^{1/3}$. Thus, the frequency-dependent pulse duration of the curvature radiation is given by
\be
T_p(\omega)\simeq\frac{\rho\theta_c(\omega)}{c}\left(1-\frac{v}{c}\right)\simeq\frac{1}{2\gamma^2}\left(\frac{3\rho^2}{c^2\omega}\right)^{1/3},\label{duration}
\ee
where the factor $(1-v/c)$ is due to the propagation time-delay effect. For $\omega\sim\omega_c$, one has $T_p\sim\rho/c\gamma^3\sim1/\omega_c$.

\section{Coherent emission by a point source charge bunch}\label{sec3}

In this section, we calculate the radiation from a point source charge bunch moving instantaneously at a constant speed on an approximate circular path. In order to satisfy the point-source approximation, one needs: 1. the system scale is much smaller than the typical curvature radius of the trajectory; 2. all electrons have nearly the same state of initial motion\footnote{For relativistic electrons with different energies, although the energy difference could be large, the velocity difference is small since the velocities are all close to the speed of light.}.
We consider two cases of point-source radiation: 1. power-law-distributed electrons; 2. particles with different charges.

\subsection{Radiation from electrons with a power-law distribution}\label{sec32}

First, we consider that the energy distribution of the electrons satisfies a power-law distribution, e.g.
\be
N_e(\gamma)d\gamma=N_{e,0}\left(\frac{\gamma}{\gamma_1}\right)^{-p}d\gamma,~~~~~\gamma_1<\gamma<\gamma_2\label{spece}
\ee
where $N_e(\gamma)d\gamma$ is the electron number in a range from $\gamma$ to $\gamma+d\gamma$, $N_{e,0}$ is the corresponding normalization, and $\gamma_1$ and $\gamma_2$ are the lower and upper limits of the electron Lorentz factor. 
In this case, a coherent sum of the amplitudes should replace the single amplitude, and one has
\be
\frac{dI}{d\omega d\Omega}&=&\frac{e^2\omega^2}{4\pi^2c}\left|-\bm{\epsilon}_\parallel \int_{\gamma_1}^{\gamma_2}N_e(\gamma){A}_\parallel(\omega,\gamma)d\gamma\right.\nonumber\\
&+&\left.\bm{\epsilon}_\perp\int_{\gamma_1}^{\gamma_2}N_e(\gamma){A}_\perp(\omega,\gamma)d\gamma\right|^2,
\ee
where $A_\parallel$ and $A_\perp$ denote the polarized component of the amplitude along $\bm{\epsilon}_\parallel$ and $\bm{\epsilon}_\perp$, respectively, see Appendix \ref{secb0}.
We consider that the observed direction is at $\theta=0$, according to Appendix \ref{secb}, one has $A_\perp(\omega,\gamma)=0$, and 
\be
A_\parallel(\omega,\gamma)&=&\frac{2i}{\sqrt{3}}\frac{\rho}{c\gamma^2}K_{2/3}\left(\frac{\omega\rho}{3c\gamma^3}\right)\nonumber\\
&\simeq&\frac{2^{4/3}i}{\sqrt{3}}\Gamma(2/3)\frac{\rho}{c\gamma^2}\left(\frac{\omega}{\omega_c}\right)^{-2/3}e^{-\omega/2\omega_c}.
\ee
For a power-law electron distribution, e.g. Eq.(\ref{spece}), the energy radiated per unit frequency interval per unit solid angle is given by
\be
\frac{dI}{d\omega d\Omega}&\simeq&\frac{e^2}{c}\frac{2^{(2p-6)/3}}{3\pi^2}\left[\Gamma\left(\frac{2}{3}\right)\Gamma\left(\frac{p-1}{3}\right)\right]^2N_{e,0}^2\gamma_1^4\nonumber\\
&\times&
\begin{dcases}
\left(\frac{\omega}{\omega_{c1}}\right)^{2/3}&\omega\ll\omega_{c1}\\
\left(\frac{\omega}{\omega_{c1}}\right)^{-(2p-4)/3}e^{-\omega/\omega_{c2}}&\omega\gg\omega_{c1}
\end{dcases},\label{ne1}
\ee
where $\omega_{c1}=\omega_c(\gamma_1)$ and $\omega_{c2}=\omega_c(\gamma_2)$.
At last, we note that the velocity spread, e.g., $\Delta v\sim\Delta\beta c\sim\Delta\gamma c/\gamma^3$, would cause a linear extent of the bunch, i.e., $\Delta l\sim\Delta vt\sim ct/\gamma^2$, where $\Delta\gamma\sim \gamma$ is assumed. For the electromagnetic wave with $\lambda\gtrsim\Delta l$, the effect of the linear extent is negligible.

\subsection{Radiation from particles with different charges}\label{sec33}

Next, we discuss the radiation from a point source with particles of different charges.
According to Eq.(\ref{multiemission}), if all the charged particles have the same trajectory, the charge term can be extracted from the integral, so that the radiation spectrum only depends on the net charge in the point source, i.e.
\be
\frac{dI}{d\omega d\Omega}=\frac{\omega^2}{4\pi^2c}\left(\sum_j^Nq_j\right)^2
\left|\int_{-\infty}^{+\infty}\bm{n}\times(\bm{n}\times\bm{\beta})e^{i\omega(t-\bm{n}\cdot\bm{r}(t)/c)}dt\right|^2.\nonumber\\
\ee
Therefore, only the net charge in the point source can contribute to curvature radiation. 

For example, in the pulsar wind, the total lepton number density is $\mu_\pm n_{\rm GJ}$, where $n_{\rm GJ}$ denotes the Goldreich-Julian density \citep{gol69}, and $\mu_\pm$ is the multiplicity resulting from the electron-positron pair cascade. However, the electron-positron pairs in a bunch do not contribute to the net charge, and hence, would not contribute to coherent emission. Only the net charge, of the order of  $n_{\rm GJ}$, in the bunch may contribute to coherent radiation.

\section{Coherent emission by charges in a spatially extended source}\label{sec4}

In the above section, the size of the emission region is considered to be much smaller than the curvature radius of the trajectory, which can be treated as a point source.
Next, we further consider the curvature radiation from an extended source, including three cases: 1. the electrons move in the same trajectory but with different delay times; 2. the electrons are in a trajectory family with the same curvature radius but different orientations; 3. the electrons are in the trajectories with different curvature radii.

\subsection{Electrons in the same trajectory}\label{sec41}

We consider that the trajectories of $N$ electrons are the same but the electrons are injected at different times. The retarded position of the $j$th electron can be written as $\bm{r}_j(t)=\bm{r}(t)+\Delta\bm{r}_j(t)$, where $\bm{r}(t)$ denotes the retarded position of the first electron, and $\Delta \bm{r}_j(t)$ denotes the relative displacement between the first electron and the $j$th electron. 
For a relativistic bunch, its radiation is beamed in a narrow cone that sweeps cross the line of sight. Therefore, if the bunch length satisfying $L\sim\Delta r_N\lesssim\rho\theta_c$, where $\theta_c$ is the spread angle of curvature radiation (see Eq.(\ref{angle})), then \emph{in the observed path} (with a length $\sim\rho\theta_c$ where the bunch velocity is almost parallel to the line of sight), the relative displacement between each electron could be considered as time-independent (the detailed discussion is shown in Appendix \ref{secd0}). 
In this case, according to Eq.(\ref{multiemission}), the total energy radiated per unit solid angle per unit frequency interval can be approximately given by
\be
\frac{dI_{(N)}}{d\omega d\Omega}&\simeq&\frac{e^2\omega^2}{4\pi^2c}\left|\int_{-\infty}^{+\infty}\bm{n}\times(\bm{n}\times\bm{\beta})e^{i\omega(t-\bm{n}\cdot\bm{r}(t)/c)}dt\right|^2\nonumber\\
&\times&\left|\sum_j^Ne^{-i\omega(\bm{n}\cdot\Delta\bm{r}_j/c)}\right|^2,
\ee
which is
\be
\frac{dI_{(N)}}{d\omega d\Omega}=\frac{dI_{(1)}}{d\omega d\Omega}F_\omega(N), \label{cohN}
\ee
where
\be
F_\omega(N)=\left|\sum_j^Ne^{-i\omega(\bm{n}\cdot\Delta\bm{r}_j/c)}\right|^2,\label{factor}
\ee
is a dimensionless parameter denoting the enhancement factor due to coherence,
and $dI_{(1)}/d\omega d\Omega$ corresponds to the radiation of the first electron.
Therefore, once $F_{\omega}(N)$ is obtained, the radiation spectrum of $N$ electrons can be calculated via Eq.(\ref{cohN}) and Eq.(\ref{factor}). For example, if all the electrons are at one point, e.g., $\Delta\bm{r}_j\simeq0$, one has $F_{\omega}(N)=N^2$, which means that the spectrum has the same shape with the single charge, but is enhanced by a factor of $N^2$.

\subsubsection{One bunch in the trajectory}\label{sec411}

\begin{figure}[H]
\centering
\includegraphics[angle=0,scale=0.3]{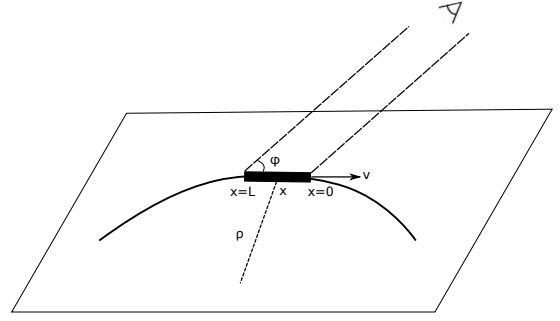}
\caption{Curvature radiation from a bunch. The dark strip denotes the bunch with length $L$ and $N$ electrons, and the curve denotes the bunch trajectory.}\label{bunch}
\end{figure}
We assume that $N$ electrons are spatially uniformly distributed in a bunch along the trajectory. In the lab frame, the intrinsic length of a bunch is $L$, the curvature radius of the trajectory is $\rho$, and the position of the $j$ electron is $x_j$ ($0<x_j<L$), as shown in Figure \ref{bunch}.  

According to the radiation theory outlined in Section \ref{sec2}, the observed radiation spectrum is taken over the path. Meanwhile, most observed energies are radiated when the angle between the line of sight and the bunch velocity is at minimum\footnote{In general, the minimum angle between the line of sight and the bunch velocity cannot be zero when the line of sight is not parallel to the trajectory plane.
For a circular trajectory, the minimum angle is equal to the angle between the line of sight and the trajectory plane.}. We define such a minimum angle as $\varphi$, as shown in Figure \ref{bunch}. If the line of sight is parallel to the trajectory plane, one has $\varphi=0$.
Thus the amplitude of the radiation from the $j$th electron is given by
\be
\bm{A}(\omega,x_j)=\bm{A}(\omega,0)e^{-ikx_j\cos\varphi},
\ee
where $k\equiv\omega/c$, $\bm{A}$ is the radiation amplitude (see Appendix \ref{seca}), and $\bm{A}(\omega,0)$ corresponds to the radiation amplitude of the first electron. 
Using Eq.(\ref{factor}), one has
\be
F_\omega(N)&=&\left|\sum_j^N e^{-ikx_j\cos\varphi}\right|^2
=\left|\frac{N}{L} \int_0^{L} e^{-ikx\cos\varphi}dx\right|^2\nonumber\\
&=&N^2\left|\frac{1-e^{-ikL\cos\varphi}}{ikL\cos\varphi}\right|^2
=N^2\left[\frac{\sin(\omega/\omega_l)}{(\omega/\omega_l)}\right]^2,\label{bunfact}
\ee
where 
\be
\omega_l=\frac{2c}{L\cos\varphi},\label{omegal}
\ee
and the second equal sign is based on the assumption that electrons are uniformly distributed in a bunch.
As shown in Eq.(\ref{bunfact}), for $\omega\ll\omega_l$, $F_m(N)\sim N^2$; for $\omega\gg\omega_l$, the local maximum value of $F_m(N)$ is proportional to $\omega^{-2}$.
If the line of sight is parallel to the trajectory plane, e.g. $\varphi\sim0$, the observed energy will reach the maximum value, and one has $\omega_l\simeq2c/L$.

On the other hand, in order to make the radiation from the electrons at $x=0$ and $x=L$ coherent, the condition $L/\rho\ll\theta_c$ needs to be satisfied, where $\theta_c\sim(3c/\omega\rho)^{1/3}$ (see Eq.(\ref{angle}) in the $\omega\ll\omega_c$ regime) is the emission angle of the relativistic electron. Therefore, the upper limit of the coherent frequency is given by
\be
\omega_m\sim\left(\frac{\rho}{L}\right)^2\omega_l.\label{omm}
\ee
Any electromagnetic wave with $\omega\gg\omega_m$ would not be coherent. In summary, one has
\be
F_\omega(N)\simeq\begin{dcases}
N^2, &\omega\ll\omega_l\\
N^2\left(\frac{\omega}{\omega_l}\right)^{-2}, &\omega_l\ll\omega\ll\omega_m
\end{dcases}.\label{cohl}
\ee
\begin{figure}[H]
\centering
\includegraphics[angle=0,scale=0.8]{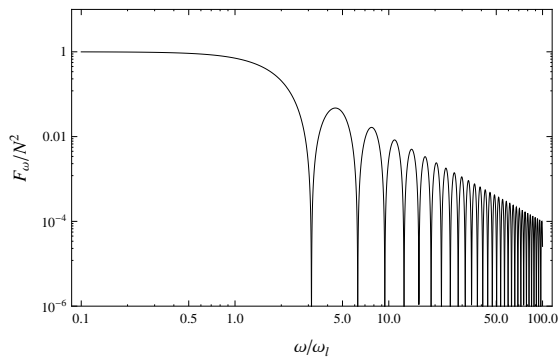}
\caption{$F_\omega$-$\omega$ relation, see Eq.(\ref{bunfact}).}\label{factl}
\end{figure}

The $F_\omega$-$\omega$ relation is shown in Figure \ref{factl}. If $\omega\ll\omega_l$, the wavelength of the electromagnetic waves will be much larger than the bunch length, which means that the radiation from each electron has almost the same phase. In this case, one has significant coherence. On the other hand, if $\omega\gtrsim\omega_l$, the factor of $\sin^2(\omega/\omega_l)/(\omega/\omega_l)^2$ will play a role to reduce coherence, which causes: 1. the maximum value of $F_\omega(N)$ proportional to $\omega^{-2}$; and 2. the spectral oscillation.
At the frequency around $\sim\omega_l$, the spectrum oscillation is significant. 
For higher frequencies, e.g., $\omega\gg 10\omega_l$, since oscillation becomes rapid, the observed spectrum would appear as a power law. 
Finally, once $\omega\gg\omega_m$ is satisfied, the radiation will become incoherent. 

\subsubsection{More than one bunch in the trajectory}\label{sec412}

Next, we consider that there are $N_B$ bunches in the trajectory. For each bunch, the length and the electron number 
are assumed to be $L$ and $N$, respectively. 
We define that $x_j$ is the distance between the $j$th electron and the first electron in each bunch,
and $s_n$ is the distance between the first electron in the $n$th bunch and the first electron in the first bunch. 
If the total length of $N_B$ bunches, including the spaces between each bunch, is much less than the curvature radius $\rho$ of the trajectory, then the angle between each bunch velocity and the line of sight is almost the same, e.g., $\varphi_n\sim\varphi$. Similar to the discussion in the above section, one has
\be
F_\omega(N,N_B)&=&\left|\sum_n^{N_B}\sum_j^N e^{-ik(x_j+s_n)\cos\varphi}\right|^2\nonumber\\
&=&\left|\sum_n^{N_B}e^{-iks_n\cos\varphi}\right|^2\left|\sum_j^N e^{-ikx_j\cos\varphi}\right|^2\nonumber\\
&=&N^2N_B^2\left[\frac{\sin(\omega/\omega_l)}{(\omega/\omega_l)}\right]^2\left[\frac{\sin(\omega/\omega_{bl})}{(\omega/\omega_{bl})}\right]^2,\label{cohfacb}
\ee
where $\omega_{bl}=2c/s_{N_B}\cos\varphi$. Define $L_s$ is the mean space between each bunch, then $s_{N_B}=(N_B-1)(L+L_s)$. 
We also define the maximum inter-bunch coherent frequency, e.g.,
\be
\omega_{bm}\sim\left(\frac{\rho}{s_{N_B}}\right)^2\omega_{bl}.\label{omegabm}
\ee
If $\omega\ll\omega_{bm}$, the superposition of the electromagnetic waves from each bunch will be coherent, and the radiation energy is corrected by the factor of Eq.(\ref{cohfacb}). However, if $\omega>\omega_{bm}$, Eq.(\ref{cohfacb}) is not applicable. In this case, the superposition of the electromagnetic waves from different bunches will not be coherent, one may have $F_\omega(N,N_B)\sim N_BF_\omega(N)$, where $F_{\omega}(N)$ corresponds to one bunch.

As shown in Figure \ref{factl}, for one bunch, the spectrum appears a significant oscillation at $\omega_l$, which can show the discrete band structure in the spectrum. Such a property might explain the narrow spectrum of the nanosecond giant pulse of the Crab pulsar \citep{han07}. However, for more than one bunch, if the time intervals of each bunch satisfies random distribution, the oscillation in the total spectrum would be smoothed.

\subsubsection{Steady current flowing in the entire trajectory}\label{sec413}

Assuming that the electrons are distributed in the entire trajectory, and the charge density and the current density are independent of time. According to Maxwell's equation, the electromagnetic field generated by a steady source is steady, which cannot contribute to radiation, i.e.,
\be
\left.\frac{dI_{\rm current}}{d\omega d\Omega}\right|_{\rm steady}=0.
\ee

In general, a current can be considered to consist of a steady component and some perturbations. Only the fluctuating part can contribute to coherent radiation.
For a rotating neutron star, there should be a background quasi-Goldreich-Julian outflow in the magnetosphere \citep{gol69}. In order to generate radiation, there must be a perturbation in the outflow so that the local charge density deviates from this Goldreich-Julian charge density. On the other hand, according to Section \ref{sec33}, only the net charge contributes to coherent radiation. Therefore, purely introducing an electron-positron pair plasma streaming in the pulsar magnetosphere may not generate coherent emission. It is the perturbation of charge density by the production of pairs that cause a deviation $\rho$ from the quasi-$\rho_{\rm GJ}$ background, and such a deviation is the source of coherent radio emission. We will briefly discuss the bunching mechanism in Section \ref{bunching}.

\subsection{Electrons in a trajectory family}\label{sec42}

\begin{figure}[H]
\centering
\includegraphics[angle=0,scale=0.35]{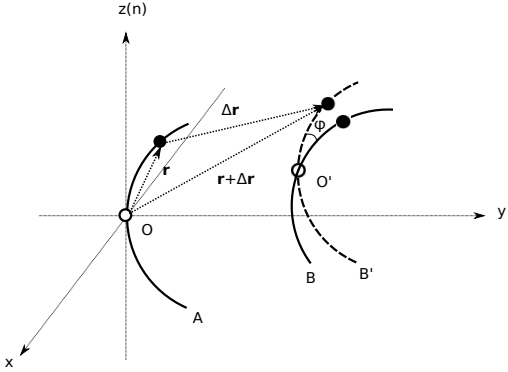}
\caption{A trajectory (B) generated via displacement ($A\rightarrow B'$) and rotation ($B'\rightarrow B$) by the ``seed'' trajectory (A).}\label{multitraj}
\end{figure}
If the charges are not in the same trajectory, a detailed calculation of the coherent emission will be complex. We consider the following ``simplified trajectory-family assumption''\footnote{For the cases not satisfying these ``simplified trajectory-family assumption'', the coherence would be weakened. For example, if the electron trajectories have different curvature radius, the ``beat'' effect will make the amplitude of the coherent wave evolve with time, and the enhanced coherence only happen in a relatively short period, see Section \ref{sec43}. On the other hand, if electrons are not in the same plane when the radiation power is maximum, the coherence will be also weakened due to the extension along the line of sight, see Section \ref{sec411}.}: 1. electrons are in the different trajectories with the same curvature radius; 2. at the retarded time $t=0$, all the electrons are in the plane perpendicular to the line of sight.
Under the above conditions, we consider the appropriate coordinate system in Figure \ref{multitraj}, where all electrons are in the $x-y$ plane at retarded time $t=0$, the direction of the line of sight is along the $z$ axis.

As shown in Figure \ref{multitraj}, for any trajectory under the above condition, e.g. the trajectory B, it can be generated via ``rotation'' (around $x$, $y$ or $z$ axis) or ``displacement'' (in the $x-y$ plane) by a ``seed'' trajectory, i.e. trajectory A in Figure \ref{multitraj}, which is in the $y-z$ plane with the corresponding electron at the origin at $t=0$. Therefore, using the seed trajectory and the transformations, we can generate a family of trajectories. We consider the following basic trajectory families: 1. generated via rotation around $z$ axis by the seed trajectory; 2. generated via rotation around $y$ axis by the seed trajectory; 3. generated via rotation around $x$ axis by the seed trajectory; 4. one of the above three cases adding a displacement.

First, we note that a displacement of a trajectory does not change its radiation spectrum. For a displacement, e.g., $\bm{r}(t)\rightarrow\bm{r}(t)+\Delta\bm{r}$ with $\bm{n}\cdot\Delta\bm{r}=0$, one has
\be
\bm{\beta}&\rightarrow&\bm{\beta},\nonumber\\
\bm{n}\cdot\bm{r}(t)&\rightarrow&\bm{n}\cdot(\bm{r}(t)+\Delta\bm{r})=\bm{n}\cdot\bm{r}(t).\label{traninv}
\ee 
According to Eq.(\ref{radiation}), $dI/d\omega d\Omega$ remains unchanged under the displacement. Therefore, one can shift one trajectory in the plane perpendicular to the line of sight and do not change its observed spectrum \footnote{Note that the displacement invariance is based on the assumption that the scale of the accelerating region is much less that the distance between the source and the observer.}.

For example, if the $N_t$ trajectories keep parallel to each other and satisfy the above simplified trajectory-family assumption, According to the displacement invariance, the corresponding energy radiated per unit solid angle per unit frequency interval is
\be 
\frac{dI_{(N_t)}}{d\omega d\Omega}=N_t^2\frac{dI_{(1)}}{d\omega d\Omega}.
\ee 
Since the displacement does not change the radiation, in the following we only need to consider the three rotation cases, i.e. the trajectory family is generated via: 1. rotation around the $z$ axis; 2. rotation around the $y$ axis; and 3. rotation around the $x$ axis.

\subsubsection{Family I: Generated via rotation around Z axis}\label{sec421}

\begin{figure}[H]
\centering
\includegraphics[angle=0,scale=0.3]{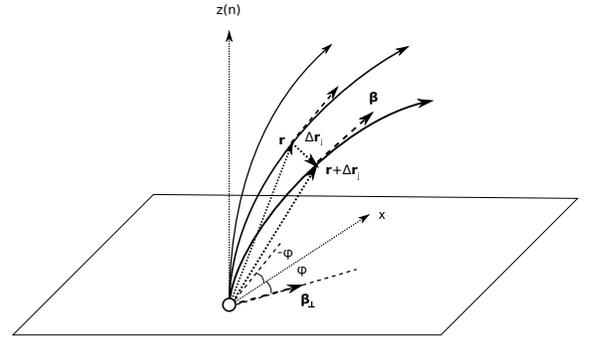}
\caption{A trajectory family generated via rotation around $z$ axis by the seed trajectory.}\label{rotz}
\end{figure}

If the trajectory family is generated via the rotation around $z$ axis, as shown in Figure \ref{rotz}, one has $\bm{n}\cdot\Delta\bm{r}_j=0$ and $\bm{n}\cdot \bm{r}_j(t)=\bm{n}\cdot \bm{r}(t)$, where $\bm{r}(t)$ denotes the seed trajectory, leading to the same exponential term in Eq.(\ref{multiemission}) for each electron.
Define $\bm{\beta}_{\perp,j}$ as the component of $\bm{\beta}_j$ in the plane that is perpendicular to the line of sight, one has,
\be
\bm{n}\times(\bm{n}\times\bm{\beta}_j)=-\bm{\beta}_{\perp,j}.
\ee
For simplicity, we assume that the trajectory family is generated by rotating the seed trajectory (the median trajectory) by $\pm \varphi$, and there are $N_t$ trajectories uniformly spaced in the opening angle $2\varphi$. $\bm{r}(t)$ corresponds to the median trajectory, and $\varphi_j$ corresponds to the angle between the $j$th trajectory and the median trajectory, as shown in Figure \ref{rotz}. 
Since the velocities of the electrons have the same $z$ component, one can define $\bm{\beta}_j=\beta(b\cos\varphi_j,b\sin\varphi_j,\sqrt{1-b^2})$, where $\beta\sqrt{1-b^2}$ corresponds to the $z$ component. Therefore, one has $\bm{n}\times(\bm{n}\times\bm{\beta}_j)=-\beta b(\cos\varphi_j,\sin\varphi_j,0)$. The sum of $\bm{n}\times(\bm{n}\times\bm{\beta}_j)$ in Eq.(\ref{multiemission}) is given by
\be
\sum_{j=1}^{N_t}\bm{n}\times(\bm{n}\times\bm{\beta}_j)&=&\left(\frac{N_t}{2\varphi}\int_{-\varphi}^{\varphi}\cos\varphi'd\varphi'\right)(-\beta b)\hat{\bm{x}}\nonumber\\
&=&\frac{\sin\varphi}{\varphi}N_t(-\bm{\beta}_{\perp}),
\ee
where $\bm{\beta}_{\perp}=-\bm{n}\times(\bm{n}\times\bm{\beta})$ is the perpendicular component of $\bm{\beta}$ of the electron in the median trajectory.
Finally, we have
\be
\frac{dI_{(N_t)}}{d\omega d\Omega}=\frac{dI_{(1)}}{d\omega d\Omega}\left(\frac{\sin\varphi}{\varphi}\right)^2N_t^2.\label{rotI}
\ee
Note that if the trajectory family is axisymmetric, e.g., $\varphi=\pi$ ($\varphi$ is the half-opening angle), the radiation energy would be zero.

\subsubsection{Family II: Generated via rotation around Y axis}\label{sec422}

\begin{figure}[H]
\centering
\includegraphics[angle=0,scale=0.45]{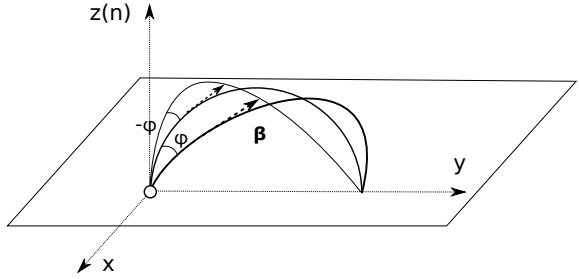}
\caption{A trajectory family generated via rotation around $y$ axis by the seed trajectory.}\label{roty}
\end{figure}

If the trajectory family is generated via the rotation around $y$ axis, the radiation amplitude of one trajectory in the trajectory family can be calculated following Appendix \ref{secb0}, with the observation angle $\theta$ replaced by $\theta+\varphi_j$, where $\varphi_j$ corresponds to the angle between the $j$th trajectory and the median trajectory. We assume that the bunch opening angle of the $N_t$ trajectories is $2\varphi$, and each trajectory is uniformly spaced within the bunch opening angle. The detailed calculation can be found in Appendix \ref{secc}. 

\begin{figure}[H]
\centering
\includegraphics[angle=0,scale=0.25]{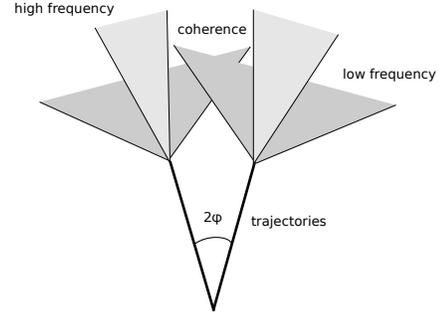}
\caption{Coherent curvature radiation from a bunch opening angle. Dark grey beam denotes low-frequency radiation, and light grey beam denotes high-frequency radiation. If $\omega>\omega_{\varphi}$, only a part of radiation with $\omega$ in the bunch opening angle will be coherent. If $\omega<\omega_{\varphi}$, the radiation from the entire bunch opening angle will be coherent.}\label{omephi}
\end{figure}

First, we consider that there is only one electron in each trajectory, and all electrons have the same Lorentz factor $\gamma$. We define the critical frequency $\omega_\varphi$ by $\theta_c(\omega_\varphi)\simeq\varphi$, where $\theta_c$ is the spread angle of curvature radiation (see Eq.(\ref{angle})). Thus, one has
\be
\omega_\varphi=\frac{3c}{\rho\varphi^3}.\label{omegaphi}
\ee
If $\omega\gg\omega_{\varphi}$, only a part of radiation within the bunch opening angle will be coherent. However, if $\omega\ll\omega_{\varphi}$, the radiation from the entire bunch opening angle will be coherent, as shown in Figure \ref{omephi}.
For $\omega_\varphi<\omega_{c}$, the energy radiated per unit frequency interval per unit solid angle is given by (see Appendix \ref{secc})
\be
\frac{dI}{d\omega d\Omega}&=&\frac{e^2}{c}\frac{3}{2^{4/3}}\left[\frac{\Gamma(2/3)}{\pi}\right]^2N_t^2\gamma^2\nonumber\\
&\times&\begin{dcases}
\left(\frac{\omega}{\omega_c}\right)^{2/3}, &\omega\ll\omega_\varphi\\
\left(\frac{\omega_\varphi}{\omega_c}\right)^{2/3}e^{-\omega/\omega_c}, &\omega\gg\omega_\varphi\label{rotym}
\end{dcases}.\nonumber\\
\ee
For $\omega_\varphi>\omega_{c}$, all the radiation energy in the bunch opening angle can be observed, the radiation energy would be given by Eq.(\ref{scr2}).

Next, we further consider that there are more than one electron in a point source in each trajectory and the electron distribution satisfies the power-law distribution, e.g. $N_e(\gamma)d\gamma=N_{e,0}(\gamma/\gamma_1)^{-p}d\gamma$ for $\gamma_1<\gamma<\gamma_2$, where $N_{e,0}$ corresponds to the normalization for all the trajectories. According to Appendix \ref{secc}, for $\omega_\varphi\ll\omega_{c1}$, the energy radiated per unit frequency interval per unit solid angel is given by
\be
\frac{dI}{d\omega d\Omega}&=&\frac{e^2}{c}\frac{2^{(2p-6)/3}}{3\pi^2}\left[\Gamma\left(\frac{2}{3}\right)\Gamma\left(\frac{p-1}{3}\right)\right]^2N_{e,0}^2\gamma_1^4\nonumber\\
&\times&
\begin{dcases}
\left(\frac{\omega}{\omega_{c1}}\right)^{2/3}, &\omega\ll\omega_\varphi\\
\left(\frac{\omega_\varphi}{\omega_{c1}}\right)^{2/3}, &\omega_\varphi\ll\omega\ll\omega_{c1}\\
\left(\frac{\omega_\varphi}{\omega_{c1}}\right)^{2/3}\left(\frac{\omega}{\omega_{c1}}\right)^{-(2p-2)/3}, &\omega\gg\omega_{c1}
\end{dcases}.\nonumber\\\label{rotIIa}
\ee
For $\omega_{\varphi}\gg\omega_{c1}$, the energy radiated per unit frequency interval per unit solid angle is given by
\be
\frac{dI}{d\omega d\Omega}&=&\frac{e^2}{c}\frac{2^{(2p-6)/3}}{3\pi^2}\left[\Gamma\left(\frac{2}{3}\right)\Gamma\left(\frac{p-1}{3}\right)\right]^2N_{e,0}^2\gamma_1^4\nonumber\\
&\times&
\begin{dcases}
\left(\frac{\omega}{\omega_{c1}}\right)^{2/3}, &\omega\ll\omega_{c1}\\
\left(\frac{\omega}{\omega_{c1}}\right)^{-(2p-4)/3}, &\omega_{c1}\ll\omega\ll\omega_{\varphi}\\
\left(\frac{\omega_\varphi}{\omega_{c1}}\right)^{-(2p-4)/3}\left(\frac{\omega}{\omega_{\varphi}}\right)^{-(2p-2)/3}, &\omega\gg\omega_{\varphi}
\end{dcases}.\nonumber\\\label{rotIIb}
\ee

\subsubsection{Family III: Generated via rotation around X axis}\label{sec423}

\begin{figure}[H]
\centering
\includegraphics[angle=0,scale=0.3]{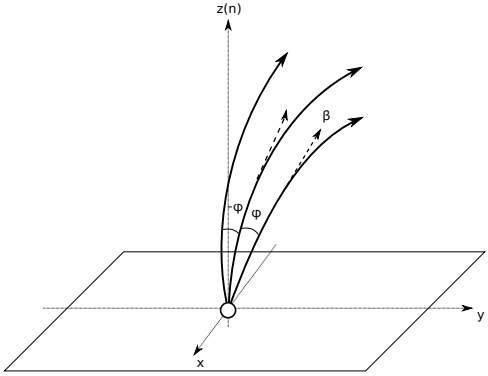}
\caption{A trajectory family generated via rotation around $x$ axis by the seed trajectory.}\label{rotx}
\end{figure}
Finally, we consider that the trajectory family generated via the rotation around $x$ axis. We also assume that the bunch opening angle of the $N_t$ trajectories is $2\varphi$, and each trajectory is uniformly spaced in the bunch opening angle, as shown in Figure \ref{rotx}.
In this case, the radiation spectrum will be as same as Family II (see Appendix \ref{secd}).
For the monoenergetic electron distribution, the radiation energy is given by Eq.(\ref{rotym}) and Eq.(\ref{scr2}). For the power-law electron distribution, the radiation energy is given by Eq.(\ref{rotIIa}) and Eq.(\ref{rotIIb}). 
The detailed calculation is shown in Appendix \ref{secd}.

In summary, for Family II and Family III, the larger the bunch opening angle, the softer the coherent spectrum. The reasons are as follows: the spread angle of the curvature radiation is $\theta_c=(3c/\omega\rho)^{1/3}$ for $\omega\ll\omega_c$. For a given bunch opening angle $2\varphi$, if $\omega>\omega_\varphi$, where $\omega_\varphi$ is defined as $\theta_c(\omega_\varphi)=\varphi$, only a part of radiation in the bunch opening angle is coherent, as shown in Figure \ref{omephi}. As a result, the flux of a high-frequency electromagnetic wave is suppressed due to incoherence. If $\omega<\omega_\varphi$, due to $\theta_c(\omega)>\varphi$, the radiation from the entire bunch opening angle is coherent. Therefore, for a mono-energetic electron distribution, one has $dI/d\omega d\Omega\propto\omega^{2/3}$ if $\omega<\omega_\varphi$, and $dI/d\omega d\Omega\propto\omega^0$ if $\omega>\omega_\varphi$, as shown in Eq.(\ref{rotym}).

\subsection{Electrons in the trajectories with different curvature radii}\label{sec43}

In the above discussions, we have assumed that all the trajectories have the same curvature radius. If the trajectories have different curvature radii, the spectrum of coherent radiation will be more complex. We discuss the simplest case: two electrons are at the origin at the retarded time $t=0$, and their trajectories lie in the same plane and have the same orientation. Assuming that the two electrons have the same energy $\gamma$ and the curvature radii of each trajectory are $\rho_1$ and $\rho_2$, respectively, the corresponding critical frequencies are $\omega_{c1}=3c\gamma^3/2\rho_1$ and $\omega_{c2}=3c\gamma^3/2\rho_2$, respectively. The electromagnetic waves at the critical frequencies are $\bm{E}_{1}=\bm{E}_{01}e^{i\omega_{c1}t}$ and $\bm{E}_{2}=\bm{E}_{02}e^{i\omega_{c2}t}$, respectively. 
We define $\Delta\rho=\rho_2-\rho_1$, $\rho=(\rho_1+\rho_2)/2$, $\Delta\omega_c=\omega_{c2}-\omega_{c1}$, $\omega_c=(\omega_{c1}+\omega_{c2})/2$, and $\bm{E}_0=(\bm{E}_{01}+\bm{E}_{02})/2$. For $\Delta\rho\lesssim\rho$, the superposition of both waves is given by
\be
\bm{E}&=&\bm{E}_{1}+\bm{E}_{2}=\bm{E}_{01}e^{i\omega_{c1}t}+\bm{E}_{02}e^{i\omega_{c2}t}\nonumber\\
&\simeq&\bm{E}_{0}\left(1+e^{i\Delta\omega_{c}t}\right)e^{i\omega_{c}t}.
\ee
Thus, the amplitude of the superposition is 
\be
\bm{E}_b\equiv\bm{E}_{0}\left(1+e^{i\Delta\omega_{c}t}\right),
\ee
with the period of
\be
T_b=\frac{2\pi}{\Delta\omega_c}\simeq\frac{4\pi\rho^2}{3c\gamma^3\Delta\rho}.\label{beatperiod}
\ee
\begin{figure}[H]
\centering
\includegraphics[angle=0,scale=0.8]{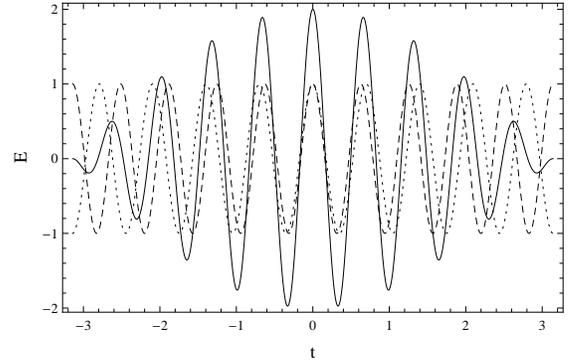} 
\caption{Wave beat: the dashed and dotted lines denote the wave with frequencies $\omega_{c1}$ and $\omega_{c2}$, respectively, and the solid line denotes the superposition of two waves.}\label{beat}
\end{figure}
Such an effect is called ``wave beat''. As shown in Figure \ref{beat}, for the first-half period with $-T_b/4<t<T_b/4$, the amplitude of the coherent wave satisfies $\bm{E}_0<\bm{E}_b<2\bm{E}_0$. For the second-half period with $T_b/4<t<T_b/2$ and $-T_b/2<t<-T_b/4$, the amplitude of the coherent wave satisfies $0<\bm{E}_b<\bm{E}_0$. During the entire beat period $T_b$, the mean amplitude of the coherent wave is $\bm{E}_0$. In order to make the superposition coherently enhanced, the pulse duration $T_p$ (See Eq.(\ref{duration})) of the curvature radiation should be much less than the half period $T_b/2$, e.g., $T_p\ll T_b/2$. 
Therefore, the coherent condition of the trajectories with different curvature radii is
\be
\Delta\rho\ll\frac{4\pi}{3}\rho\left(\frac{\omega}{2\omega_c}\right)^{1/3}.\label{deltarho}
\ee

\section{Curvature radiation from a three-dimensional bunch}\label{sec5}

\begin{figure}[H]
\centering
\includegraphics[angle=0,scale=0.4]{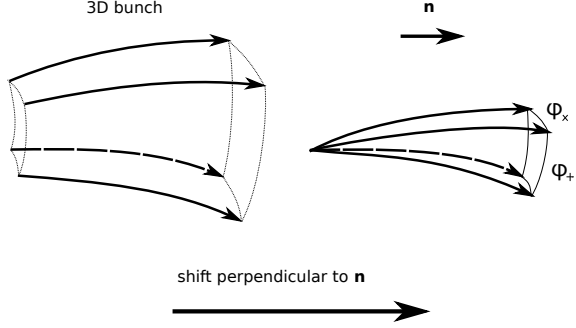}
\caption{Left panel: A three-dimensional bunch characterized by: bunch length, curvature radius and bunch opening angles of the trajectory family. Right panel: Displacement transformation (perpendicular to the line of sight) from the left panel.}\label{3dbun}
\end{figure}

In this section, we consider that the curvature radiation from a three-dimensional bunch  characterized by the following parameters: the electron energy distribution $N_e(\gamma)d\gamma$, the curvature radius of the trajectory family $\rho$ (in order to make the wave coherent, the difference of the curvature radii of the trajectories is required to be very small, see Section \ref{sec43}), the bunch length $L$, and a pair of the orthogonal bunch opening angles $(\varphi_\times, \varphi_+)$ with their centers pointing to the observer, as shown in Figure \ref{3dbun}. 
According to the displacement invariance. e.g., Eq.(\ref{traninv}), we can gather the trajectories in the plane perpendicular to the line of sight, see Figure \ref{3dbun}.
In the simplest case, we assume that the electrons are uniformly distributed in $L$ and $(\varphi_\times, \varphi_+)$, then such a three-dimensional bunch can be treated as the combination of one-dimensional bunch (see Section \ref{sec41}) and two of the three rotation cases (see Section \ref{sec42}). 
\begin{figure}[H]
\centering
\includegraphics[angle=0,scale=0.4]{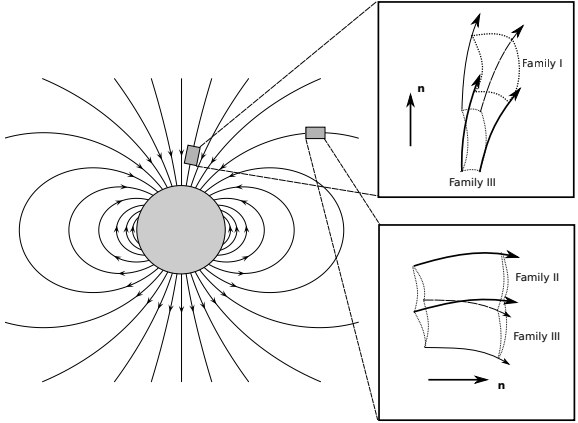}
\caption{Two typical magnetic field configures of a three-dimensional bunch in the magnetosphere. Top-right panel: the field configure in the bunch is consist of Family I and Family III. Bottom-right panel: the field configure in the bunch is consist of Family II and Family III.}\label{dip123}
\end{figure}

For curvature radiation, due to the strong magnetic fields in the magnetosphere near a neutron star, the bunch will move along with the field line. Thus, the trajectories overlap with the field lines. We consider two typical magnetic field configurations for a three-dimensional bunch, as shown in Figure \ref{dip123}. In the case of the top-right panel in Figure \ref{dip123}, the emission region is close to the magnetic axis of the dipole field, and the field configuration in the bunch consists of Family I and Family III, as discussed in Section \ref{sec42}\footnote{Strictly speaking  the field configuration consists of Family I and Family III, only when the emission region is at the center of the dipole field. However, for an emission region close to the magnetic axis, the approximation is reasonable.}. 
In the case of the bottom-right panel in Figure \ref{dip123}, the emission region is near the region where the field is perpendicular to the magnetic axis. The field configuration in the bunch consists of Family II and Family III. 
In the above two cases, $\varphi_\times$ and $\varphi_+$ correspond to the bunch opening angle of the corresponding cases, respectively.

\subsection{Family A: combination of Family I and Family III}\label{sec51}

At first, we consider the case that the emission region is close to the magnetic axis, as shown in the top-right panel of Figure \ref{dip123}.
In this case, the pair of the orthogonal bunch opening angles $(\varphi_\times, \varphi_+)$ are $(2\varphi_1,2\varphi)$, where $\varphi_1$ and $\varphi$ are the bunch half-opening angles of Family I and Family III field lines, respectively.
We also define
\be
K(p)\equiv\frac{2^{(2p-6)/3}}{3\pi^2}\left[\Gamma\left(\frac{2}{3}\right)\Gamma\left(\frac{p-1}{3}\right)\right]^2.
\ee
Without loss of generality, we consider that the upper limit of the frequency, e.g., $\min(\omega_m,\omega_{c2})$, is much larger than other typical frequencies.
According to Section \ref{sec411}, Section \ref{sec421} and Section \ref{sec423}, using Eq.(\ref{cohN}), Eq.(\ref{cohl}), Eq.(\ref{rotI}), Eq.(\ref{rotIIa}) and Eq.(\ref{rotIIb}), the energy radiated per unit frequency interval per unit solid angle could be given by the following formulas.

\begin{figure}[h]
\centering
\includegraphics[angle=0,scale=0.35]{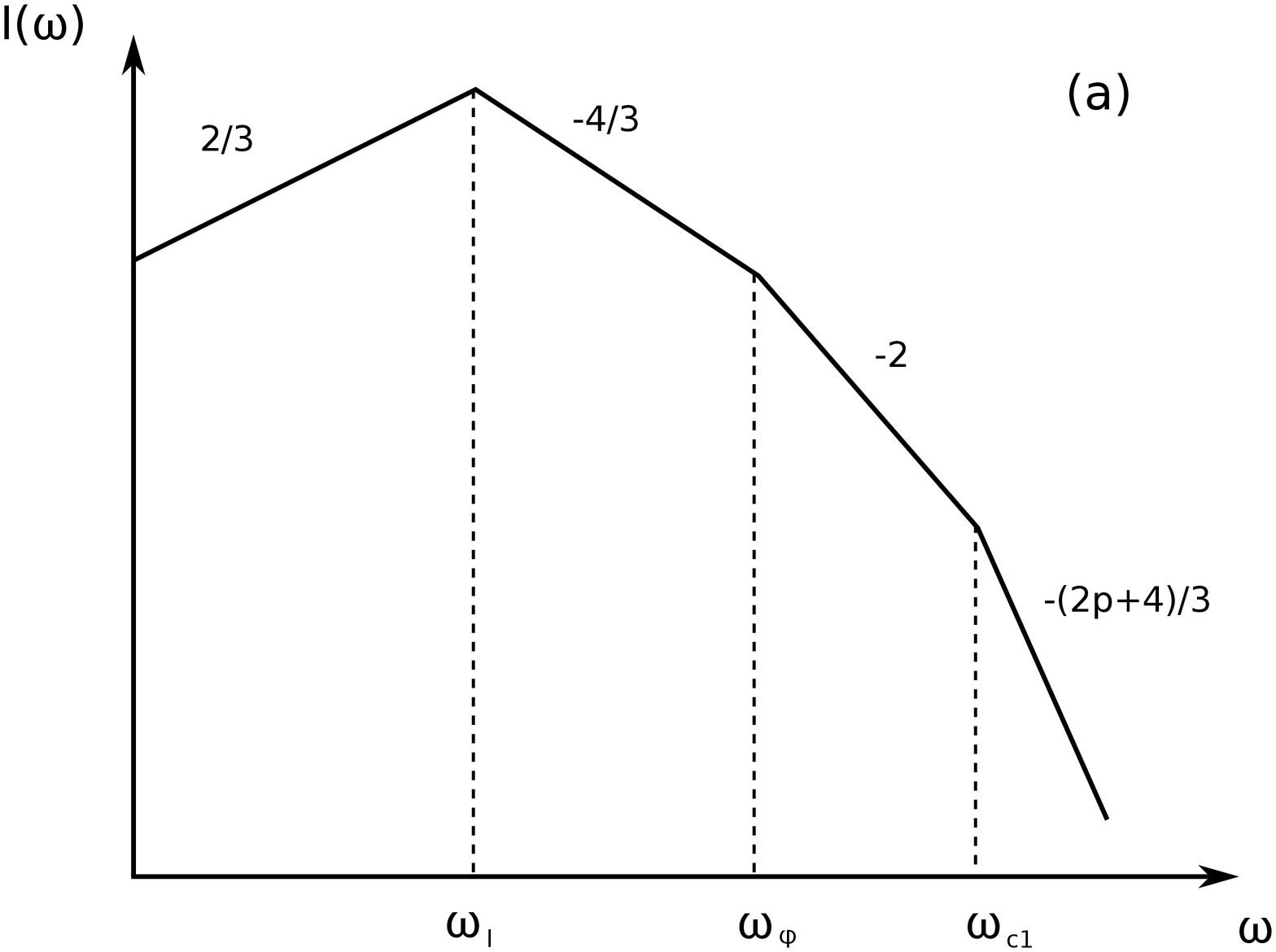}
\includegraphics[angle=0,scale=0.35]{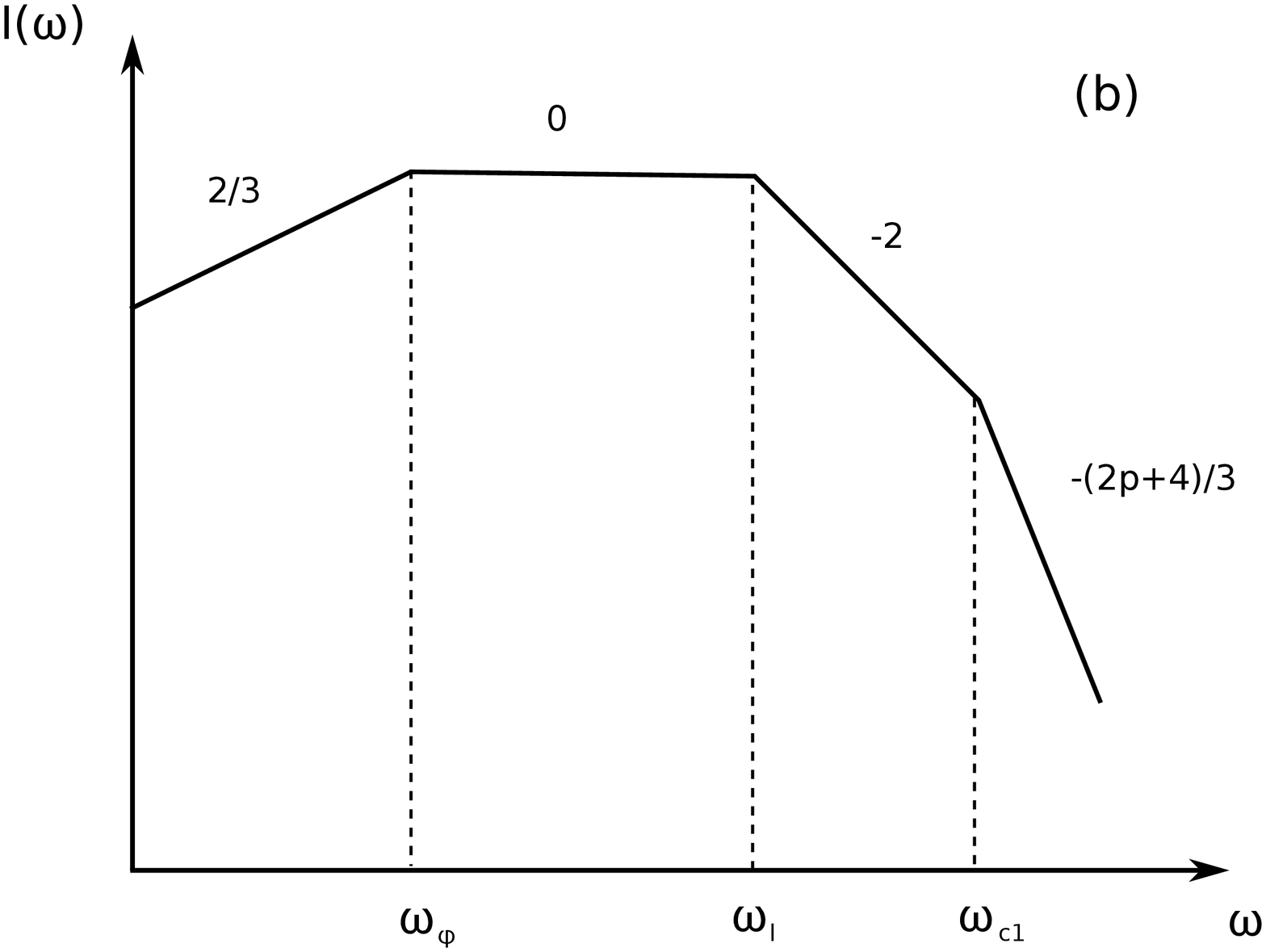}
\includegraphics[angle=0,scale=0.35]{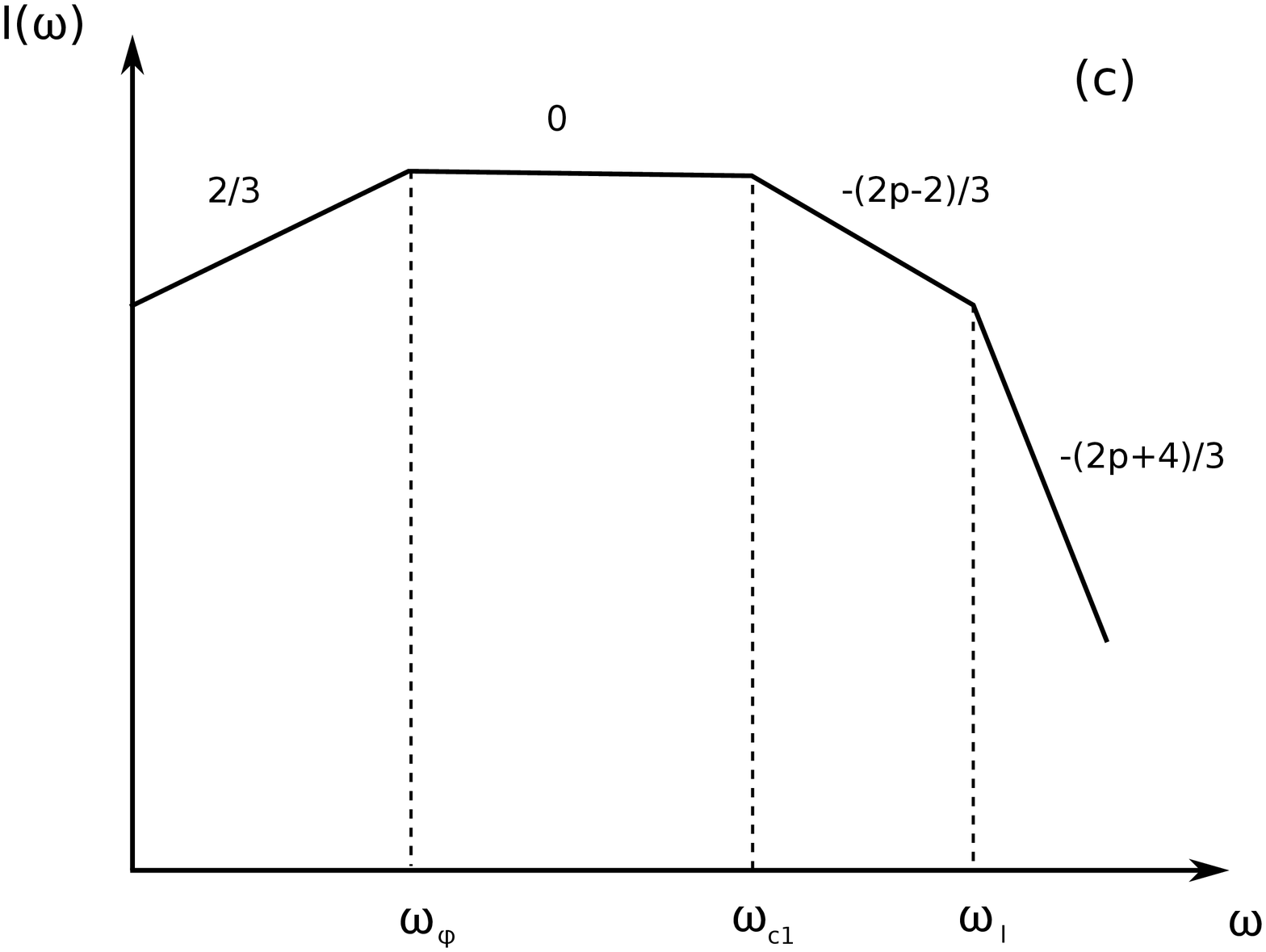}
\caption{The spectra of the coherent curvature radiation from a three-dimensional bunch: (a) the spectrum with $\omega_l\ll\omega_\varphi\ll\omega_{c1}$; (b) the spectrum with $\omega_\varphi\ll\omega_l\ll\omega_{c1}$; (c) the spectrum with $\omega_\varphi\ll\omega_{c1}\ll\omega_l$.}\label{spectra1}
\end{figure}

A. For $\omega_\varphi\ll\omega_{c1}$:

(a) If $\omega_l\ll\omega_\varphi\ll\omega_{c1}$, the spectrum is shown in the panel (a) of Figure \ref{spectra1}, and one has

\be
\frac{dI}{d\omega d\Omega}&=&K(p)\frac{e^2}{c}N_{e,0}^2\gamma_1^4\left(\frac{\sin\varphi_1}{\varphi_1}\right)^2\nonumber\\
&\times&
\begin{dcases}
\left(\frac{\omega}{\omega_{c1}}\right)^{2/3},~~~~~~~~~~~~~~~~~~~~~~~~~~~~~~~~~~~~~~~\omega\ll\omega_l\\
\left(\frac{\omega_l}{\omega_{c1}}\right)^{2/3}\left(\frac{\omega}{\omega_l}\right)^{-4/3},~~~~~~~~~~~~~~~~~~\omega_l\ll\omega\ll\omega_\varphi\\
\left(\frac{\omega_l}{\omega_{c1}}\right)^{2/3}\left(\frac{\omega_\varphi}{\omega_l}\right)^{-4/3}\left(\frac{\omega}{\omega_\varphi}\right)^{-2},~~~\omega_\varphi\ll\omega\ll\omega_{c1}\\
\left(\frac{\omega_l}{\omega_{c1}}\right)^{2/3}\left(\frac{\omega_\varphi}{\omega_l}\right)^{-4/3}\left(\frac{\omega_{c1}}{\omega_\varphi}\right)^{-2}\left(\frac{\omega}{\omega_{c1}}\right)^{-(2p+4)/3},\\
~~~~~~~~~~~~~~~~~~~~~~~~~~~~~~~~~~~~~~~~~~~~~~~~~~~~\omega\gg\omega_{c1}
\end{dcases}\nonumber\\
\ee

(b) If $\omega_\varphi\ll\omega_l\ll\omega_{c1}$, the spectrum is shown in the panel (b) of Figure \ref{spectra1}, and one has
\be
\frac{dI}{d\omega d\Omega}&=&K(p)\frac{e^2}{c}N_{e,0}^2\gamma_1^4\left(\frac{\sin\varphi_1}{\varphi_1}\right)^2\nonumber\\
&\times&
\begin{dcases}
\left(\frac{\omega}{\omega_{c1}}\right)^{2/3},
~~~~~~~~~~~~~~~~~~~~~~~~~~~~~~~~~~~\omega\ll\omega_\varphi\\
\left(\frac{\omega_\varphi}{\omega_{c1}}\right)^{2/3},
~~~~~~~~~~~~~~~~~~~~~~~~~~~~\omega_\varphi\ll\omega\ll\omega_l\\
\left(\frac{\omega_\varphi}{\omega_{c1}}\right)^{2/3}\left(\frac{\omega}{\omega_l}\right)^{-2},
~~~~~~~~~~~~~~~~\omega_l\ll\omega\ll\omega_{c1}\\
\left(\frac{\omega_\varphi}{\omega_{c1}}\right)^{2/3}\left(\frac{\omega_{c1}}{\omega_l}\right)^{-2}\left(\frac{\omega}{\omega_{c1}}\right)^{-(2p+4)/3},\\
~~~~~~~~~~~~~~~~~~~~~~~~~~~~~~~~~~~~~~~~~~~~~~~~~\omega\gg\omega_{c1}
\end{dcases}\nonumber\\
\ee

(c) If $\omega_\varphi\ll\omega_{c1}\ll\omega_l$, the spectrum is shown in the panel (c) of Figure \ref{spectra1}, and one has
\be
\frac{dI}{d\omega d\Omega}&=&K(p)\frac{e^2}{c}N_{e,0}^2\gamma_1^4\left(\frac{\sin\varphi_1}{\varphi_1}\right)^2\nonumber\\
&\times&
\begin{dcases}
\left(\frac{\omega}{\omega_{c1}}\right)^{2/3},
~~~~~~~~~~~~~~~~~~~~~~~~~~~~~~~~~~\omega\ll\omega_\varphi\\
\left(\frac{\omega_\varphi}{\omega_{c1}}\right)^{2/3},
~~~~~~~~~~~~~~~~~~~~~~~~~~\omega_\varphi\ll\omega\ll\omega_{c1}\\
\left(\frac{\omega_\varphi}{\omega_{c1}}\right)^{2/3}\left(\frac{\omega}{\omega_{c1}}\right)^{-(2p-2)/3},
~~~~~\omega_{c1}\ll\omega\ll\omega_l\\
\left(\frac{\omega_\varphi}{\omega_{c1}}\right)^{2/3}\left(\frac{\omega_l}{\omega_{c1}}\right)^{-(2p-2)/3}\left(\frac{\omega}{\omega_l}\right)^{-(2p+4)/3},\\
~~~~~~~~~~~~~~~~~~~~~~~~~~~~~~~~~~~~~~~~~~~~~~~~~\omega\gg\omega_l
\end{dcases}\nonumber\\
\ee

\begin{figure}[h]
\centering
\includegraphics[angle=0,scale=0.35]{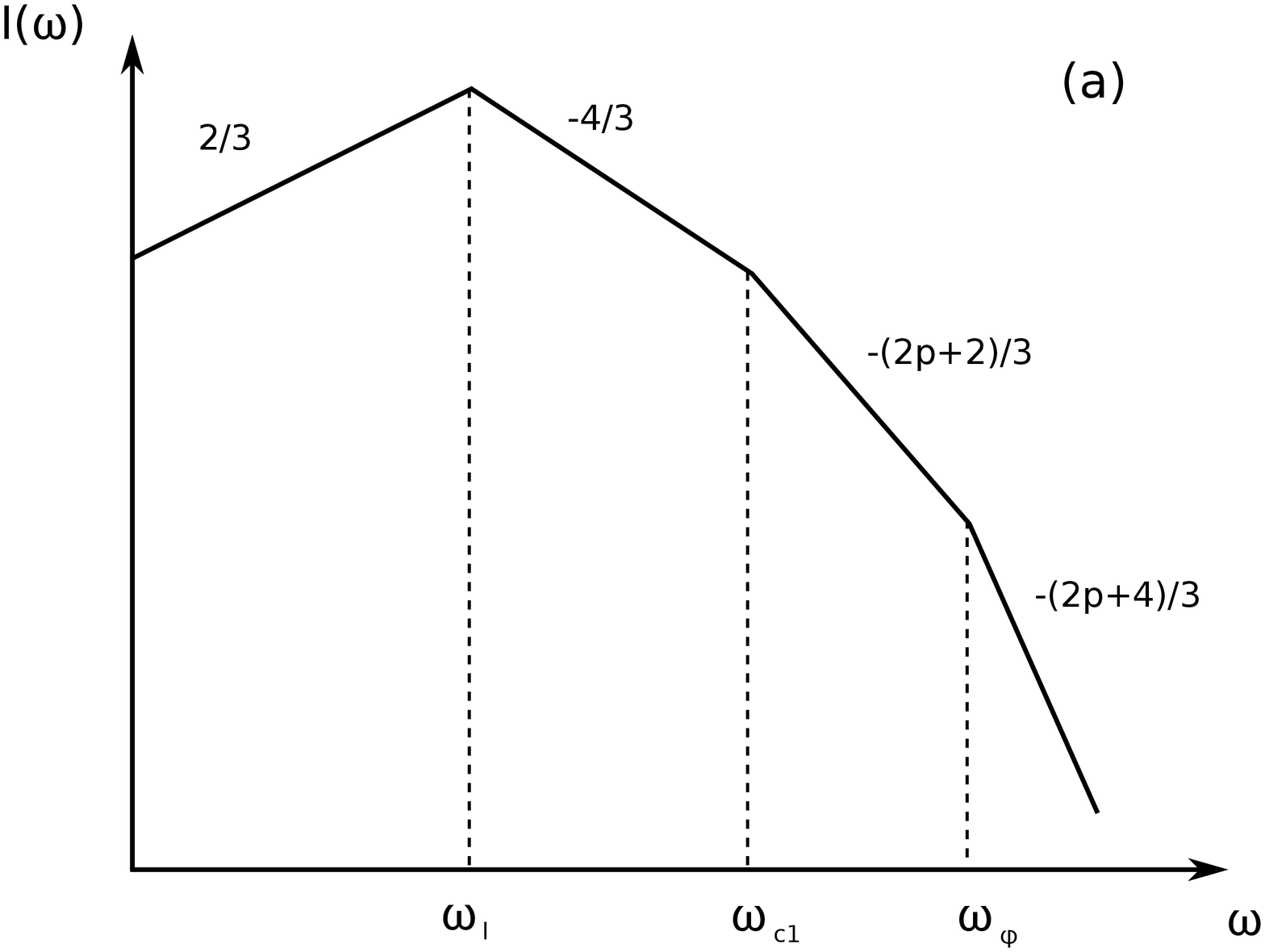}
\includegraphics[angle=0,scale=0.35]{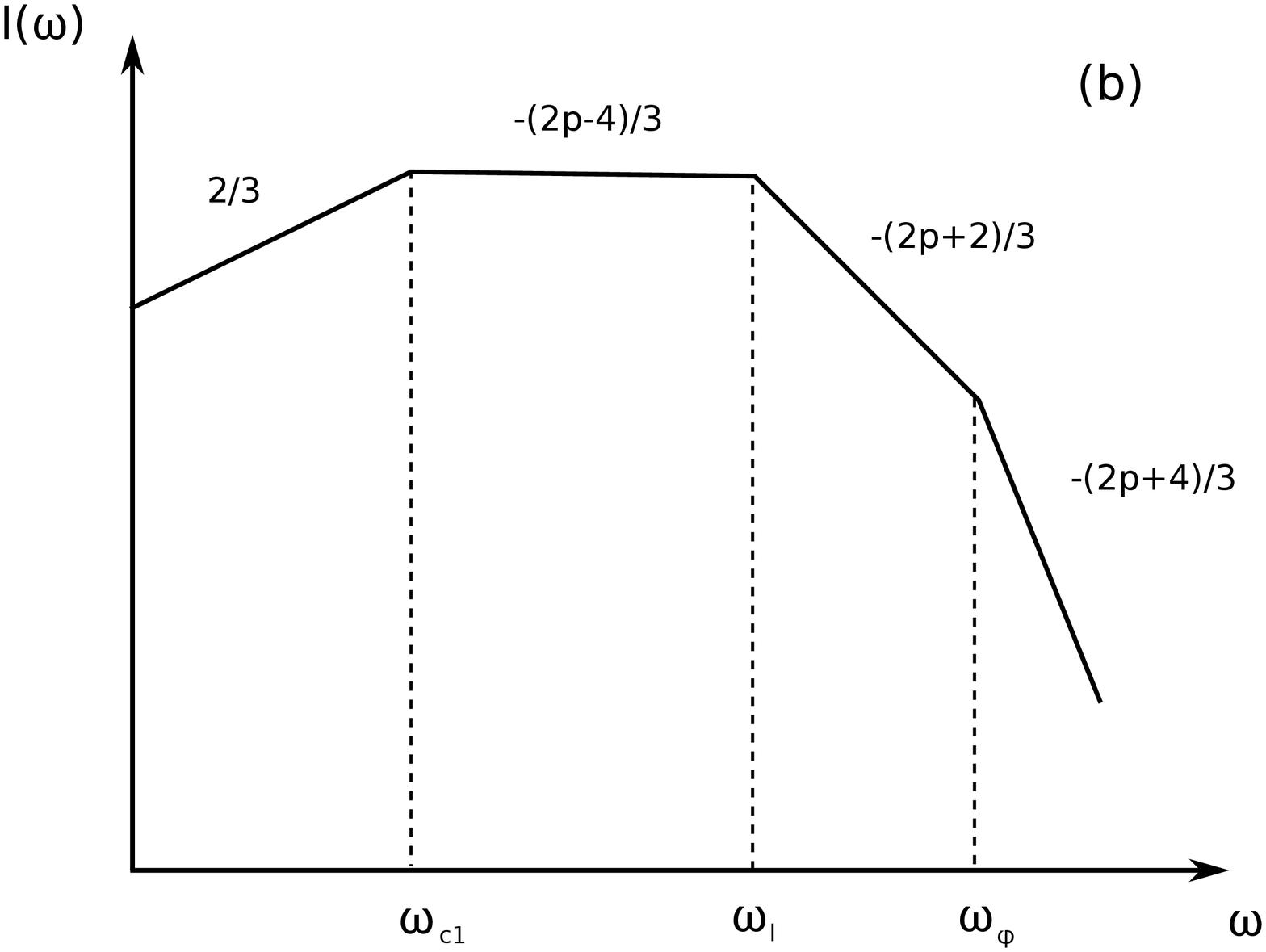}
\includegraphics[angle=0,scale=0.35]{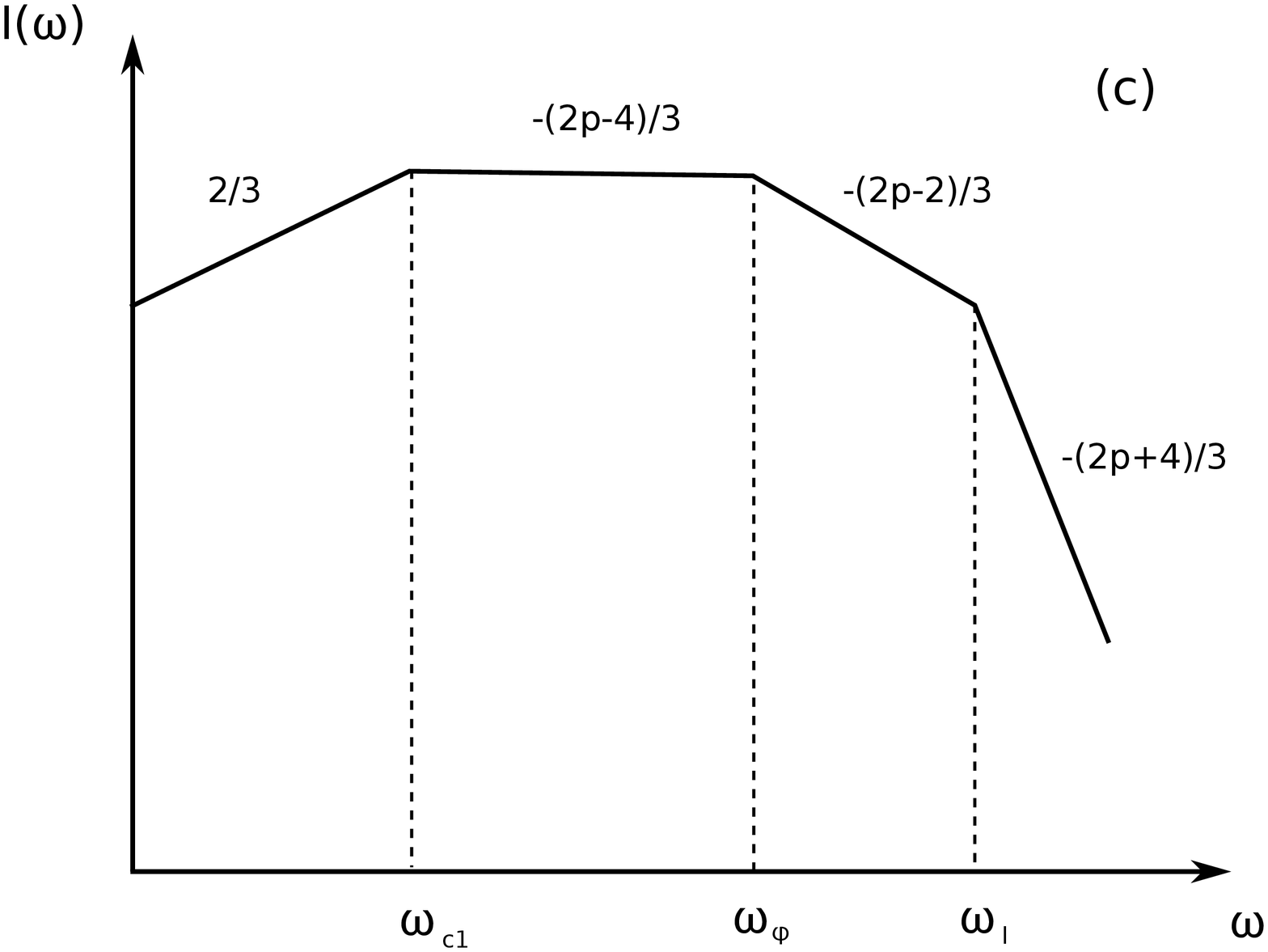}
\caption{The spectra of the coherent curvature radiation from a three-dimensional bunch: (a) the spectrum with $\omega_l\ll\omega_{c1}\ll\omega_\varphi$; (b) the spectrum with $\omega_{c1}\ll\omega_l\ll\omega_\varphi$; (c) the spectrum with $\omega_{c1}\ll\omega_\varphi\ll\omega_l$.}\label{spectra2}
\end{figure}

B. For $\omega_\varphi\gg\omega_{c1}$:

(a) If $\omega_l\ll\omega_{c1}\ll\omega_\varphi$, the spectrum is shown in the panel (a) of Figure \ref{spectra2}, and one has
\be
\frac{dI}{d\omega d\Omega}&=&K(p)\frac{e^2}{c}N_{e,0}^2\gamma_1^4\left(\frac{\sin\varphi_1}{\varphi_1}\right)^2\nonumber\\
&\times&
\begin{dcases}
\left(\frac{\omega}{\omega_{c1}}\right)^{2/3},
~~~~~~~~~~~~~~~~~~~~~~~~~~~~~~~~~\omega\ll\omega_l\\
\left(\frac{\omega_l}{\omega_{c1}}\right)^{2/3}\left(\frac{\omega}{\omega_l}\right)^{-4/3},
~~~~~~~~~~~\omega_l\ll\omega\ll\omega_{c1}\\
\left(\frac{\omega_l}{\omega_{c1}}\right)^{2}\left(\frac{\omega}{\omega_{c1}}\right)^{-(2p+2)/3},
~~~~~\omega_{c1}\ll\omega\ll\omega_\varphi\\
\left(\frac{\omega_l}{\omega_{c1}}\right)^{2}\left(\frac{\omega_\varphi}{\omega_{c1}}\right)^{-(2p+2)/3}\left(\frac{\omega}{\omega_{\varphi}}\right)^{-(2p+4)/3},\\
~~~~~~~~~~~~~~~~~~~~~~~~~~~~~~~~~~~~~~~~~~~~~~~\omega\gg\omega_{\varphi}\\
\end{dcases}\nonumber\\
\ee

(b) If $\omega_{c1}\ll\omega_l\ll\omega_\varphi$, the spectrum is shown in the panel (b) of Figure \ref{spectra2}, and one has
\be
\frac{dI}{d\omega d\Omega}&=&K(p)\frac{e^2}{c}N_{e,0}^2\gamma_1^4\left(\frac{\sin\varphi_1}{\varphi_1}\right)^2\nonumber\\
&\times&
\begin{dcases}
\left(\frac{\omega}{\omega_{c1}}\right)^{2/3},
~~~~~~~~~~~~~~~~~~~~~~~~~~~~~~~~~~~~~~~\omega\ll\omega_{c1}\\
\left(\frac{\omega}{\omega_{c1}}\right)^{-(2p-4)/3},
~~~~~~~~~~~~~~~~~~~~~~~\omega_{c1}\ll\omega\ll\omega_l\\
\left(\frac{\omega_l}{\omega_{c1}}\right)^{-(2p-4)/3}\left(\frac{\omega}{\omega_l}\right)^{-(2p+2)/3},
~~~~\omega_l\ll\omega\ll\omega_\varphi\\
\left(\frac{\omega_l}{\omega_{c1}}\right)^{-(2p-4)/3}\left(\frac{\omega_\varphi}{\omega_l}\right)^{-(2p+2)/3}\left(\frac{\omega}{\omega_\varphi}\right)^{-(2p+4)/3},\\
~~~~~~~~~~~~~~~~~~~~~~~~~~~~~~~~~~~~~~~~~~~~~~~~~~~~~\omega\gg\omega_\varphi
\end{dcases}\nonumber\\
\ee

(c) If $\omega_{c1}\ll\omega_\varphi\ll\omega_l$, the spectrum is shown in the panel (c) of Figure \ref{spectra2}, and one has
\be
\frac{dI}{d\omega d\Omega}&=&K(p)\frac{e^2}{c}N_{e,0}^2\gamma_1^4\left(\frac{\sin\varphi_1}{\varphi_1}\right)^2\nonumber\\
&\times&
\begin{dcases}
\left(\frac{\omega}{\omega_{c1}}\right)^{2/3},
~~~~~~~~~~~~~~~~~~~~~~~~~~~~~~~~~~~~~~~\omega\ll\omega_{c1}\\
\left(\frac{\omega}{\omega_{c1}}\right)^{-(2p-4)/3},
~~~~~~~~~~~~~~~~~~~~~~~\omega_{c1}\ll\omega\ll\omega_\varphi\\
\left(\frac{\omega_\varphi}{\omega_{c1}}\right)^{-(2p-4)/3}\left(\frac{\omega}{\omega_\varphi}\right)^{-(2p-2)/3},
~~~\omega_\varphi\ll\omega\ll\omega_l\\
\left(\frac{\omega_\varphi}{\omega_{c1}}\right)^{-(2p-4)/3}\left(\frac{\omega_l}{\omega_\varphi}\right)^{-(2p-2)/3}\left(\frac{\omega}{\omega_l}\right)^{-(2p+4)/3},\\
~~~~~~~~~~~~~~~~~~~~~~~~~~~~~~~~~~~~~~~~~~~~~~~~~~~~~~\omega\gg\omega_l\\
\end{dcases}\nonumber\\
\ee

\subsection{Family B: combination of Family II and Family III}\label{sec52}

Next, we consider the emission region is near the region where the field is perpendicular to the magnetic axis, as shown in the bottom-right panel of Figure \ref{dip123}.
In this case, the pair of the orthogonal bunch opening angles $(\varphi_\times, \varphi_+)$ are $(2\varphi_2,2\varphi_3)$, where $\varphi_2$ and $\varphi_3$ are the bunch half-opening angles of Family II and Family III, respectively. Noticing that the spectral properties of Family II and Family III field lines are the same, we define\footnote{Note that the quantity $\varphi$ here is different from that of Family A in Section \ref{sec51} (for Family A, $\varphi$ is defined as the half-opening angle of Family III).}
\be
\varphi=\max(\varphi_2,\varphi_3).\label{openangleb}
\ee
According to Section \ref{sec411}, Section \ref{sec422} and Section \ref{sec423}, using Eq.(\ref{cohN}), Eq.(\ref{cohl}), Eq.(\ref{rotIIa}) and Eq.(\ref{rotIIb}), the energy radiated per unit frequency interval per unit solid angle has the same form with Family A but without the factor $(\sin\varphi_1/\varphi_1)^{2}$, i.e.
\be
\left.\frac{dI}{d\omega d\Omega}\right|_{\rm Family~B}=\left(\frac{\sin\varphi_1}{\varphi_1}\right)^{-2}\left.\frac{dI}{d\omega d\Omega}\right|_{\rm Family~A}
\ee
where the subscript ``Family A'' corresponds to the results in Section \ref{sec51}. 

\section{Application to pulsars}\label{sec6}

\subsection{General considerations}\label{sec61}

In general, the observed duration of one sub-pulse from a pulsar (also the duration of an FRB),
e.g., $T_{\rm obs}\sim1~\unit{ms}$, is much longer than the pulse duration of the curvature radiation, e.g., $T_p\sim1/\nu_c\sim1~\unit{ns}~(\nu_c/1~\unit{GHz})$ (see Eq.(\ref{duration})). This means that there must be numerous bunches sweeping cross the line of sight during the observed duration $T_{\rm obs}$. 
The distance between the first bunch and the last bunch is $s_{N_B}\sim cT_{\rm obs}\sim 3\times10^7~\unit{cm}(T_{\rm obs}/1~\unit{ms})$. According to Section \ref{sec412}, the maximum inter-bunch coherent frequency becomes $\omega_{bm}\sim(\rho/s_{N_B})^2\omega_{bl}\sim2\times10^8~\unit{rad~s^{-1}}(T_{\rm obs}/1~\unit{ms})^{-3}(\rho/10^{10}~\unit{cm})^2$, where  $\omega_{bl}\sim2c/s_{N_B}$, which is smaller than the maximum coherent frequency $\omega_{bl}$ of individual bunches. For the GHz radio waves, one has $\omega>\omega_{bm}$, which means that the superposition of the electromagnetic waves from each bunch is not coherent. 
Therefore, according to Eq.(\ref{power}), the observed flux, the energy received per unit time per unit frequency per unit area, is given by 
\be
F_{\nu}=\frac{2\pi}{TD^2}\frac{dI}{d\omega d\Omega},\label{flux}
\ee
where the factor of $2\pi$ is from the relation $F_\nu=2\pi F_{\omega}$, $D$ is the distance between source and observer, and $T$ is the mean time interval between adjacent bunches. 
We assume that the distance scale of the gap between adjacent bunches is of the order of the bunch scale itself, which gives $T\sim L/c$.

\subsection{Bunching mechanism}\label{bunching} 

Before performing a more quantitative calculation of the curvature radiation spectra, we briefly summarize the possible mechanisms to form bunches in the magnetosphere of a pulsar (and an FRB if it originates from the magnetosphere of a rotating neutron star). 

The outflow from the pulsar polar cap region is likely unsteady. Due to interplay between near-surface parallel electric field and binding of particles from the surface, the pulsar inner gap likely produce non-stationary sparks \citep{rud75,zhang96,gil00,gil06}. These sparks of electron-positron pairs have spatial and temporal structures. The two-stream instability would be triggered in the inhomogeneous pulsar plasma when the outflowing plasma clouds disperse and overlap with each other \citep{mel81,uso87,urs88}, and electrostatic Langmuir waves are further triggerred in the magnetosphere. The nonlinear evolution of the unstable electrostatic oscillations results in the formation of plasma solitons \citep{pat80}. Due to relative streaming of electrons and positrons and the corresponding difference in relativistic masses, the net charge of plasma solitons will result from the ponderomotive Miller force that acts on them at different rates, which give rise to coherent curvature radiation \citep{mel00}. According to \cite{mel00}, each soliton consists of three bunches with charges of opposite signs. This is because the excess of one charge is  compensated by the lack of this charge in the nearby regions. \cite{mel00} estimated that $\sim 10^5$ solitons in $\sim 25$ sparks can produce a brightness temperature of the typical radio pulsar. In this estimate, the solitons are regarded as one single charge without considering the spatial dimension. The spectrum of the radio emission is also not calculated. In the following, we will apply the three-dimensional coherent curvature radiation theory developed in \S\ref{sec5} to study radio pulsar emission in detail.

\subsection{Curvature radiation from a dipole magnetosphere}\label{sec63}

\begin{figure}[H]
\centering
\includegraphics[angle=0,scale=0.25]{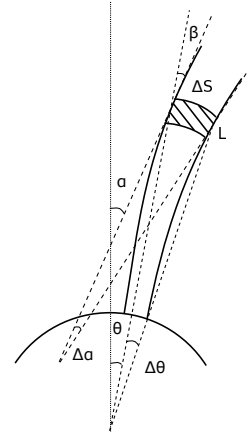}
\caption{Curvature radiation from the dipole magnetosphere. The shadow area denotes a bunch with length $L$ and cross section $\Delta S$. $\theta$ is the poloidal angle of the dipole field, and $\alpha$ denotes the angel between the magnetic axis and the magnetic field. The emission region is close to the magnetic axis.}\label{pulsarbunch}
\end{figure}

First, we consider that the emission region is close to the magnetic axis of a dipole field, as shown in Figure \ref{pulsarbunch}. A bunch, with the length $L$ and the bunch opening angles $(\Delta\phi,\Delta\alpha)$, moves along the field lines. The bunch opening angles $(\Delta\phi,\Delta\alpha)$ are determined by the magnetic field configuration in the bunch, where $\alpha$ denotes the angel between the magnetic axis and the magnetic field, and $\phi$ denotes the toroidal angle of the dipole field. In this case, the magnetic field configuration is consistent with  Family A, and the spectrum is described in Section \ref{sec51}.  
For a given observed direction $\theta$ (the poloidal angle of the dipole field) and curvature radius $\rho$, the emission region $(r,\theta)$ can be determined by the magnetosphere geometry (see Appendix \ref{sece}), one has
\be
r\simeq\frac{3}{4}\rho\sin\theta,~~~~~{\rm for}~~\theta\lesssim0.5
\ee

In this case, according to Section \ref{sec51} and Eq.(\ref{flux}), the observed flux, the energy received per unit time per unit frequency per unit area, at the peak frequency $\nu_{\rm peak}$ is given by
\be
F_{\nu,\max}&=&\frac{2\pi}{TD^2}\left.\frac{dI}{d\omega d\Omega}\right|_{\max}\nonumber\\
&\simeq&\frac{2\pi e^2}{c} K(p) \frac{N_{e,0}^2\gamma_{1}^4}{D^2T}\left(\frac{\sin\Delta\phi}{\Delta\phi}\right)^2\left(\frac{\nu_{\rm peak}}{\nu_{c1}}\right)^{2/3},\nonumber\\\label{fmax}
\ee
where the peak frequency is given by $\nu_{\rm peak}=\min(\nu_l,\nu_\varphi,\nu_{c1})$, where
\be
\nu_l=\frac{c}{\pi L},~~~\nu_\varphi=\frac{3c}{2\pi\rho(\Delta\alpha/2)^3},~~~\nu_{c1}=\frac{3c\gamma_1^3}{4\pi\rho}.\label{chafre}
\ee
Note that $\nu_\varphi$ is defined in Family A.
As discussed in Section \ref{sec33} and Section \ref{sec413}, only the fluctuating net charges in the Goldreich-Julian outflow can make the contribution to coherent radiation.
We define the effective electron number as $N_{e,\rm{eff}}$, which corresponds to the fluctuating net charge number in a bunch.
The net charge density of the bunch is $n_{\rm bun}=n_{\rm GJ}+\delta n_{\rm GJ}\equiv(1+\mu_c)n_{\rm GJ}$, where $n_{\rm GJ}=(\Omega B_p/2\pi ec)(r/R)^{-3}$ is the Goldreich-Julian density \citep{gol69}, $B_p$ is the magnetic field strength at the polar cap, $R$ is the neutron star radius, $\Omega=2\pi/P$ is the angular velocity of the neutron star, $P$ is the rotation period, and 
\be
\mu_c \equiv \frac{\delta n_{\rm GJ}}{n_{\rm GJ}}\label{muc}
\ee
is the normalized fluctuating net charge density $\delta n_{\rm GJ}$ that contributes to coherent radiation. Here, we have assumed that the magnetic axis is parallel to the rotation axis.
Therefore, for a power-law distribution of the effective electron number, e.g., $N_{e}(\gamma)d\gamma=N_{e,0}(\gamma/\gamma_1)^{-p}d\gamma$ with $N_{e,{\rm eff}}=\int N_e(\gamma) d\gamma$, the effective electron number in the bunch volume $V$ is given by
\be
N_{e,{\rm eff}}=\gamma_1N_{e,0}/(p-1)=\mu_c n_{\rm GJ}V.
\ee
The normalization of the electron distribution is given by
\be
N_{e,0}=(p-1)\gamma_{1}^{-1}\mu_c n_{\rm GJ}V.
\ee
As shown in Figure \ref{pulsarbunch}, according to the dipole magnetosphere geometry (see Appendix \ref{sece}), the bunch volume with the length $L$ and the bunch opening angles $(\Delta\phi,\Delta\alpha)$ can be approximately given by
\be
V=L\Delta S\simeq Lr^2\frac{\sin\theta}{\cos\beta}\Delta\theta\Delta\phi\simeq\frac{2}{3}Lr^2\sin\theta\Delta\alpha\Delta\phi,\label{volume}
\ee
where $\Delta S$ denotes the cross section, and $\beta$ is the angle between the radial direction and the magnetic field, which is given by Eq.(\ref{beta}). The relation between $\Delta\alpha$ and $\Delta\theta$ is given by Eq.(\ref{alpthe}). 

Next, we constrain $\Delta\alpha$ and $\Delta\phi$.
In order to make the electromagnetic waves from the field lines with different curvature radii coherent, according to Eq.(\ref{deltarho}), the difference of the curvature radii should satisfy $\Delta\rho\ll\rho$ for $\omega\lesssim\omega_c$. For a given field line length $l$ (from the dipole center to the emission region), the curvature radius at $\theta$ is approximately $\rho\sim8l/3\theta$ (see Eq.(\ref{rhoL}) in Appendix \ref{sece}), leading to $\Delta\theta/\theta\sim\Delta\rho/\rho\ll1$. Using the relation $\Delta\alpha\sim(3/2)\Delta\theta$, the bunch opening angle $\Delta\alpha$ can be adopted as 
\be
\Delta\alpha\lesssim0.1\alpha\sim0.15\theta.
\ee
For $\Delta\alpha>0.1\alpha\sim0.15\theta$, the curvature radius would significantly change, which means that the electromagnetic waves from different field lines will not be coherent.
On the other hand, different from $\Delta\alpha$, $\Delta\phi$ can be relatively larger. There are two reasons: 1. for a given poloidal angle $\theta$, the field lines with different toroidal angle $\phi$ have the same curvature radius; 2. for a given $(r,\theta)$, the angle between the magnetic field direction and the line of sight (the line of sight is taken to be tangent with the intermediate field line in the bunch opening angle), $\psi$, is always small, e.g.
\be
\cos\psi=\cos\frac{\Delta\phi}{2}\sin^2\alpha+\cos^2\alpha
\simeq1-\alpha^2(1-\cos\frac{\Delta\phi}{2}).\nonumber\\
\ee
Therefore, even for a relatively large $\Delta\phi$, one always has $\psi\ll1$ for $\alpha\simeq(3/2)\theta\ll1$, so that all the approximate conditions in the curvature radiation ($\psi$ corresponds to $\theta$ in Section \ref{sec2}) are satisfied.
Finally, the peak flux can be written as
\be
F_{\nu,\max}&=&\frac{32(p-1)^2K(p)}{81\pi c^2}\frac{\mu_c^2\Omega^2B_p^2R^6L\gamma_1^2\Delta\alpha^2\Delta\phi^2}{\rho^2D^2}\nonumber\\
&\times&\left(\frac{\sin\Delta\phi}{\Delta\phi}\right)^2\left(\frac{\nu_{\rm peak}}{\nu_{c1}}\right)^{2/3},
\ee
where
\be
\frac{\nu_{\rm peak}}{\nu_{c1}}=\min\left(\frac{4\rho}{3L\gamma_1^3},\frac{16}{\Delta\alpha^3\gamma_1^3},1\right).
\ee
There are three cases for the peak flux:
\begin{itemize}
\item Case I: for $\nu_{\rm peak}=\nu_l$, one has
\be
F_{\nu,\max}&=&\frac{32(p-1)^2K(p)}{81\pi c^2}\left(\frac{4}{3}\right)^{2/3}\nonumber\\
&\times&\frac{\mu_c^2\Omega^2B_p^2R^6L^{1/3}\Delta\alpha^2\Delta\phi^2}{\rho^{4/3}D^2}\left(\frac{\sin\Delta\phi}{\Delta\phi}\right)^2.
\ee
\item Case II: for $\nu_{\rm peak}=\nu_\varphi$, one has
\be
F_{\nu,\max}&=&\frac{2\cdot16^{5/3}(p-1)^2K(p)}{81\pi c^2}\nonumber\\
&\times&\frac{\mu_c^2\Omega^2B_p^2R^6L\Delta\phi^2}{\rho^2D^2}\left(\frac{\sin\Delta\phi}{\Delta\phi}\right)^2.
\ee
\item Case III: for $\nu_{\rm peak}=\nu_{c1}$, one has
\be
F_{\nu,\max}&=&\frac{32(p-1)^2K(p)}{81\pi c^2}\nonumber\\
&\times&\frac{\mu_c^2\Omega^2B_p^2R^6L\gamma_1^2\Delta\alpha^2\Delta\phi^2}{\rho^2D^2}\left(\frac{\sin\Delta\phi}{\Delta\phi}\right)^2.\label{pc3}
\ee
\end{itemize}

\subsection{Model confronting pulsar data}

Observationally, the spectra of pulsars can be fitted by a single power law, a two-segment broken power law, or a multi-segment broken power law (or log-parabolic) form. Eighty percent of pulsars appear to have a single power-law spectrum. The spectral index is around $-(3-0)$ with the mean value of $-1.6$ \citep{lor95,jan17}. 
Seven percent of pulsars appear to have a two-segment broken power law. They show the mean spectral indices of $-1.55$ and $-2.72$ , respectively, before and after the spectral break at $\sim 1~\unit{GHz}$, with both indices having large scatter \citep{xil96}.
Ten percent of pulsars can be fitted via a multi-segment broken power law or a log-parabolic spectral model \citep{jan17}. These spectra appear more complex than the above two classes.

As discussed in Section \ref{sec5}, the spectrum of the curvature radiation from a bunch moving in a three-dimensional field naturally predicts a multi-segment broken power law spectrum. The break frequencies are determined by the curvature radius, bunch length, and bunch opening angles. In general, the observed frequency band is narrow, which means that the observed spectra might be a part of a multi-segment broken power law. Thus, our model can naturally explain the observed spectra of pulsars. 

For example, we consider the following typical parameters of a pulsar: $D=5~\unit{kpc}$, $R=10^6~\unit{cm}$, $B_p=10^{12}~\unit{G}$, and $P=0.1~\unit{s}$. We introduce a moderate fluctuation parameter $\mu_c=0.1$. Other parameters are adopted as $L=10~\unit{cm}$, $\gamma_1=1000$, $\rho=10^{10}~\unit{cm}$, $p=3$,, $\theta=0.01$, and $\Delta\phi=0.1$.
In this case, the distance from the emission region to the dipole field center is $r=7.5\times10^7~\unit{cm}$, the typical frequencies are $\nu_{c1}=0.7~\unit{GHz}$, $\nu_{l}=0.9~\unit{GHz}$, $\nu_{\varphi}=3.4~\unit{GHz}$, and the observed flux is $F_{\nu,\max}=2~\unit{mJy}$. 
Due to $\nu_{c1}<\nu_{l}<\nu_{\varphi}$, the predict spectral index $\alpha_{\rm index}$ is shown in the panel (b) of Figure \ref{spectra2}: the spectral index is $\alpha_{\rm index}\sim0.7$ for $\nu\lesssim0.7~\unit{GHz}$; $\alpha_{\rm index}\sim-0.7$ for $0.7~\unit{GHz}\lesssim\nu\lesssim0.9~\unit{GHz}$; $\alpha_{\rm index}\sim-2.7$ for $0.9~\unit{GHz}\lesssim\nu\lesssim3.4~\unit{GHz}$; $\alpha_{\rm index}\sim-3.3$ for $\nu\gtrsim3.4~\unit{GHz}$. Thus, for the above parameters, the spectrum shows a multi-segment broken law near $\sim1~\unit{GHz}$.   

These are generally consistent with the pulsar data. For the pulsars with the observed spectra having a single power law and a two-segment broken power law, they can also be explained by this model as long as the break frequencies, e.g., $\nu_{c1}$, $\nu_l$ and $\nu_\varphi$, have relatively large separations so that within the observed frequency band only zero or one break are observable. These can be achieved with reasonable pulsar parameters.

\section{Application to fast radio bursts}\label{sec7}

\subsection{Model A: Spindown powered scenario}\label{sec71}

Fast radio bursts (FRBs) are mysterious radio transients characterized by millisecond-duration durations, large dispersion measure, and extremely high brightness temperature \citep[e.g.][]{lor07,tho13,cha17}.
Thanks to multi-wavelength follow-up observations and a precise localization \citep{cha17,mar17}, the repeating FRB, FRB 121102, was identified in a dwarf galaxy at $z=0.19273$ \citep{ten17} surrounded by a persistent radio counterpart \citep{cha17,mar17}.
The observation of nine VLA bursts from FRB 121102 showed that the spectra of FRB 121102 are narrow, which are characterized by a $\sim3~\unit{GHz}$ peak frequency width of roughly $\sim500~\unit{MHz}$ \citep{law17}. 

Since the coherent curvature radiation by bunches always emits a wide intrinsic spectrum, e.g., $\Delta\nu/\nu\sim1$, the observed narrow spectra might result from the absorption of low-frequency radio emission. As discussed in Section \ref{sec5}, at high frequencies, e.g. $\nu\gtrsim\max(\nu_{c1},\nu_{l},\nu_{\varphi})$, the spectral index is approximately $-(2p+4)/3$. Thus, if $\nu_a\gtrsim\max(\nu_{c1},\nu_{l},\nu_{\varphi})$, where $\nu_a$ is the absorption frequency, the observed spectra would be narrow.

Since FRBs have a much higher brightness temperature than that of pulsar, the fluctuating net charge number in a bunch need be much larger. Given the abrupt nature of FRBs, it is not unreasonable to introduce $\mu_c = \delta n_{\rm GJ}/n_{\rm GJ} \sim 1$ or even larger. If one limits $\mu_c = 1$, the extremely high brightness temperature of FRBs still require a neutron star with a stronger magnetic field and a faster rotation than normal pulsars, with the emission region close to the neutron star. This conclusion is similar to \cite{kum17}, although the details to achieve this conclusion are different.
We adopt the following typical parameters: $D=1~\unit{Gpc}$, $R=10^6~\unit{cm}$, $B_p=10^{14}~\unit{G}$, and $P=10~\unit{ms}$\footnote{Notice that we did not invoke an even stronger magnetic field or an even shorter spin period. This is because the spindown time scale of those rapidly spinning magnetars would be shorter than the observation time of FRB 121102, which is of the order of several years.}.
The model parameters are assumed as: $L=10~\unit{cm}$, $\gamma_1=200$, $\rho=3\times10^{7}~\unit{cm}$, $p=3$,  $\theta=0.1$, and $\Delta\phi=0.1$.
In this case, the distance from the emission region to the dipole field center is $r=2.2\times10^6~\unit{cm}$, the typical frequencies are $\nu_{c1}=1.9~\unit{GHz}$, $\nu_{l}=0.9~\unit{GHz}$, $\nu_{\varphi}=1.1~\unit{GHz}$, and the intrinsic maximum flux is $F_{\nu,\max}=1.6~\unit{Jy}$. 
Due to $\nu_l<\nu_{\varphi}<\nu_{c1}$, the intrinsic spectral index $\alpha_{\rm index}$ is shown in the panel (a) of Figure \ref{spectra1}: $\alpha_{\rm index}\sim0.7$ for $\nu\lesssim0.9~\unit{GHz}$; $\alpha_{\rm index}\sim-1.3$ for $0.9~\unit{GHz}\lesssim\nu\lesssim1.1~\unit{GHz}$; $\alpha_{\rm index}\sim-2$ for $1.1~\unit{GHz}\lesssim\nu\lesssim1.9~\unit{GHz}$; $\alpha_{\rm index}\sim-3.3$ for $\nu\gtrsim1.9~\unit{GHz}$. 

Observations showed that there is a persistent radio counterpart around the repeating FRB source FRB 121102 \citep{cha17,mar17,ten17}. According to \citet{yan16}, if the FRB frequency is below the synchrotron self-absorption (SSA) frequency of the nebula, electrons in the nebula would absorb the FRB photons, leading to enhanced self-absorbed synchrotron emission, which might explain the persist radio emission of FRB 121102. For such a synchrotron nebula, its luminosity is approximately $\mathcal{L}\simeq4\pi r^2[\nu_a\pi I_\nu(\nu_a)]$, where $r$ is the nebula radius \citep{yan16}, $I_\nu\simeq(2m_e/\sqrt{3}\nu_B^{1/2})\nu^{5/2}(1-\exp(-\tau_\nu))$ is the SSA intensity, $\nu_B=eB/2\pi m_ec$ is the electron cyclotron frequency, and $\nu_a$ is the SSA frequency, which is defined by $\tau_\nu(\nu_a)=1$.
On the other hand, the bursts with $\nu<\nu_a$ will be absorbed by the nebula, leading to a low-frequency cutoff. 
Thus, the synchrotron self-absorption luminosity $\mathcal{L}$ and the nebula magnetic field $B$ can be constrained via \citep{yan16}
\be
\nu_{\rm obs}>\nu_a&\simeq&1.8~\unit{GHz}\left(\frac{B}{1~\unit{\mu~G}}\right)^{1/7}\nonumber\\
&\times&\left(\frac{\mathcal{L}}{10^{39}~\unit{erg~s^{-1}}}\right)^{2/7}\left(\frac{r}{0.01~\unit{pc}}\right)^{-4/7}, \label{nebula}
\ee
and the observed peak flux is $F_{\nu,{\rm obs}}\sim F_{\nu}(\nu_a)\simeq0.3~\unit{Jy}$.
We note that the luminosity $\mathcal{L}\sim10^{39}~\unit{erg~s^{-1}}$ is just the order of the luminosity of the persistent radio emission of FRB 121102 \citep{cha17,mar17,ten17}. Therefore, the FRB-heated synchrotron nebulae can well explain the narrow spectrum of the bursts and the persistent radio emission.

\subsection{Model B: Cosmic comb scenario}\label{sec72}

\subsubsection{Curvature radiation from a combed magnetosphere}\label{sec721}

\begin{figure}[H]
\centering
\includegraphics[angle=0,scale=0.25]{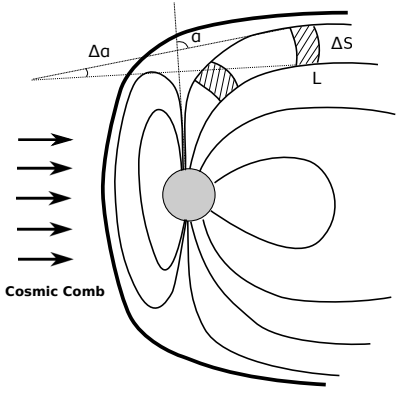}
\caption{Curvature radiation from the comb magnetosphere. The shadow area denotes a bunch with length $L$ and cross section $\Delta S$. $\alpha$ denotes the angel between the magnetic axis and the magnetic field. The emission region is somewhat insider the light cylinder.}\label{comb}
\end{figure}

\citet{zha17} proposed that an FRB might be produced via the interaction between a nearby astrophysical plasma stream (from e.g. a nearby AGN flare, a GRB, a supernova or an outburst of a binary companion) and a foreground regular pulsar, the so called ``cosmic comb''. Due to the ram pressure of the stream, the magnetic field configuration of a pulsar would deviate from the dipole field configuration, meanwhile, the Goldreich-Julian outflow would be suddenly compressed, which would cause a large fluctuation of the net charge density, producing coherent bunches. When these bunches sweep across the line of sight as they are combed towards the anti-stream-source direction, they would make a detectable FRB. Such a model recently gains more motivation \citep{zha18} in view of the large and variable rotation measure observed in the repeating FRB 121102 \citep{michilli18}.

Within this model, the field configuration in a bunch is similar to the Family B field lines discussed in Section \ref{sec5}, and the bunch would have a larger curvature radius and a larger cross section than that in the dipole magnetosphere (see Section \ref{sec51}), since the field lines are combed to be nearly parallel to each other by the cosmic stream. Let us assume that the bunch opening angles are $(\Delta\phi,\Delta\alpha)$, where $\phi$ denotes the toroidal angle around the magnetic axis, and $\alpha$ denotes the angel between the magnetic axis and the magnetic field. We notice that in a combed magnetosphere, $\Delta\phi$ and $\Delta\alpha$ would be very small, since the field lines are combed to be nearly parallel to each other.
The observed flux at the peak frequency $\nu_{\rm peak}$ is given by
\be
F_{\nu,\max}=\frac{2\pi}{TD^2}\left.\frac{dI}{d\omega d\Omega}\right|_{\max}\simeq\frac{2\pi e^2}{c} K(p) \frac{N_{e,0}^2\gamma_{1}^4}{D^2T}\left(\frac{\nu_{\rm peak}}{\nu_{c1}}\right)^{2/3}.\nonumber\\
\ee
The peak frequency is given by $\nu_{\rm peak}=\min(\nu_l,\nu_\varphi,\nu_{c1})$, where
\be
\nu_l=\frac{c}{\pi L},~~~\nu_\varphi=\frac{3c}{2\pi\rho\varphi^3},~~~\nu_{c1}=\frac{3c\gamma_1^3}{4\pi\rho},
\ee
where $\varphi=\max(\Delta\alpha/2,\Delta\phi/2)$ is defined in Family B. 

For a violent combing event, within the short period of time of interest, the original Goldreich-Julian charge flow density would not be directly relevant, since the field line configuration is abruptly modified. For an easy description, we relate the net charge density of a bunch with the compressed Goldreich-Julian density, i.e. $n_{\rm bun}=(1+\mu_c)n_{\rm GJ}'$, where $n_{\rm GJ}'$ is the compressed Goldreich-Julian density, and $\mu_c n_{\rm GJ}'$ denotes the fluctuation of the net charge density of a bunch, which contributes to the coherent radiation. 
Similar to Section \ref{sec63}, we define the effective electron number as $N_{e,\rm{eff}}$, which corresponds to the fluctuating net charge number in a bunch.
For a power-law distribution of the effective electron number, e.g., $N_{e}=N_{e,0}(\gamma/\gamma_1)^{-p}$ with $N_{e,{\rm eff}}=\int N_e d\gamma$, the effective electron number in the compressed volume $V'$ of a bunch is given by
\be
N_{e,{\rm eff}}=\gamma_1N_{e,0}/(p-1)=\mu_c n_{\rm GJ}'V'.
\ee
As shown in Figure \ref{comb}, somewhat inside the light cylinder $R_{\rm LC}=c/\Omega$, the field lines are compressed by the stream, thus the net electron number density of a bunch is of the order of that at the light cylinder with a compression factor $\xi_c > 1$, e.g., $n_{\rm GJ}'\sim \xi_c n_{\rm GJ}(R_{\rm LC})$. 
Consider that the field lines are combed to parallel to each other, the cross section of a bunch may be taken as $\eta R_{\rm LC}^2$, where $\eta$ is a parameter describing the cross section. One has $n_{\rm GJ}'V'\sim\xi_c n_{\rm GJ}(R_{\rm LC})(\eta R_{\rm LC}^2L)$. Thus, the normalization of the effective electron distribution is given by
\be
N_{e,0}=(p-1)\gamma_{1}^{-1}\mu_c n_{\rm GJ}'V'
=\frac{(p-1)\mu_c \xi_c \eta\Omega^2 B_pR^3L}{2\pi ec^2\gamma_1}.\nonumber\\
\ee
Since the emission is somewhat inside the light cylinder \citep[e.g.][]{zha17}, we approximately take the curvature radius as $\rho\sim R_{\rm LC}$.
Still assuming that the time interval between each bunch is $T\sim L/c$, one can write the peak flux as 
\be
F_{\nu,\max}=\frac{(p-1)^2K(p)}{2\pi c^4}\frac{\mu_c^2 \xi_c^2 \eta^2\Omega^4B_p^2R^6L\gamma_1^2}{D^2}\left(\frac{\nu_{\rm peak}}{\nu_{c1}}\right)^{2/3},\nonumber\\
\ee
where
\be
\frac{\nu_{\rm peak}}{\nu_{c1}}=\min\left(\frac{4c}{3\Omega L\gamma_1^3},\frac{2}{\varphi^3\gamma_1^3},1\right).
\ee
There are three cases for the peak flux:
\begin{itemize}
\item Case I: for $\nu_{\rm peak}=\nu_l$, one has
\be
F_{\nu,\max}=\frac{(p-1)^2K(p)}{2\pi c^{10/3}}\left(\frac{4}{3}\right)^{2/3}\frac{\mu_c^2 \xi_c^2 \eta^2\Omega^{10/3}B_p^2R^6L^{1/3}}{D^2}.\nonumber\\
\ee
\item Case II: for $\nu_{\rm peak}=\nu_\varphi$, one has
\be
F_{\nu,\max}=\frac{(p-1)^2K(p)}{2^{1/3}\pi c^4}\frac{\mu_c^2 \xi_c^2 \eta^2\Omega^4B_p^2R^6L}{D^2\varphi^2}.
\ee
\item Case III: for $\nu_{\rm peak}=\nu_{c1}$, one has
\be
F_{\nu,\max}=\frac{(p-1)^2K(p)}{2\pi c^4}\frac{\mu_c^2 \xi_c^2 \eta^2\Omega^4B_p^2R^6L\gamma_1^2}{D^2}.
\ee
\end{itemize}

\subsubsection{Model confronting FRB data}\label{sec722}
 
Let us take a conservative approach\footnote{In principle, the fluctuation parameter $\mu_c$ could exceed unity given the abruptness of the combing event, and the compression parameter $\xi_c$ should be at least a few times greater than unity.} by adopting $\mu_c = 1$ and $\xi_c=10$.
For the cosmic comb model for FRBs \citep{zha17}, we adopt the following typical parameters: $D=1~\unit{Gpc}$, $R=10^6~\unit{cm}$, $B_p=10^{13}~\unit{G}$, and $P=0.1~\unit{s}$.
The model parameter is assumed to be: $L=10~\unit{cm}$, $\gamma_1=600$, $\rho=c/\Omega=4.8\times10^{8}~\unit{cm}$, $p=3$, $\eta=0.1$, and $\varphi=\max(\Delta\alpha/2,\Delta\phi/2)=0.003$.
In this case, the typical frequencies are $\nu_{c1}=3.2~\unit{GHz}$, $\nu_{l}=0.9~\unit{GHz}$, $\nu_{\varphi}=1.1~\unit{GHz}$, and the intrinsic maximum flux is $F_{\nu,\max}=2.3~\unit{Jy}$. Due to $\nu_l<\nu_{\varphi}<\nu_{c1}$, the predicted spectral index $\alpha_{\rm index}$ is shown in the panel (a) of Figure \ref{spectra1}: $\alpha_{\rm index}\sim0.7$ for $\nu\lesssim0.9~\unit{GHz}$; $\alpha_{\rm index}\sim-1.3$ for $0.9~\unit{GHz}\lesssim\nu\lesssim1.1~\unit{GHz}$; $\alpha_{\rm index}\sim-2$ for $1.1~\unit{GHz}\lesssim\nu\lesssim3.2~\unit{GHz}$; $\alpha_{\rm index}\sim-3.3$ for $\nu\gtrsim3.2~\unit{GHz}$. 

In the cosmic comb model, the ram pressure $P_r$ should exceed the magnetic pressure $P_B$ at the light cylinder $R_{\rm LC}=c/\Omega\simeq4.8\times10^8~\unit{cm}(P/0.1~\unit{s})$ \citep{zha17}.
In the above parameter, the magnetic pressure at $R_{\rm LC}$ is 
\be
P_B&\simeq&\frac{B_p^2}{8\pi}\left(\frac{\Omega R}{c}\right)^6\nonumber\\
&\simeq&3.4\times10^8~\unit{erg~cm^{-3}}\left(\frac{B_p}{10^{13}~\unit{G}}\right)^{2}\left(\frac{P}{0.1~\unit{s}}\right)^{-6}.\label{pb}
\ee
To overcome such a pressure, one needs a strong stream from a nearby source so that the ram pressure of the continuous wind at the interaction region is, e.g.
\be
P_s&\simeq&\frac{\dot{M}v}{4\pi r^2}\nonumber\\
&\simeq&3.4\times10^8~\unit{erg~cm^{-3}}\left(\frac{\dot M}{M_\odot\unit{yr^{-1}}}\right)\left(\frac{\beta}{0.5}\right)\left(\frac{r}{\unit{AU}}\right)^{-2}\nonumber\\\label{ps}
\ee
where $\dot M$ is the wind mass-loss rate, $r$ is the distance from the source that produce the stream, and $v=\beta c$ is the wind velocity. 
Notice again that here we have adopted a conservative value of $\mu_c= 1$ in the above discussion. It is quite possible $\mu_c$ could be (much) greater than unity given the violent combing process. If so, the required combing condition could be (much) less stringent than derived here.

At last, in order to explain the narrow spectrum of FRB 121102, we also consider the synchrotron self-absorption from the FRB-heated synchrotron nebula \citep{yan16}, as discussed in Section \ref{sec71}. Thus, the observed peak flux is $F_{\nu,{\rm obs}}\sim F_{\nu}(\nu_a)\simeq0.3~\unit{Jy}$. This scenario also requires that the combed pulsar is within the nebula (not necessarily at the center). 
As shown in Eq.(\ref{nebula}), the radius of the nebula is larger than the separation between the combing source and the combed pulsar, consistent with our expectation.

\section{Conclusions and Discussion}\label{sec8} 

In this work, we developed a general radiation theory of coherent curvature radiation by bunches under the three-dimensional magnetic field geometry from the first principles and apply the model to interpret coherent radio emission from radio pulsars and FRBs. Following new conclusions are obtained:
\begin{itemize}
\item {Different from previous works \citep[e.g.][]{stu75,els76} that assumed that electron spatial distribution is stationary, we considered a more general scenario that the trajectories (field lines) are not parallel to each other. As a result, the opening angle of the bunch enters the problem. For particles streams out from an open field line region, a bunch slightly expands when it moves away from the dipole center, and the coherent radiation depends on the opening angle of the bunch (see Section \ref{sec42}). Since the electromagnetic waves with higher frequencies have smaller spread angles, coherence is not effective for high-frequency electromagnetic waves. According to Section \ref{sec422} and Section \ref{sec423}, for both Family II and Family III, the larger the bunch opening angle, the softer the coherent spectrum.}
\item {Another important ingredient introduced in our theory is the power law distribution of electron energy. Combining with the three-dimensional magnetic field configuration considered, one can quantify coherent curvature radiation of bunches and calculate the predicted radiation spectrum for the first time. We consider a bunch, consisting of a trajectory family, that is characterized by the following parameters: bunch length $L$, curvature radius $\rho$ of the trajectory family,  bunch opening angles  $(\varphi_\times, \varphi_+)$, and electron energy distribution $N_e(\gamma)d\gamma$ with $\gamma_1<\gamma<\gamma_2$. The predicted radiation spectrum shows a multi-segment broken power law with the break frequencies $\nu_{c1}=3c\gamma_1^3/4\pi\rho$, $\nu_l=c/\pi L$, and $\nu_\varphi=3c/2\pi\rho\varphi^3$, where $\varphi$ depends on $(\varphi_\times, \varphi_+)$ and the field line configuration in the bunch. The spectral indices depend on the relative order of these characteristic frequencies. The detailed spectra are presented in Sections \ref{sec51} and \ref{sec52}. }
\item {We emphasize that coherent emission in a pulsar magnetosphere is generated by the ``fluctuation'' of the net charge with respect the background Goldreich-Julian charge density, as discussed in Sections \ref{sec33} and \ref{sec413}.}
We find that with the ``bunches'' whose net charge density slightly deviated from the Goldreich-Julian density (e.g. $\mu_c = \delta n_{\rm GJ}/n_{\rm GJ} \sim 0.1$), the observed high brightness temperature of radio pulsars can be reproduced.  
Even though the total lepton number density, $n_\pm$, in the magnetosphere is greatly increased with respect to the Goldreich-Julian density, $n_{\rm GJ}$, in a pair-dominated magnetosphere, i.e.
$n_\pm \sim\mathcal{M}n_{\rm GJ}$ with $\mathcal{M}\gg1$, the net charge density remains close to the Goldreich-Julian, and the radiation of the pairs would essentially cancel out if they are spatially bunched together. 
Observationally, pulsars that emit radio emission seems to follow the condition of pair production \citep{rud75,zhang00}. The connection between coherent radiation and pair production might be indirect. For example, violent pair production and their spatial separation (in order to ``screen'' process to the parallel electric field in the gap region) would induce deviations of the local net charge densities from the Goldreich-Julian density to produce coherent radiation. 
Notice that these conclusions are different from some recent works on coherent curvature radiation by bunches from FRBs \citep[e.g.][]{kum17,lu17}, in which the total number of electron-positron pairs are introduced to calculate the luminosity of bunching coherent curvature radiation. According to our theory, the coherent radiation luminosity is greatly overestimated in those investigations.
\item {The coherent mechanism of pulsar radio emission has been subject to debate over the years \citep{mel17}. Our study suggests that coherent curvature radiation by bunches remains a promising candidate to interpret the observations. In particular, the observed spectra of pulsars, which can be fitted by either a single power law, or two-segment or multi-segment broken power laws \citep[e.g.][]{lor95,xil96,jan17}, are  naturally interpreted for the first time, given the typical pulsar parameters (e.g. $B_p=10^{12}~\unit{G}$ and $P=0.1~\unit{s}$). The required fluctuation is only moderate (e.g. $\mu_c = \delta n_{\rm GJ} / n_{\rm GJ} \sim 0.1$).}
\item The physical origin of FRBs is mysterious. Many FRB models invoked coherent curvature radiation by bunches to explain their extremely brightness temperature, e.g., pulsar-like activities \citep{connor16,cordes16,metzger17,kashiyama17}, mergers of compact binaries \citep{kas13,tot13,wan16,zha16,liu16}, collapse of supra-massive neutron stars to black holes \citep{fal14,zha14}, collisions between a neutron star and a comet or asteroids \citep{gen15,dai16}, cosmic combs \citep{zha17}, and so on. However, most FRB models mainly focus on the released energy and duration, with the description of coherent radiation overly simplified. With the theory developed in this paper, coherent bunching mechanism of FRBs can be quantitatively discussed in great detail.
Due to their extremely high brightness temperatures, FRBs have a much larger fluctuating net charge density compared with pulsars. Several factors may contribute to such a large fluctuating net charges: 1. The FRB source may involve a neutron star with a stronger magnetic field and faster rotation \citep[e.g.][]{mur16,met17} with the emission region close to the stellar surface \cite[e.g.][]{kum17}.  
2. Due to the abrupt nature of FRBs, the normalized fluctuation of the net charge density $\mu_c$ for FRBs may reach or even exceed unity. 
3. In the cosmic comb scenario \citep{zha17,zha18}, the magnetosphere may be suddenly compressed by an astrophysical stream so that the 
effective fluctuation $\mu_c$ would exceed unity.  
Meanwhile, since the field lines are combed to nearly parallel to each other, the cross section of a
bunch could be very large, leading to more significant coherent emission. 
The bunching coherent mechanism proposed in this paper can interpret the steep negative spectral index observed in the bursts detected from the repeating source FRB 121102. In order to account for the narrowness of the spectrum, one needs to introduce 
synchrotron self-absorption from the FRB-heated synchrotron nebula \citep{yan16}. The required nebula luminosity of this model coincides with the observed luminosity of the persistent radio emission of FRB 121102 \citep{cha17}. 
\end{itemize}

When we apply our model to radio emission of pulsars and FRBs, as discussed in Section \ref{sec5}, we have assumed that the bunch opening angle is mainly defined by the magnetic field geometry of the bunch. In general, a curvature drift that is perpendicular to the plane that contains the field lines, e.g. $v_d=\pm m_ecv_\varphi^2\gamma/eB\rho$, where $v_\varphi\sim c$ is the velocity along the field lines, is required, since an electron/positron moving along a field line must be subject to a Lorentz force that causes it to follow the curved path \citep[e.g.][]{zhe79}. As a result, curvature drift would make electrons and positrons to drift in the opposite directions across the field lines, and cause more energetic particles to drift faster than less energetic particles. All these tend to disperse the bunch. However, the drifting angle due to the curvature drift effect, e.g., $\varphi_d\sim v_d/v_\varphi\sim\gamma m_ec^2/eB\rho\sim 10^{-10}(B_p/10^{12}~\unit{G})^{-1}(\rho/10^{10}~\unit{cm})^{-1}(\gamma/10^3)(r/10^8~\unit{cm})^3$, is much smaller than the opening angle of field lines.  
Therefore, it is reasonable to ignore this curvature drift effect and assume that the bunch opening angle mainly depends on the field geometry.

On the other hand, we have assumed that the emission region of the curvature radiation is in the open field lines rather than the closed field lines. The main reason is that bunches from the open field line region would move along the field lines that curve away from the emitted coherent radio waves, so that they are not subject to further absorption by the proceeding bunches. Emission from the bunches moving in closed field line regions may be subject to further absorption by bunches moving along adjacent field lines. Even if it may not be absorbed, the emission in these cases would not be narrowly beamed, which might give rise to smoother lightcurves than observed. Pulsar radio emission is known to originate from the open field line regions of pulsars \citep[e.g.][]{rankin83}. For FRBs, models that invoke open field lines are favored. Those invoking closed field lines require further justification regarding the propagation of the coherent radio waves across the magnetosphere of the source.

Finally, we'd like to comment on that some basic conditions of the classical formula, e.g., Eq.(\ref{power}) and Eq.(\ref{syn}), have been omitted in some previous works when applied to study curvature radiation \citep[e.g.][]{ghi17a}. Below are some examples:
1. Equation (\ref{syn}) was often used to describe the spectrum of curvature radiation. However, one should note that the classical $\nu^{1/3}$ spectrum corresponds to the total radiation spectrum in all directions rather than the direction along the line of sight \citep{yan18}. For curvature radiation, the radiation of one bunch is beamed in a narrow cone that sweeps cross the line of sight, thus one should consider the radiation per unit solid angle, rather than the total radiation spectrum. Even considering more than one bunch with different motion directions, a coherent sum of amplitudes should be considered, rather than the simple integration over angles. 
2. The definition of radiation power should be based on the average time scale, $T$, of pulse repetition, as shown in Eq.(\ref{power}). For synchrotron radiation, pulses repeat naturally with the gyration period. However, for one-time curvature radiation, the average time scale of  pulse repetition depends on the average time interval between the bunches, instead of the gyration period. 
3. Since the basic formulae about the radiation of moving charges, e.g., Eq.(\ref{multiemission}), can be applied to relativistic charged particles, it is unnecessary to repeatedly apply some relativistic effects, such as time delay, beaming effect and so on, to the derive $dI/ d\omega d\Omega$, etc. (cf. Eq.(\ref{multiemission})).

\acknowledgments
We thank G. I. Melikidze for helpful discussion. This work is partially supported by Project funded by the Initiative Postdocs Supporting Program (No. BX201600003), the National Basic Research Program (973 Program) of China (No. 2014CB845800), and the China Postdoctoral Science Foundation (No. 2016M600851). Y.-P.Y. is supported by a KIAA-CAS Fellowship. 


\begin{thebibliography}{}
\expandafter\ifx\csname natexlab\endcsname\relax\def\natexlab#1{#1}\fi

\bibitem[{{Beloborodov}(2017)}]{bel17}
{Beloborodov}, A.~M. 2017, \apjl, 843, L26

\bibitem[{{Beloborodov} \& {Thompson}(2007)}]{bel07}
{Beloborodov}, A.~M., \& {Thompson}, C. 2007, \apj, 657, 967

\bibitem[{{Benford} \& {Buschauer}(1977)}]{ben77}
{Benford}, G., \& {Buschauer}, R. 1977, \mnras, 179, 189

\bibitem[{{Blandford}(1975)}]{bla75}
{Blandford}, R.~D. 1975, \mnras, 170, 551

\bibitem[{{Buschauer} \& {Benford}(1976)}]{bus76}
{Buschauer}, R., \& {Benford}, G. 1976, \mnras, 177, 109

\bibitem[{{Chatterjee} {et~al.}(2017){Chatterjee}, {Law}, {Wharton},
  {Burke-Spolaor}, {Hessels}, {Bower}, {Cordes}, {Tendulkar}, {Bassa},
  {Demorest}, {Butler}, {Seymour}, {Scholz}, {Abruzzo}, {Bogdanov}, {Kaspi},
  {Keimpema}, {Lazio}, {Marcote}, {McLaughlin}, {Paragi}, {Ransom}, {Rupen},
  {Spitler}, \& {van Langevelde}}]{cha17}
{Chatterjee}, S., {Law}, C.~J., {Wharton}, R.~S., {et~al.} 2017, ArXiv
  e-prints, arXiv:1701.01098

\bibitem[{{Cheng} \& {Ruderman}(1977)}]{chen77}
{Cheng}, A.~F., \& {Ruderman}, M.~A. 1977, \apj, 212, 800

\bibitem[{{Cocke}(1973)}]{coc73}
{Cocke}, W.~J. 1973, \apj, 184, 291

\bibitem[{{Connor} {et~al.}(2016){Connor}, {Sievers}, \& {Pen}}]{connor16}
{Connor}, L., {Sievers}, J., \& {Pen}, U.-L. 2016, \mnras, 458, L19

\bibitem[{{Cordes} \& {Wasserman}(2016)}]{cordes16}
{Cordes}, J.~M., \& {Wasserman}, I. 2016, \mnras, 457, 232

\bibitem[{{Dai} {et~al.}(2016){Dai}, {Wang}, {Wu}, \& {Huang}}]{dai16}
{Dai}, Z.~G., {Wang}, J.~S., {Wu}, X.~F., \& {Huang}, Y.~F. 2016, \apj, 829, 27

\bibitem[{{Egorenkov} {et~al.}(1983){Egorenkov}, {Lominadze}, \&
  {Mamradze}}]{ego83}
{Egorenkov}, V.~D., {Lominadze}, D.~G., \& {Mamradze}, P.~G. 1983,
  Astrophysics, 19, 426

\bibitem[Elsaesser \& Kirk(1976)]{els76} Elsaesser, K., \& Kirk, J.\ 1976, \aap, 52, 449 

\bibitem[{{Falcke} \& {Rezzolla}(2014)}]{fal14}
{Falcke}, H., \& {Rezzolla}, L. 2014, \aap, 562, A137

\bibitem[{{Gedalin} {et~al.}(2002){Gedalin}, {Gruman}, \& {Melrose}}]{ged02}
{Gedalin}, M., {Gruman}, E., \& {Melrose}, D.~B. 2002, \mnras, 337, 422

\bibitem[{{Geng} \& {Huang}(2015)}]{gen15}
{Geng}, J.~J., \& {Huang}, Y.~F. 2015, \apj, 809, 24

\bibitem[{{Ghisellini}(2017)}]{ghi17b}
{Ghisellini}, G. 2017, \mnras, arXiv:1609.04815

\bibitem[{{Ghisellini} \& {Locatelli}(2017)}]{ghi17a}
{Ghisellini}, G., \& {Locatelli}, N. 2017, ArXiv e-prints, arXiv:1708.07507

\bibitem[{{Gil} {et~al.}(2004){Gil}, {Lyubarsky}, \& {Melikidze}}]{gil04}
{Gil}, J., {Lyubarsky}, Y., \& {Melikidze}, G.~I. 2004, \apj, 600, 872

\bibitem[{{Gil} {et~al.}(2006){Gil}, {Melikidze}, \& {Zhang}}]{gil06}{Gil}, J., {Melikidze}, G., \& {Zhang}, B. 2006, \apj, 650, 1048

\bibitem[{{Gil} \& {Sendyk}(2000)}]{gil00}
{Gil}, J.~A., \& {Sendyk}, M. 2000, \apj, 541, 351

\bibitem[{{Ginzburg} \& {Zhelezniakov}(1975)}]{gin75}
{Ginzburg}, V.~L., \& {Zhelezniakov}, V.~V. 1975, \araa, 13, 511

\bibitem[{{Goldreich} \& {Julian}(1969)}]{gol69}
{Goldreich}, P., \& {Julian}, W.~H. 1969, \apj, 157, 869

\bibitem[{{Gunn} \& {Ostriker}(1971)}]{gun71}
{Gunn}, J.~E., \& {Ostriker}, J.~P. 1971, \apj, 165, 523

\bibitem[{{Hankins} \& {Eilek}(2007)}]{han07}
{Hankins}, T.~H., \& {Eilek}, J.~A. 2007, \apj, 670, 693

\bibitem[{{Jackson}(1998)}]{jac62}
{Jackson}, J.~D. 1998, {Classical Electrodynamics, 3rd Edition (New York: Wiley)}, 832

\bibitem[{{Jankowski} {et~al.}(2017){Jankowski}, {van Straten}, {Keane},
  {Bailes}, {Barr}, {Johnston}, \& {Kerr}}]{jan17}
{Jankowski}, F., {van Straten}, W., {Keane}, E.~F., {et~al.} 2017, ArXiv
  e-prints, arXiv:1709.08864

\bibitem[{{Kashiyama} {et~al.}(2013){Kashiyama}, {Ioka}, \&
  {M{\'e}sz{\'a}ros}}]{kas13}
{Kashiyama}, K., {Ioka}, K., \& {M{\'e}sz{\'a}ros}, P. 2013, \apjl, 776, L39

\bibitem[{{Kashiyama} \& {Murase}(2017)}]{kashiyama17}
{Kashiyama}, K., \& {Murase}, K. 2017, \apjl, 839, L3

\bibitem[{{Katz}(2014)}]{kat14}
{Katz}, J.~I. 2014, \prd, 89, 103009

\bibitem[Katz(2018a)]{kat18a} Katz, J.~I.\ 2018, arXiv:1803.01938

\bibitem[Katz(2018b)]{kat18b} Katz, J.~I.\ 2018, arXiv:1804.09092 

\bibitem[{{Kazbegi} {et~al.}(1991){Kazbegi}, {Machabeli}, \&
  {Melikidze}}]{kaz91}
{Kazbegi}, A.~Z., {Machabeli}, G.~Z., \& {Melikidze}, G.~I. 1991, \mnras, 253,
  377

\bibitem[{{Kellermann} \& {Pauliny-Toth}(1969)}]{kel69}
{Kellermann}, K.~I., \& {Pauliny-Toth}, I.~I.~K. 1969, \apjl, 155, L71

\bibitem[{{Kroll} \& {McMullin}(1979)}]{kro79}
{Kroll}, N.~M., \& {McMullin}, W.~A. 1979, \apj, 231, 425

\bibitem[{{Kumar} {et~al.}(2017){Kumar}, {Lu}, \& {Bhattacharya}}]{kum17}
{Kumar}, P., {Lu}, W., \& {Bhattacharya}, M. 2017, \mnras, 468, 2726

\bibitem[{{Law} {et~al.}(2017){Law}, {Abruzzo}, {Bassa}, {Bower},
  {Burke-Spolaor}, {Butler}, {Cantwell}, {Carey}, {Chatterjee}, {Cordes},
  {Demorest}, {Dowell}, {Fender}, {Gourdji}, {Grainge}, {Hessels}, {Hickish},
  {Kaspi}, {Lazio}, {McLaughlin}, {Michilli}, {Mooley}, {Perrott}, {Ransom},
  {Razavi-Ghods}, {Rupen}, {Scaife}, {Scott}, {Scholz}, {Seymour}, {Spitler},
  {Stovall}, {Tendulkar}, {Titterington}, {Wharton}, \& {Williams}}]{law17}
{Law}, C.~J., {Abruzzo}, M.~W., {Bassa}, C.~G., {et~al.} 2017, \apj, 850, 76

\bibitem[{{Levinson} {et~al.}(2005){Levinson}, {Melrose}, {Judge}, \&
  {Luo}}]{lev05}
{Levinson}, A., {Melrose}, D., {Judge}, A., \& {Luo}, Q. 2005, \apj, 631, 456

\bibitem[{{Liu} {et~al.}(2016){Liu}, {Romero}, {Liu}, \& {Li}}]{liu16}
{Liu}, T., {Romero}, G.~E., {Liu}, M.-L., \& {Li}, A. 2016, \apj, 826, 82

\bibitem[{{Lorimer} {et~al.}(2007){Lorimer}, {Bailes}, {McLaughlin},
  {Narkevic}, \& {Crawford}}]{lor07}
{Lorimer}, D.~R., {Bailes}, M., {McLaughlin}, M.~A., {Narkevic}, D.~J., \&
  {Crawford}, F. 2007, Science, 318, 777

\bibitem[{{Lorimer} {et~al.}(1995){Lorimer}, {Yates}, {Lyne}, \&
  {Gould}}]{lor95}
{Lorimer}, D.~R., {Yates}, J.~A., {Lyne}, A.~G., \& {Gould}, D.~M. 1995,
  \mnras, 273, 411

\bibitem[Lu \& Kumar(2018)]{lu17} Lu, W., \& Kumar, P.\ 2018, \mnras, 477, 2470 


\bibitem[{{Luo} \& {Melrose}(2008)}]{luo08}
{Luo}, Q., \& {Melrose}, D. 2008, \mnras, 387, 1291

\bibitem[{{Luo} \& {Melrose}(1992)}]{luo92}
{Luo}, Q., \& {Melrose}, D.~B. 1992, \mnras, 258, 616

\bibitem[{{Luo} \& {Melrose}(1995)}]{luo95}
---. 1995, \mnras, 276, 372

\bibitem[{{Lyubarsky}(2014)}]{lyu14}
{Lyubarsky}, Y. 2014, \mnras, 442, L9

\bibitem[{{Lyutikov} {et~al.}(1999{\natexlab{a}}){Lyutikov}, {Blandford}, \&
  {Machabeli}}]{lyu99a}
{Lyutikov}, M., {Blandford}, R.~D., \& {Machabeli}, G. 1999{\natexlab{a}},
  \mnras, 305, 338

\bibitem[{{Lyutikov} {et~al.}(1999{\natexlab{b}}){Lyutikov}, {Machabeli}, \&
  {Blandford}}]{lyu99b}
{Lyutikov}, M., {Machabeli}, G., \& {Blandford}, R. 1999{\natexlab{b}}, \apj,
  512, 804

\bibitem[{{Machabeli} \& {Usov}(1979)}]{mac79}
{Machabeli}, G.~Z., \& {Usov}, V.~V. 1979, Soviet Astronomy Letters, 5, 445

\bibitem[{{Marcote} {et~al.}(2017){Marcote}, {Paragi}, {Hessels}, {Keimpema},
  {van Langevelde}, {Huang}, {Bassa}, {Bogdanov}, {Bower}, {Burke-Spolaor},
  {Butler}, {Campbell}, {Chatterjee}, {Cordes}, {Demorest}, {Garrett}, {Ghosh},
  {Kaspi}, {Law}, {Lazio}, {McLaughlin}, {Ransom}, {Salter}, {Scholz},
  {Seymour}, {Siemion}, {Spitler}, {Tendulkar}, \& {Wharton}}]{mar17}
{Marcote}, B., {Paragi}, Z., {Hessels}, J.~W.~T., {et~al.} 2017, \apjl, 834, L8

\bibitem[{{McCray}(1966)}]{mcc66}
{McCray}, R. 1966, Science, 154, 1320

\bibitem[Melikidze et al.(2000)]{mel00} Melikidze, G.~I., Gil, J.~A., \& Pataraya, A.~D.\ 2000, \apj, 544, 1081 

\bibitem[{{Melrose}(1978)}]{mel78}
{Melrose}, D.~B. 1978, \apj, 225, 557

\bibitem[Melrose(1981)]{mel81} Melrose, D.~B.\ 1981, Pulsars: 13 Years of Research on Neutron Stars, 95, 133 

\bibitem[{{Melrose}(2017)}]{mel17}
---. 2017, ArXiv e-prints, arXiv:1707.02009

\bibitem[{{Melrose} \& {Gedalin}(1999)}]{mel99}
{Melrose}, D.~B., \& {Gedalin}, M.~E. 1999, \apj, 521, 351

\bibitem[{{Metzger} {et~al.}(2017{\natexlab{a}}){Metzger}, {Berger}, \&
  {Margalit}}]{metzger17}
{Metzger}, B.~D., {Berger}, E., \& {Margalit}, B. 2017{\natexlab{a}}, \apj,
  841, 14

\bibitem[{{Metzger} {et~al.}(2017{\natexlab{b}}){Metzger}, {Berger}, \&
  {Margalit}}]{met17}
---. 2017{\natexlab{b}}, \apj, 841, 14

\bibitem[{{Michilli} {et~al.}(2018){Michilli}, {Seymour},
  {Hessels} {et~al}}]{michilli18}
{Michilli}, D., {Seymour}, A., {Hessels}, J. W. T. et al. 2018, \nat, 553, 182

\bibitem[{{Murase} {et~al.}(2016){Murase}, {Kashiyama}, \&
  {M{\'e}sz{\'a}ros}}]{mur16} 
{Murase}, K., {Kashiyama}, K., \& {M{\'e}sz{\'a}ros}, P. 2016, \mnras, 461,
  1498

\bibitem[Pataraia \& Melikidze(1980)]{pat80} Pataraia, A., \& Melikidze, G.\ 1980, \apss, 68, 61 

\bibitem[{{Rankin}(1983)}]{rankin83}
{Rankin}, J.~M. 1983, \apj, 274, 333

\bibitem[{{Ruderman} \& {Sutherland}(1975)}]{rud75}
{Ruderman}, M.~A., \& {Sutherland}, P.~G. 1975, \apj, 196, 51

\bibitem[{{Rybicki} \& {Lightman}(1979)}]{ryb79}
{Rybicki}, G.~B., \& {Lightman}, A.~P. 1979, {Radiative processes in
  astrophysics (New York: Wiley-Interscience)}

\bibitem[{{Sturrock}(1971)}]{stu71}
{Sturrock}, P.~A. 1971, \apj, 164, 529

\bibitem[Sturrock et al.(1975)]{stu75} Sturrock, P.~A., Petrosian, V., \& Turk, J.~S.\ 1975, \apj, 196, 73 

\bibitem[{{Tendulkar} {et~al.}(2017){Tendulkar}, {Bassa}, {Cordes}, {Bower},
  {Law}, {Chatterjee}, {Adams}, {Bogdanov}, {Burke-Spolaor}, {Butler},
  {Demorest}, {Hessels}, {Kaspi}, {Lazio}, {Maddox}, {Marcote}, {McLaughlin},
  {Paragi}, {Ransom}, {Scholz}, {Seymour}, {Spitler}, {van Langevelde}, \&
  {Wharton}}]{ten17}
{Tendulkar}, S.~P., {Bassa}, C.~G., {Cordes}, J.~M., {et~al.} 2017, \apjl, 834,
  L7

\bibitem[{{Thornton} {et~al.}(2013){Thornton}, {Stappers}, {Bailes},
  {Barsdell}, {Bates}, {Bhat}, {Burgay}, {Burke-Spolaor}, {Champion}, {Coster},
  {D'Amico}, {Jameson}, {Johnston}, {Keith}, {Kramer}, {Levin}, {Milia}, {Ng},
  {Possenti}, \& {van Straten}}]{tho13}
{Thornton}, D., {Stappers}, B., {Bailes}, M., {et~al.} 2013, Science, 341, 53

\bibitem[{{Totani}(2013)}]{tot13}
{Totani}, T. 2013, \pasj, 65, L12

\bibitem[{{Twiss}(1958)}]{twi58}
{Twiss}, R.~Q. 1958, Australian Journal of Physics, 11, 564

\bibitem[Ursov \& Usov(1988)]{urs88} Ursov, V.~N., \& Usov, V.~V.\ 1988, \apss, 140, 325 


\bibitem[Usov(1987)]{uso87} Usov, V.~V.\ 1987, \apj, 320, 333 

\bibitem[{{Wang} {et~al.}(2016){Wang}, {Yang}, {Wu}, {Dai}, \& {Wang}}]{wan16}
{Wang}, J.-S., {Yang}, Y.-P., {Wu}, X.-F., {Dai}, Z.-G., \& {Wang}, F.-Y. 2016,
  \apjl, 822, L7

\bibitem[{{Waxman}(2017)}]{wax17}
{Waxman}, E. 2017, \apj, 842, 34

\bibitem[{{Weatherall}(1998)}]{wea98}
{Weatherall}, J.~C. 1998, \apj, 506, 341

\bibitem[Westfold(1959)]{wes59} Westfold, K.~C.\ 1959, \apj, 130, 241 

\bibitem[{{Xilouris} {et~al.}(1996){Xilouris}, {Kramer}, {Jessner},
  {Wielebinski}, \& {Timofeev}}]{xil96}
{Xilouris}, K.~M., {Kramer}, M., {Jessner}, A., {Wielebinski}, R., \&
  {Timofeev}, M. 1996, \aap, 309, 481

\bibitem[{{Yang} {et~al.}(2016){Yang}, {Zhang}, \& {Dai}}]{yan16}
{Yang}, Y.-P., {Zhang}, B., \& {Dai}, Z.-G. 2016, \apjl, 819, L12

\bibitem[Yang \& Zhang(2018)]{yan18} Yang, Y.-P., \& Zhang, B.\ 2018, arXiv:1808.05170 

\bibitem[{{Zhang}(2014)}]{zha14}
{Zhang}, B. 2014, \apjl, 780, L21

\bibitem[{{Zhang}(2016)}]{zha16}
---. 2016, \apjl, 827, L31

\bibitem[{{Zhang}(2017)}]{zha17}
---. 2017, \apjl, 836, L32

\bibitem[{{Zhang}(2018)}]{zha18}
---. 2018, \apjl, 841, L21

\bibitem[Zhang et al.(2000)]{zhang00} Zhang, B., Harding, A.~K., \& Muslimov, A.~G.\ 2000, \apjl, 531, L135 

\bibitem[{{Zhang} \& {Qiao}(1996)}]{zhang96}{Zhang}, B., \& {Qiao}, G.~J. 1996, \aap, 310, 135

\bibitem[{{Zhang} {et~al.}(1997){Zhang}, {Qiao}, {Lin}, \& {Han}}]{zhang97}{Zhang}, B., {Qiao}, G.~J., {Lin}, W.~P., \& {Han}, J.~L. 1997, \apj, 478, 313

\bibitem[{{Zhelezniakov} \& {Shaposhnikov}(1979)}]{zhe79}
{Zhelezniakov}, V.~V., \& {Shaposhnikov}, V.~E. 1979, Australian Journal of
  Physics, 32, 49

\end{thebibliography}

\appendix

\section{A. Radiation by moving charges}\label{seca}

In this section, we briefly summarize the radiation from moving charges. 
The fields at a point $\bm{x}$ at time $t$ is determined by the retarded position $\bm{r}(t')$ and time $t'$ of the charged particle. Defining $\bm{\mathcal{R}}\equiv\bm{x}-\bm{r}(t')$, $\bm{n}\equiv\bm{\mathcal{R}}/\mathcal{R}$ and $\bm{\beta}\equiv\dot{\bm{r}}(t')/c$, the electromagnetic fields are given by \citep[e.g.][]{jac62,ryb79}
\be
\bm{B}(\bm{x},t)&=&\left[\bm{n}\times\bm{E}(\bm{x},t)\right]_{\rm ret}\nonumber\\
\bm{E}(\bm{x},t)&=&e\left[\frac{\bm{n}-\bm{\beta}}{\gamma^2(1-\bm{n}\cdot\bm{\beta})^3\mathcal{R}^2}\right]_{\rm ret}+\frac{e}{c}\left[\frac{\bm{n}\times\{(\bm{n}-\bm{\beta})\times\bm{\dot\beta}\}}{(1-\bm{n}\cdot\bm{\beta})^3\mathcal{R}}\right]_{\rm ret},
\label{EMfield}
\ee
where the subscript ``ret'' means that the quantities in the square brackets are all evaluated at the retarded time $t'$.
As shown in the above equations, the electric field is composed of two terms: (1) the velocity field, which is the generalization of the Coulomb law to a moving charge and falls off as $1/\mathcal{R}^2$; and (2) the acceleration field, which, is proportional to the particle's acceleration, constitutes the radiation field falling off as $1/\mathcal{R}$.

First, the power radiated per unit solid angle has a general form, i.e.,
\be
\frac{dP(t)}{d\Omega}=|\bm{A}(t)|^2,
~~~~~~
{\rm where} 
~~~
\bm{A}(t)\equiv\left(\frac{c}{4\pi}\right)^{1/2}[\mathcal{R}\bm{E}]_{\rm ret},\label{amplitude}
\ee
and $t$ is the observed time at the field point, $\mathcal{R}$ is the distance between the field point and the retarded position of the charged particle. In general, the observed point is far enough away from the source, thus the velocity-field term in Eq.(\ref{EMfield}) could be ignored in the ``far zone''.
Based on the time-dependent electromagnetic field of a single moving charge, e.g., Eq.(\ref{EMfield}), the radiation frequency spectrum can be calculated by the Fourier transformation, e.g.,
\be
\bm{A}(\omega)=\left(\frac{e^2}{8\pi^2c}\right)^{1/2}(-i\omega)\int_{-\infty}^{+\infty}\bm{n}\times(\bm{n}\times\bm{\beta})e^{i\omega(t'-\bm{n}\cdot\bm{r}(t')/c)}dt',\label{a3}
\ee
where $t'$ is the retarded time, $\bm{r}(t')$ denotes the retarded position of the charged particle, $\bm{\beta}$ and $\bm{n}$ are defined as $\bm{\beta}=\dot{\bm{r}}(t')/c$ and $\bm{n}=\bm{\mathcal{R}}/\mathcal{R}$. Here $\bm{\mathcal{R}}$ is defined as $\bm{\mathcal{R}}=\bm{x}-\bm{r}(t')$ and $\bm{x}$ is the field point. For $r(t')\ll x$, one has $\mathcal{R}(t')\simeq x-\bm{n}\cdot\bm{r}(t')$, which has been considered in the above integral. Note that in Eq.(\ref{a3}) one has used the identity $\bm{n}\times\left[(\bm{n}-\bm{\beta})\times\bm{\dot\beta}\right]/(1-\bm{n}\cdot\bm{\beta})^2=d/dt\left[\bm{n}\times(\bm{n}\times\bm{\beta})/(1-\bm{n}\cdot\bm{\beta})\right]$.
On the other hand, the total energy radiated per unit solid angle is given by \citep{jac62}
\be
\frac{dW}{d\Omega}&=&\int_{-\infty}^{+\infty}\frac{dP(t)}{d\Omega}dt=\int_{0}^{+\infty}\frac{dI(\omega,\bm{n})}{d\omega d\Omega}d\omega\nonumber\\
&=&\int_{-\infty}^{+\infty}|\bm{A}(t)|^2dt=\int_{-\infty}^{+\infty}|\bm{A}(\omega)|^2d\omega,
\ee
where $dI/d\omega d\Omega$ denotes the energy radiated per unit solid angle per unit frequency interval. Due to $dI/d\omega d\Omega=2|\bm{A}(\omega)|^2$, one has finally \citep[e.g.][]{jac62,ryb79}
\be
\frac{dI}{d\omega d\Omega}=\frac{e^2\omega^2}{4\pi^2c}\left|\int_{-\infty}^{+\infty}\bm{n}\times(\bm{n}\times\bm{\beta})e^{i\omega(t-\bm{n}\cdot\bm{r}(t)/c)}dt\right|^2.\label{radiation1}
\ee
For brevity, the primes on the retarded time have been omitted.

If there is more than one charged particle, a coherent sum of the amplitudes should replace the single amplitude in the above equation.
In this case, the energy radiated per unit solid angle per unit frequency interval is given by
\be
\frac{dI}{d\omega d\Omega}=\frac{\omega^2}{4\pi^2c}\left|\int_{-\infty}^{+\infty}\sum_j^Nq_j\bm{n}\times(\bm{n}\times\bm{\beta}_j)e^{i\omega(t-\bm{n}\cdot\bm{r}_j(t)/c)}dt\right|^2,
\ee
where $j$ represents the identifier of each charged particle, and $q_j$ is the corresponding charge.

\section{B. curvature radiation from instantaneously circular motion}\label{secb0}

\begin{figure}[H]
\centering
\includegraphics[angle=0,scale=0.3]{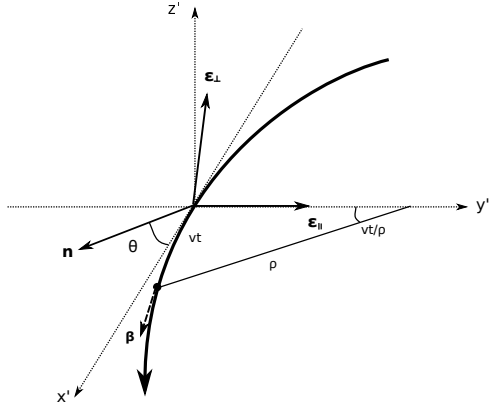}
\caption{Geometry for an instantaneously circular motion. The trajectory lies in the $x'-y'$ plane. At the retarded time $t=0$, the electron is at the origin, and the velocity is along the $x'$ axis.}\label{CR0}
\end{figure}

In this section, we briefly summarize the curvature radiation of a single electron during instantaneously circular motion \citep[e.g.][]{jac62}.
Consider the appropriate coordinate system in Figure \ref{CR0}, where the origin is the location of the electron at the retarded time $t=0$, and the instantaneously-circular trajectory lies in the $x'-y'$ plane with a curvature radius $\rho$. The electron velocity is along $x'$ axis at $t=0$. 
Since the integral in Eq.(\ref{radiation}) is taken along the trajectory, $\bm{n}$ can be chosen to lie in the $x'-z'$ plane without losing generality. $\bm{\epsilon}_\parallel$ is the unit vector pointing to the center of the instantaneous circle, which is set to the $y'$ direction, and $\bm{\epsilon}_\perp=\bm{n}\times\bm{\epsilon}_\parallel$ is defined.
The energy radiated per unit frequency interval per unit solid angle is given by \citep[e.g.][]{jac62}
\be
\frac{dI}{d\omega d\Omega}&=&\frac{e^2\omega^2}{4\pi^2c}\left|-\bm{\epsilon}_\parallel{A}_\parallel+\bm{\epsilon}_\perp{A}_\perp\right|^2\nonumber\\
&=&\frac{e^2}{3\pi^2c}\left(\frac{\omega\rho}{c}\right)^2
\left(\frac{1}{\gamma^2}+\theta^2\right)^2\left[K_{2/3}^2(\xi)+\frac{\theta^2}{(1/\gamma^2)+\theta^2}K_{1/3}^2(\xi)\right],\nonumber\\\label{CRspec1}
\ee
where $A_\perp$ and $A_\parallel$ are the polarized components of the amplitude along $\bm{\epsilon}_\perp$ and $\bm{\epsilon}_\parallel$, respectively, which are given by
\be
A_\parallel&=&\frac{2i}{\sqrt{3}}\frac{\rho}{c}\left(\frac{1}{\gamma^2}+\theta^2\right)K_{2/3}(\xi),\nonumber\\
A_\perp&=&\frac{2}{\sqrt{3}}\frac{\rho\theta}{c}\left(\frac{1}{\gamma^2}+\theta^2\right)^{1/2}K_{1/3}(\xi).\label{amp}
\ee
The argument $\xi$ in the modified Bessel function is defined as
\be
\xi=\frac{\omega\rho}{3c}\left(\frac{1}{\gamma^2}+\theta^2\right)^{3/2}.\label{xi}
\ee
According to the properties of the modified Bessel function, i.e., $K_{\nu}(\xi)\rightarrow0$ for $\xi\gg1$, the radiation intensity is negligible for $\xi\gg1$.
As shown in Eq.(\ref{xi}), $\xi\gg1$ is satisfied at large angles. On the other hand, if $\omega$ becomes too large, $\xi$ will be large at all angles. Therefore, one can define
the critical frequency by $\xi=1/2$ for $\theta=0$, beyond which the radiation can be negligible at all angles \citep[e.g.][]{jac62}. Such a critical frequency is given by
\be
\omega_c=\frac{3}{2}\gamma^3\left(\frac{c}{\rho}\right).\label{omegac1}
\ee
For an accelerated relativistic electron, its radiation is beamed in a narrow cone that sweeps cross the line of sight, which means that the radiation concentrates around $\theta=0$, and the parallel polarized component is dominant.
According to Eq.(\ref{CRspec1}) and the properties of the modified Bessel function, e.g., $K_\nu(x)\rightarrow (\Gamma(\nu)/2)(x/2)^{-\nu}$ for $x\ll1$ and $\nu\neq0$, and $K_\nu(x)\rightarrow\sqrt{\pi/2x}\exp(-x)$ for $x\gg1$ and $\nu\neq0$, one has \citep[e.g.][]{jac62}
\be
\left.\frac{dI}{d\omega d\Omega}\right|_{\theta=0}\simeq
\begin{dcases}
\frac{e^2}{c}\left[\frac{\Gamma(2/3)}{\pi}\right]^2\left(\frac{3}{4}\right)^{1/3}\left(\frac{\omega\rho}{c}\right)^{2/3}, &\omega\ll\omega_c, \\
\frac{3}{4\pi}\frac{e^2}{c}\gamma^2\frac{\omega}{\omega_c}e^{-\omega/\omega_c}, &\omega\gg\omega_c.
\end{dcases} \nonumber\\\label{scr1}
\ee
For simplicity, we use the following approximation:
\be
\frac{dI}{d\omega d\Omega}\simeq\frac{e^2}{c}\left[\frac{\Gamma(2/3)}{\pi}\right]^2\left(\frac{3}{4}\right)^{1/3}\left(\frac{\omega\rho}{c}\right)^{2/3}e^{-\omega/\omega_c}\label{scr21}
\ee
where the subscript $\theta=0$ has been omitted for brevity. 
Eq.(\ref{scr21}) has an uncertainty of less than 50\% over the range $0<\omega/2\omega_c<10^{0.5}$.
For a given frequency $\omega$, the spread in angle can be estimated by determining the angle $\theta_c$ at which $\xi (\theta_c)=\max(\xi(0),1)$. One has \citep{jac62}
\be
\theta_c(\omega)\simeq
\begin{dcases}
\frac{1}{\gamma}\left(\frac{2\omega_c}{\omega}\right)^{1/3}=\left(\frac{3c}{\omega\rho}\right)^{1/3}, &\omega\ll\omega_c\\
\frac{1}{\gamma}\left(\frac{2\omega_c}{3\omega}\right)^{1/2}, &\omega\gg\omega_c
\end{dcases}.\label{angle1}
\ee
As shown in the above equation, for frequencies comparable to $\omega_c$, the radiation is confined to angles of the order $\sim 1/\gamma$; for lower frequencies, the angular spread is larger. Note that for frequencies higher than $\omega_c$, one has $\theta_c\propto\omega^{-1/2}$. However, the radiation has become negligible due to the exponential term, see Eq.(\ref{scr1}) or Eq.(\ref{scr21}). 

Finally, the spectrum of the total energy emitted by the electron can be found by integrating Eq.(\ref{CRspec1}) over angle \citep{wes59}
\be
\frac{dI}{d\omega}=\sqrt{3}\frac{e^2}{c}\gamma\frac{\omega}{\omega_c}\int_{\omega/\omega_c}^\infty K_{5/3}(x)dx.\label{syn1}
\ee 
However, one must note that it is the total radiation spectrum in all directions rather than the direction along the line of sight. 

\section{C. Radiation from a point source with power-law distributed electrons}\label{secb}

We consider that the radiation from a point containing relativistic electrons with different energies. The energy distribution of the electrons is assumed to be a power-law distribution, i.e.
\be
N_e(\gamma)=N_{e,0}\left(\frac{\gamma}{\gamma_1}\right)^{-p},~~~~~\gamma_1<\gamma<\gamma_2
\ee
where $N_e(\gamma)d\gamma$ is the electron number in a range from $\gamma$ to $\gamma+d\gamma$, $N_{e,0}$ is the corresponding normalization, $\gamma_1$ and $\gamma_2$ are the lower and upper limits of Lorentz factor. 
The energy radiated per unit frequency interval per unit solid angle is given by
\be
\frac{dI}{d\omega d\Omega}=\frac{e^2\omega^2}{4\pi^2c}\left|-\bm{\epsilon}_\parallel \int_{\gamma_1}^{\gamma_2}N_e(\gamma){A}_\parallel(\omega,\gamma)d\gamma+\bm{\epsilon}_\perp\int_{\gamma_1}^{\gamma_2}N_e(\gamma){A}_\perp(\omega,\gamma)d\gamma\right|^2.
\ee
If the observed direction is in the trajectory plane, e.g., $\theta=0$, the perpendicular polarized component is zero, e.g., $A_\perp(\omega,\gamma)=0$. Thus, one has
\be
\frac{dI}{d\omega d\Omega}=\frac{e^2\omega^2}{4\pi^2c}\left|\int_{\gamma_1}^{\gamma_2}N_e(\gamma){A}_\parallel(\omega,\gamma)d\gamma\right|^2,
\ee
where the parallel polarized amplitude is given by Eq.(\ref{amp}), i.e.,
\be
A_\parallel(\omega,\gamma)=\frac{2i}{\sqrt{3}}\frac{\rho}{c\gamma^2}K_{2/3}\left(\frac{\omega\rho}{3c\gamma^3}\right)\simeq\frac{2^{4/3}i}{\sqrt{3}}\Gamma(2/3)\frac{\rho}{c\gamma^2}\left(\frac{\omega}{\omega_c}\right)^{-2/3}e^{-\omega/2\omega_c}.\label{b4}
\ee
Here, we use the approximation $K_{\nu}(x)\sim(\Gamma(\nu)/2)(x/2)^{-\nu}e^{-x}$.
The coherent sum of the amplitudes is given by
\be
\int_{\gamma_1}^{\gamma_2}N_e(\gamma){A}_\parallel(\omega,\gamma)d\gamma=\frac{2^{4/3}i}{3^{3/2}}\Gamma(2/3)N_{e,0}\frac{\rho}{c\gamma_1}\left(\frac{\omega}{\omega_{c1}}\right)^{-(p+1)/3}\int_{x_2}^{x_1}x^{(p-4)/3}e^{-x/2}dx,
\ee
where $x_1=\omega/\omega_{c1}$, $x_2=\omega/\omega_{c2}$, $x=\omega/\omega_{c}$, $\omega_{c1}=\omega_c(\gamma_1)$, and $\omega_{c2}=\omega_c(\gamma_2)$.
The radiation energy satisfies
\be
\frac{dI}{d\omega d\Omega}\simeq\frac{e^2}{c}\left[\frac{\Gamma(2/3)}{\pi}\right]^2\frac{1}{3\cdot2^{4/3}}N_{e,0}^2\gamma_1^4\left(\frac{\omega}{\omega_{c1}}\right)^{-(2p-4)/3}\left(\int_{x_2}^{x_1}x^{(p-4)/3}e^{-x/2}dx\right)^2.
\ee
If $x_2\ll 1$ and $x_2\ll x_1$, according to the property of Gamma function, one has
\be
\int_{x_2}^{x_1}x^{(p-4)/3}e^{-x/2}dx=2^{(p-1)/3}\left[\Gamma\left(\frac{p-1}{3}\right)-\Gamma\left(\frac{p-1}{3},\frac{x_1}{2}\right)\right].\label{gammaf}
\ee
If $x_1\rightarrow\infty$, $\Gamma((p-1)/3,x_1/2)\rightarrow0$, for the power-law distribution of electrons, the energy radiated per unit frequency interval per unit solid angle is given by
\be
\frac{dI}{d\omega d\Omega}\simeq\frac{e^2}{c}\frac{2^{(2p-6)/3}}{3\pi^2}\left[\Gamma\left(\frac{2}{3}\right)\Gamma\left(\frac{p-1}{3}\right)\right]^2N_{e,0}^2\gamma_1^4
\begin{dcases}
\left(\frac{\omega}{\omega_{c1}}\right)^{2/3},&\omega\ll\omega_{c1},\\
\left(\frac{\omega}{\omega_{c1}}\right)^{-(2p-4)/3}e^{-\omega/\omega_{c2}},&\omega\gg\omega_{c1}.
\end{dcases}\label{ne10}
\ee

\section{D. Spectrum of one-dimensional bunch}\label{secd0}

First, we consider that the electron distribution is stationary. The retarded position of the $j$th electron can be written as $\bm{r}_j(t)=\bm{r}(t)+\Delta\bm{r}_j$, where $\bm{r}(t)$ denotes the retarded position of the first electron, and $\Delta \bm{r}_j$ denotes the relative displacement between the first electron and the $j$th electron, which is time-independent. According to Eq.(\ref{multiemission}), the total energy radiated per unit solid angle per unit frequency interval is given by
\be
\frac{dI_{(N)}}{d\omega d\Omega}=\frac{e^2\omega^2}{4\pi^2c}\left|\int_{-\infty}^{+\infty}\bm{n}\times(\bm{n}\times\bm{\beta})e^{i\omega(t-\bm{n}\cdot\bm{r}(t)/c)}dt\right|^2\left|\sum_j^Ne^{-i\omega(\bm{n}\cdot\Delta\bm{r}_j/c)}\right|^2=\frac{dI_{(1)}}{d\omega d\Omega}F_\omega(N),\label{cohN1}
\ee
where
\be
F_\omega(N)=\left|\sum_j^Ne^{-i\omega(\bm{n}\cdot\Delta\bm{r}_j/c)}\right|^2,\label{factor1}
\ee
is a dimensionless parameter denoting the enhancement factor due to coherence,
and $dI_{(1)}/d\omega d\Omega$ corresponds to the radiation of the first electron.

\begin{figure}[H]
\centering
\includegraphics[angle=0,scale=0.5]{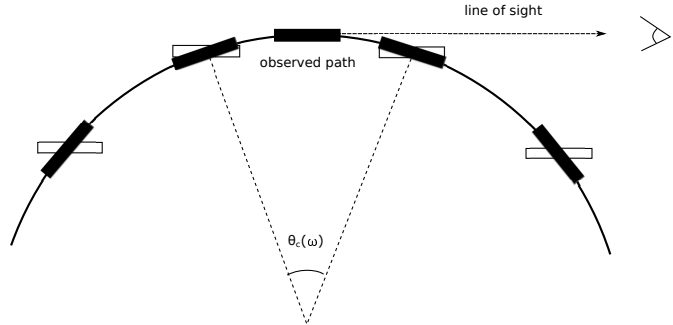}
\caption{Curvature radiation from an one-dimensional bunch. The open strip denote a bunch with stationary-distributed electrons. The solid strip denotes a bunch with all electrons having the same trajectory. The former corresponds to the displacement of the spatial distribution of electrons, and the latter corresponds to rotation of the spatial distribution of electrons around the center of instantaneously circular. The solid line denotes the trajectory of the central electron. The observed path corresponds to a path in the spread angle $\theta_c$, e.g., Eq.(\ref{angle})}\label{rbun}
\end{figure}

Next, we are interesting in that $N$ electrons have the same trajectory but are injected at different times. In this case, the retarded position of the $j$th electron can be written as $\bm{r}_j(t)=\bm{r}(t)+\Delta\bm{r}_j(t)$. We note that $\Delta\bm{r}_j (t)$ must change with time, even if $|\Delta \bm{r}_j(t)|$ is assumed to be time-independent. 
As shown in Figure \ref{rbun}, for stationary-distributed electrons, their motions correspond to the displacement of the spatial distribution of electrons. However, for the electrons lying in the same trajectory, their motions correspond to rotation of the spatial distribution of electrons around the center of instantaneously circular.

Although the above two motion modes have significantly difference, we can prove that the latter can be approximately equal to the former when the bunch is relativistic and its length is enough small:
For a relativistic bunch, its radiation is beamed in a narrow cone that sweeps cross the line of sight. Therefore, if the bunch length satisfying $L\sim\Delta r_N\lesssim\rho\theta_c$, where $\theta_c$ is the spread angle of curvature radiation ( for $\omega\sim\omega_c$, $\theta_c\sim1/\gamma$, see Eq.(\ref{angle})), then in the observed path (with a length $\sim\rho\theta_c$ where the bunch velocity is almost parallel to the line of sight), the relative displacement between each electron could be considered as time-independent.
Note that although Eq.(\ref{radiation}) shows that the observed spectrum is determined by the electron trajectory over a period of time, for a relativistic charged particle, the major contribution of the spectrum is from a path with $\sim\rho\theta_c$.
Once outside the observed path, the radiation contributing to the line of sight could be ignored.

\section{E. Trajectory family II: Generated via rotation around Y axis}\label{secc}

\begin{figure*}[h]
\centering
\includegraphics[angle=0,scale=0.3]{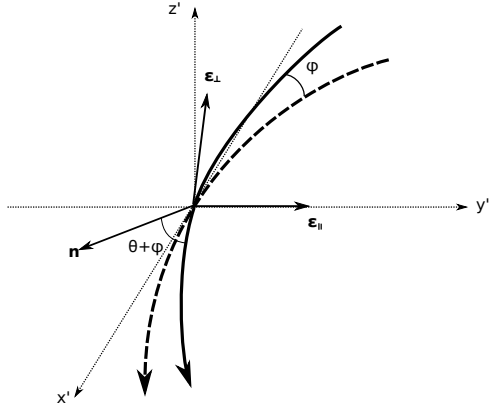}
\caption{Geometry for instantaneously circular motion for Family II in the local frame. The dashed line denotes the trajectory lies in the $x'-y'$ plane. At retarded time $t=0$, the electron is at the origin, and the velocity is along $x'$ axis. The solid line denotes a trajectory generated via the rotation around $y'$ axis by the dashed line (corresponding to a trajectory generated via the rotation around $y$ axis in Figure \ref{roty} when $\bm{n}$ is parallel to the $x'$ axis.).}\label{rotya}
\end{figure*} 
 
In this section, we consider that the trajectory family is generated via the rotation around $y$ axis in Figure \ref{roty}. In the local frame, as shown in Figure \ref{rotya}, the amplitude of one in the trajectory family can be calculated following Appendix \ref{secb0}, with the observation angle $\theta$ replaced by $\theta+\varphi_j$, where $\varphi_j$ corresponds to the angle between the $j$th trajectory and the median trajectory.
The energy radiated per unit frequency interval per unit solid angle is given by
\be
\frac{dI}{d\omega d\Omega}=\frac{e^2\omega^2}{4\pi^2c}\left|-\bm{\epsilon}_\parallel\sum_j^{N_t}{A}_{\parallel,j}(\omega)+\bm{\epsilon}_\perp\sum_j^{N_t}{A}_{\perp,j}(\omega)\right|^2. \label{syn23}
\ee
First, we assume that the bunch opening angle of the $N_t$ trajectories is $2\varphi$, each trajectory is uniformly spaced in the bunch opening angle, and there is only one electron in each trajectory.
Then the amplitudes in the above equation are given by
\be
\sum_j^{N_t}A_{\parallel,j}(\omega)&=&\frac{N_t}{2\varphi}\int_{-\varphi}^{\varphi}\frac{\rho}{c}\left[\frac{1}{\gamma^2}+(\theta+\varphi')^2\right]\frac{2i}{\sqrt{3}}K_{2/3}(\xi)d\varphi', \nonumber\\
\sum_j^{N_t}A_{\perp,j}(\omega)&=&\frac{N_t}{2\varphi}\int_{-\varphi}^{\varphi}\frac{\rho(\theta+\varphi')}{c}\left[\frac{1}{\gamma^2}+(\theta+\varphi')^2\right]^{1/2}\frac{2}{\sqrt{3}}K_{1/3}(\xi)d\varphi',
\ee
where
\be
\xi=\frac{\omega\rho}{3c}\left[\frac{1}{\gamma^2}+(\theta+\varphi')^2\right]^{3/2}.
\ee
Since the radiation is beamed in a narrow cone that sweeps cross the observation point, we are only interested in the case with $\theta=0$. One has
\be
\sum_j^{N_t}A_{\parallel,j}(\omega)&=&\frac{N_t}{2\varphi}\int_{-\varphi}^{\varphi}\frac{\rho}{c}\left(\frac{1}{\gamma^2}+\varphi'^2\right)\frac{2i}{\sqrt{3}}K_{2/3}(\xi)d\varphi', \nonumber\\
\sum_j^{N_t}A_{\perp,j}(\omega)&=&0.\label{c4}
\ee
We define
\be
\varphi_c\equiv\theta_c(\omega)=\frac{1}{\gamma}\left(\frac{2\omega_c}{\omega}\right)^{1/3}=\left(\frac{3c}{\omega\rho}\right)^{1/3}.
\ee
For any $\varphi'\gg\varphi_c$, one has $\xi\gg1$, leading to $K_{2/3}(\xi)\rightarrow0$.
Therefore, one has approximately
\be
\sum_j^{N_t}A_{\parallel,j}(\omega)=\frac{N_t}{2\varphi}\int_{-\varphi}^{\varphi}\frac{\rho}{c}\left(\frac{1}{\gamma^2}+\varphi'^2\right)\frac{2i}{\sqrt{3}}K_{2/3}(\xi)d\varphi'\simeq\frac{N_t}{2\varphi}\frac{\rho}{c\gamma^2}\frac{2i}{\sqrt{3}}K_{2/3}\left(\frac{\omega\rho}{3c\gamma^3}\right)(2\Delta\varphi),
\ee
where
\be
\Delta\varphi\simeq\begin{dcases}
\varphi,&\varphi\ll\varphi_c,\\
\varphi_c,&\varphi\gg\varphi_c.\\
\end{dcases}
\ee
For $\omega_\varphi\ll\omega_c$, the sum of the parallel amplitudes is
\be
\sum_j^{N_t}A_{\parallel,j}(\omega)=\frac{2^{4/3}i}{\sqrt{3}}\Gamma(2/3)N_t\frac{\rho}{c\gamma^2}\begin{dcases}
\left(\frac{\omega}{\omega_c}\right)^{-2/3},&\omega\ll\omega_\varphi,\\
\left(\frac{\omega_\varphi}{\omega_c}\right)^{1/3}\left(\frac{\omega}{\omega_c}\right)^{-1}e^{-\omega/2\omega_c},&\omega\gg\omega_\varphi,
\end{dcases}
\ee
where 
\be
\omega_\varphi=\frac{3c}{\rho\varphi^3},
\ee
which is defined as $\varphi_c(\omega_\varphi)=\varphi$. 
Therefore, the energy radiated per unit frequency interval per unit solid angle is given by
\be
\frac{dI}{d\omega d\Omega}=\frac{e^2}{c}\frac{3}{2^{4/3}}\left[\frac{\Gamma(2/3)}{\pi}\right]^2N_t^2\gamma^2\begin{dcases}
\left(\frac{\omega}{\omega_c}\right)^{2/3},&\omega\ll\omega_\varphi,\\
\left(\frac{\omega_\varphi}{\omega_c}\right)^{2/3}e^{-\omega/\omega_c},&\omega\gg\omega_\varphi.
\end{dcases}
\ee
On the other hand, for $\omega_\varphi\gg\omega_c$, the radiation from the entire bunch opening angle can be observed, as shown in Figure \ref{omephi}. In this case, the sum of the parallel amplitudes is given by $\sum_j^{N_t}A_{\parallel,j}(\omega)=N_tA_{\parallel}(\omega,\gamma)$, where $A_{\parallel}(\omega,\gamma)$ is given by Eq.(\ref{b4}).
Thus the radiation energy is given by Eq.(\ref{scr2}).

Next, we further consider that there are more than one electrons in a point source in each trajectory and the electron distribution satisfies the power-law distribution, e.g. $N_e(\gamma)d\gamma=N_{e,0}(\gamma/\gamma_1)^{-p}d\gamma$ for $\gamma_1<\gamma<\gamma_2$, where $N_{e,0}$ corresponds to the normalization for all the trajectories. 
In the case that $\omega_\varphi\ll\omega_{c1}$, if $\omega\ll\omega_\varphi$, since $\omega_\varphi$ is independent of $\gamma$, one always has $dI/d\omega d\Omega\propto\omega^{2/3}$;
if $\omega\gg\omega_\varphi$, one has
\be
\sum_j^{N_t}A_{\parallel,j}(\omega)&=&\frac{2^{5/3}i}{\sqrt{3}}\Gamma(2/3)\frac{\rho}{c\varphi}\int_{\gamma_1}^{\gamma_2}\frac{N_{e}(\gamma)}{\gamma^3}\left(\frac{\omega}{\omega_c}\right)^{-1}e^{-\omega/2\omega_c}d\gamma\nonumber\\
&=&\frac{2^{5/3}i}{3^{3/2}}\Gamma(2/3)\frac{N_{e,0}}{\varphi}\frac{\rho}{c\gamma_1^2}\left(\frac{\omega}{\omega_{c1}}\right)^{-(p+2)/3}\int_{x_2}^{x_1}x^{(p-4)/3}e^{-x/2}dx\label{c11},
\ee
and the radiation energy is given by
\be
\frac{dI}{d\omega d\Omega}&=&\frac{e^2}{c}\left[\frac{\Gamma(2/3)}{\pi}\right]^2\frac{1}{3\cdot2^{2/3}}\frac{N_{e,0}^2\gamma_1^2}{\varphi^2}\left(\frac{\omega}{\omega_{c1}}\right)^{-(2p-2)/3}\left(\int_{x_2}^{x_1}x^{(p-4)/3}e^{-x/2}dx\right)^2\nonumber\\
&\simeq&\frac{e^2}{c}\frac{2^{(2p-4)/3}}{3\pi^2}\left[\Gamma\left(\frac{2}{3}\right)\Gamma\left(\frac{p-1}{3}\right)\right]^2\frac{N_{e,0}^2\gamma_1^2}{\varphi^2}\begin{dcases}
1,&\omega\ll\omega_{c1},\\
\left(\frac{\omega}{\omega_{c1}}\right)^{-(2p-2)/3}e^{-\omega/\omega_{c2}},&\omega\gg\omega_{c1}.
\end{dcases}\nonumber\\\label{c12}
\ee
Note that in the above equation, we have used Eq.(\ref{gammaf}). In the case that $\omega_{c1}\ll\omega_{\varphi}$, if $\omega\ll\omega_\varphi$, the radiation energy is directly given by Eq.(\ref{ne1}); if $\omega\gg\omega_\varphi$, similar to the calculation process of Eq.(\ref{c11}) and Eq.(\ref{c12}), one has $dI/d\omega d\Omega\propto\omega^{-(2p-2)/3}$.

In summary, if $\omega_\varphi\ll\omega_{c1}$, the energy radiated per unit frequency interval per unit solid angle is given by
\be
\frac{dI}{d\omega d\Omega}=\frac{e^2}{c}\frac{2^{(2p-6)/3}}{3\pi^2}\left[\Gamma\left(\frac{2}{3}\right)\Gamma\left(\frac{p-1}{3}\right)\right]^2N_{e,0}^2\gamma_1^4
\begin{dcases}
\left(\frac{\omega}{\omega_{c1}}\right)^{2/3},&\omega\ll\omega_\varphi,\\
\left(\frac{\omega_\varphi}{\omega_{c1}}\right)^{2/3},&\omega_\varphi\ll\omega\ll\omega_{c1},\\
\left(\frac{\omega_\varphi}{\omega_{c1}}\right)^{2/3}\left(\frac{\omega}{\omega_{c1}}\right)^{-(2p-2)/3},&\omega\gg\omega_{c1}.
\end{dcases}\nonumber\\
\ee
If $\omega_{\varphi}\gg\omega_{c1}$, the energy radiated per unit frequency interval per unit solid angle is given by
\be
\frac{dI}{d\omega d\Omega}=\frac{e^2}{c}\frac{2^{(2p-6)/3}}{3\pi^2}\left[\Gamma\left(\frac{2}{3}\right)\Gamma\left(\frac{p-1}{3}\right)\right]^2N_{e,0}^2\gamma_1^4
\begin{dcases}
\left(\frac{\omega}{\omega_{c1}}\right)^{2/3}, &\omega\ll\omega_{c1}, \\
\left(\frac{\omega}{\omega_{c1}}\right)^{-(2p-4)/3}, &\omega_{c1}\ll\omega\ll\omega_{\varphi}, \\
\left(\frac{\omega_\varphi}{\omega_{c1}}\right)^{-(2p-4)/3}\left(\frac{\omega}{\omega_{\varphi}}\right)^{-(2p-2)/3}, &\omega\gg\omega_{\varphi}.
\end{dcases}\nonumber\\
\ee

\section{F. Trajectory family III: Generated via rotation around X axis}\label{secd}

\begin{figure*}[h]
\centering 
\includegraphics[angle=0,scale=0.3]{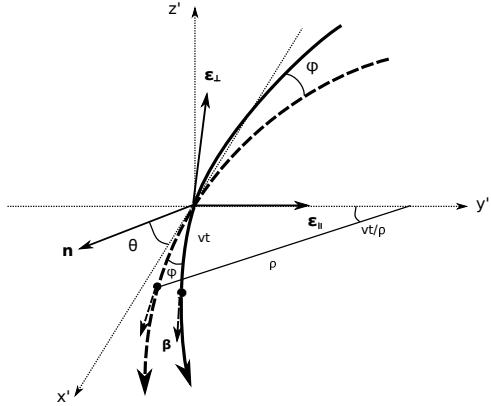}
\caption{Geometry for instantaneously circular motion for Family III in the local frame. The dashed line denotes the trajectory lies in the $x'-y'$ plane. At retarded time $t=0$, the electron is at the origin, and the velocity is along $x'$ axis. The solid line denotes a trajectory generated via the rotation around $z'$ axis by the dashed line (corresponding to a trajectory generated via the rotation around $x$ axis in Figure \ref{rotx} when $\bm{n}$ is parallel to the $x'$ axis.), which is also in the $x'-y'$ plane.}\label{rotxa}
\end{figure*}
For the trajectory family generated via the rotation around $x$ axis, as shown Figure \ref{rotx}, we need to consider a more general situation to calculate the amplitude from each trajectory than that in Appendix \ref{secb0}. In the local frame, as shown in Figure \ref{rotxa}, all the trajectories are in the $x'-y'$ plane, and the angle between the electron velocity direction and $x'$ axis at $t=0$, is defined as $\varphi_j$, where $\varphi=0$ corresponds to the case of Appendix \ref{secb0}.
The vector term in the integrand Eq.(\ref{multiemission}) can be written as
\be
\bm{n}\times(\bm{n}\times\bm{\beta}_j)=\beta\left[-\bm{\epsilon}_\parallel\sin\left(\frac{vt}{\rho}+\varphi_j\right)+\bm{\epsilon}_\perp\cos\left(\frac{vt}{\rho}+\varphi_j\right)\sin\theta\right].
\ee
The exponential term in the integrand Eq.(\ref{multiemission}) is given by
\be
\omega\left(t-\frac{\bm{n}\cdot\bm{r}_j(t)}{c}\right)&=&\omega\left[t-\frac{2\rho}{c}\sin\left(\frac{vt}{2\rho}\right)\cos\left(\frac{vt}{2\rho}+\varphi_j\right)\cos\theta\right]\nonumber\\
&\simeq&\frac{\omega}{2}\left[\left(\frac{1}{\gamma^2}+\theta^2+\varphi_j^2\right)t+\frac{c^2t^3}{3\rho^2}+\frac{ct^2}{\rho}\varphi_j\right].
\ee
Therefore, the amplitudes are given by
\be
A_{\parallel,j}&\simeq&\int_{-\infty}^{\infty}\left(\frac{ct}{\rho}+\varphi_j\right)\exp\left(i\frac{\omega}{2}\left[\left(\frac{1}{\gamma^2}+\theta^2+\varphi_j^2\right)t+\frac{c^2t^3}{3\rho^2}+\frac{ct^2}{\rho}\varphi_j\right]\right)dt,\\
A_{\perp,j}&\simeq&\theta\int_{-\infty}^{\infty}\exp\left(i\frac{\omega}{2}\left[\left(\frac{1}{\gamma^2}+\theta^2+\varphi_j^2\right)t+\frac{c^2t^3}{3\rho^2}+\frac{ct^2}{\rho}\varphi_j\right]\right)dt.
\ee
Since the radiation is beamed in a narrow cone that sweeps cross the observation point, we are only interested in the case with $\theta=0$. Thus
\be
A_{\parallel,j}&\simeq&\int_{-\infty}^{\infty}\left(\frac{ct}{\rho}+\varphi_j\right)\exp\left(i\frac{\omega}{2}\left[\left(\frac{1}{\gamma^2}+\varphi_j^2\right)t+\frac{c^2t^3}{3\rho^2}+\frac{ct^2}{\rho}\varphi_j\right]\right)dt,\nonumber\\
A_{\perp,j}&\simeq&0.
\ee
Let us define
\be
x&=&\frac{ct}{\rho}\left(\frac{1}{\gamma^2}+\varphi_j^2\right)^{-1/2},\\
\xi&=&\frac{\omega\rho}{3c}\left(\frac{1}{\gamma^2}+\varphi_j^2\right)^{3/2},
\ee
one then has
\be
A_{\parallel,j}\simeq\frac{\rho}{c}\left(\frac{1}{\gamma^2}+\varphi_j^2\right)\int_{-\infty}^{\infty}\left(x+\frac{\varphi_j}{\sqrt{1/\gamma^2+\varphi_j^2}}\right)\exp\left(i\frac{3}{2}\xi\left(x+\frac{1}{3}x^3+\frac{\varphi_j}{\sqrt{1/\gamma^2+\varphi_j^2}}x^2\right)\right)dx.
\ee
Note that $x+x^3/3+(\varphi_j/\sqrt{1/\gamma^2+\varphi_j^2})x^2\rightarrow x$ for $x\rightarrow0$ and $x+x^3/3+(\varphi_j/\sqrt{1/\gamma^2+\varphi_j^2})x^2\rightarrow x^3/3$ for $x\rightarrow\pm\infty$.
Therefore, the following approximations are reasonable, e.g.
\be
\int_{-\infty}^{\infty}x\exp\left(i\frac{3}{2}\xi\left(x+\frac{1}{3}x^3+\frac{\varphi_j}{\sqrt{1/\gamma^2+\varphi_j^2}}x^2\right)\right)\simeq\int_{-\infty}^{\infty}x\exp\left(i\frac{3}{2}\xi\left(x+\frac{1}{3}x^3\right)\right)&=&\frac{2i}{\sqrt{3}}K_{2/3}(\xi),\nonumber\\
\int_{-\infty}^{\infty}\exp\left(i\frac{3}{2}\xi\left(x+\frac{1}{3}x^3+\frac{\varphi_j}{\sqrt{1/\gamma^2+\varphi_j^2}}x^2\right)\right)\simeq\int_{-\infty}^{\infty}\exp\left(i\frac{3}{2}\xi\left(x+\frac{1}{3}x^3\right)\right)&=&\frac{2}{\sqrt{3}}K_{1/3}(\xi).\nonumber\\
\ee
Finally, one has
\be
A_{\parallel,j}&\simeq&\frac{2i}{\sqrt{3}}\frac{\rho}{c}\left(\frac{1}{\gamma^2}+\varphi_j^2\right)K_{2/3}(\xi)+\frac{2}{\sqrt{3}}\frac{\rho}{c}\varphi_j\left(\frac{1}{\gamma^2}+\varphi_j^2\right)^{1/2}K_{1/3}(\xi),\nonumber\\
A_{\perp,j}&\simeq&0.
\ee
We assume that the bunch opening angle of the $N_t$ trajectories is $2\varphi$, and each trajectory is uniformly spaced in the bunch opening angle. In this case, the second term in $A_{\parallel,j}$ will be zero. Thus, one has
\be
\sum_j^{N_t}A_{\parallel,j}(\omega)&=&\frac{N_t}{2\varphi}\int_{-\varphi}^{\varphi}\frac{\rho}{c}\left(\frac{1}{\gamma^2}+\varphi'^2\right)\frac{2i}{\sqrt{3}}K_{2/3}(\xi)d\varphi', \nonumber\\
\sum_j^{N_t}A_{\perp,j}(\omega)&=&0.
\ee
This result is as same as Eq.(\ref{c4}), and the next calculation about the radiation will be as same as Appendix \ref{secc}.
We consider that the energy distribution of the electrons satisfies the power-law distribution, e.g. $N_e(\gamma)d\gamma=N_{e,0}(\gamma/\gamma_1)^{-p}d\gamma$ for $\gamma_1<\gamma<\gamma_2$.
If $\omega_\varphi\ll\omega_{c1}$, the energy radiated per unit frequency interval per unit solid angle is given by
\be
\frac{dI}{d\omega d\Omega}=\frac{e^2}{c}\frac{2^{(2p-6)/3}}{3\pi^2}\left[\Gamma\left(\frac{2}{3}\right)\Gamma\left(\frac{p-1}{3}\right)\right]^2N_{e,0}^2\gamma_1^4
\begin{dcases}
\left(\frac{\omega}{\omega_{c1}}\right)^{2/3}, &\omega\ll\omega_\varphi,\\
\left(\frac{\omega_\varphi}{\omega_{c1}}\right)^{2/3}, &\omega_\varphi\ll\omega\ll\omega_{c1},\\
\left(\frac{\omega_\varphi}{\omega_{c1}}\right)^{2/3}\left(\frac{\omega}{\omega_{c1}}\right)^{-(2p-2)/3}, &\omega\gg\omega_{c1}
\end{dcases}. \nonumber\\\label{rotIIIa}
\ee
If $\omega_{\varphi}\ll\omega_{c}$, the energy radiated per unit frequency interval per unit solid angle is given by
\be
\frac{dI}{d\omega d\Omega}=\frac{e^2}{c}\frac{2^{(2p-6)/3}}{3\pi^2}\left[\Gamma\left(\frac{2}{3}\right)\Gamma\left(\frac{p-1}{3}\right)\right]^2N_{e,0}^2\gamma_1^4
\begin{dcases}
\left(\frac{\omega}{\omega_{c1}}\right)^{2/3}, & \omega\ll\omega_{c1}, \\
\left(\frac{\omega}{\omega_{c1}}\right)^{-(2p-4)/3}, & \omega_{c1}\ll\omega\ll\omega_{\varphi}, \\
\left(\frac{\omega_\varphi}{\omega_{c1}}\right)^{-(2p-4)/3}\left(\frac{\omega}{\omega_{\varphi}}\right)^{-(2p-2)/3}, & \omega\gg\omega_{\varphi}.
\end{dcases}\nonumber\\\label{rotIIIb}
\ee

\section{G. Dipole magnetosphere geometry}\label{sece}

In this section, we give a brief summary about the dipole magnetosphere geometry. For a magnetic dipole field, the field line in the polar coordinates $(r,\theta)$ is given by
\be
r=R_{\max}\sin^2\theta,\label{dline}
\ee
where $R_{\max}$ denotes the distance at which the field line crosses the equator.
For a field line with a certain $R_{\max}$, the curvature radius at $(r,\theta)$ is given
\be
\rho&=&\frac{(r^2+r'^2)^{3/2}}{|r^2+2r'^2-rr''|}=\frac{1}{3}R_{\max}\sin^4\theta\frac{(1+4\cot^2\theta)^{3/2}}{1+\cos^2\theta}
=\frac{1}{3}R_{\max}(1-\mu^2)^{1/2}\frac{(1+3\mu^2)^{3/2}}{1+\mu^2}\nonumber\\
&\simeq&\frac{4}{3}R_{\max}\sin\theta\simeq\frac{4r}{3\sin\theta},~~~~~~~~~\rm{for}~\theta\lesssim0.5,\label{currad}
\ee
where $r'$ and $r''$ denote the first and second derivatives to $\theta$, and $\mu$ is defined as $\mu=\cos\theta$. According to Eq.(\ref{dline}), the differential length of the dipole field line is given by
\be
dl=-R_{\max}\sqrt{1+3\cos^2\theta}d(\cos\theta).
\ee
Therefore, the total length from the origin to a point $(r,\theta)$ is given by
\be
l&=&R_{\max}\left[1+\frac{\ln(2+\sqrt{3})}{2\sqrt{3}}-\left(\frac{1}{2}\mu\sqrt{1+3\mu^2}+\frac{{\rm arcsinh}(\sqrt{3}\mu)}{2\sqrt{3}}\right)\right]\nonumber\\
&\simeq&R_{\max}(1-\cos\theta)=r\frac{1-\cos\theta}{\sin^2\theta},~~~~~~~\rm{for}~\theta\lesssim1.\label{length}
\ee
According Eq.(\ref{currad}) and Eq.(\ref{length}), for a given length, the curvature radius satisfies
\be
\frac{\rho}{l}&=&\frac{1}{3}(1-\mu^2)^{1/2}\frac{(1+3\mu^2)^{3/2}}{1+\mu^2}\left[1+\frac{\ln(2+\sqrt{3})}{2\sqrt{3}}-\left(\frac{1}{2}\mu\sqrt{1+3\mu^2}+\frac{{\rm arcsinh}(\sqrt{3}\mu)}{2\sqrt{3}}\right)\right]^{-1}\nonumber\\
&\simeq&\frac{4}{3}\frac{\sin\theta}{1-\cos\theta},~~~~~~~{\rm for}~\theta\ll1.\label{rhoL}
\ee
Note that the above formula is independent of $R_{\max}$, which means that it is applicable for all dipole field lines; and that due to $\delta\rho/\rho\sim\delta\theta/\sin\theta$, the curvature radius does not significantly change for $\delta\theta\lesssim0.1\theta$. 

Next, we define $\beta$ as the angle between the radial direction and the magnetic field, which is given by
\be
\cos\beta=\frac{2\cos\theta}{\sqrt{1+3\cos^2\theta}}.\label{beta}
\ee
The difference of $\beta$ satisfies
\be
\frac{d\beta}{d\theta}=\frac{2}{1+3\cos^2\theta}\simeq\frac{1}{2},~~~~~~~{\rm for}~\theta\ll1.
\ee
Then the angle between the magnetic axis and the magnetic field is $\alpha=\theta+\beta$, e.g.,
\be
\alpha=\theta+\arccos\left(\frac{2\cos\theta}{\sqrt{1+3\cos^2\theta}}\right),
\ee
and its corresponding difference reads
\be
\frac{d\alpha}{d\theta}=\frac{3(1+\cos^2\theta)}{1+3\cos^2\theta}
\simeq\frac{3}{2},~~~~~~~{\rm for}~\theta\ll1.\label{alpthe}
\ee

\clearpage

\section{H. Notation list}

\begin{tabular}{lll}
\hline
\hline
Symbol & Definition & First Appear\\
\hline
subscript $j$ & The identifier of each charged particle & Section \ref{sec2} \\
subscript ret & A quantities evaluated at the retarded time & Section \ref{sec2}\\
$c$ & Speed of light & Section \ref{sec1}\\
$dI/d\omega d\Omega$ & Energy radiated per unit solid angle per unit frequency interval & Eq.(\ref{radiation}), Section \ref{sec2} \\
$e$ & Elementary charge & Section \ref{sec2}\\
$k_{\rm B}$ & Boltzmann constant & Section \ref{sec1}\\
$l$ & Length of a field line & Eq.(\ref{length}), Appendix \ref{sece} \\
$m_e$ & Electron mass & Section \ref{sec1}\\
$\bm{n}$ & Unit vector of the line of sight & Section \ref{sec2} \\
$n_{\rm GJ}$ & Goldreich-Julian density & Section \ref{sec33} \\
$n_{\pm}$ & Number density of electron-positron pair & Section \ref{sec8} \\
$p$ & Index of electron distribution & Eq.(\ref{spece}), Section \ref{sec32} \\
$q$ & Charge of particle & Section \ref{sec2}\\
$\bm{r}$ & Retarded position of an electron & Section \ref{sec2}\\
 & Distance between emission region and neutron star center & Section \ref{sec63}\\
$v_d$ & Curvature drift & Section \ref{sec8}\\
$v_\varphi$ & Velocity along the field lines & Section \ref{sec8}\\
$\bm{A}$ & Amplitude of electromagnetic wave & Appendix \ref{seca}\\
$A_\parallel$ & Parallel component of amplitude & Appendix \ref{secb0}\\
$A_\perp$ & Perpendicular component of amplitude & Appendix \ref{secb0}\\
$\bm{B}$ & Magnetic field & Eq.(\ref{EMfield}), Section \ref{sec2}\\
$B_p$ & Strength of magnetic field at polar cap of a neutron star & Section \ref{sec63}\\
$D$ & Distance between source and observer & Section \ref{sec61}\\
$\bm{E}$ & Electric field & Eq.(\ref{EMfield}), Section \ref{sec2}\\
$F_\nu$ & Flux at frequency $\nu$ & Eq.(\ref{flux}), Section \ref{sec61}\\
$F_{\nu,\max}$ & The maximum flux & Eq.(\ref{fmax}), Section \ref{sec63}\\
$F_{\omega}(N)$ & The fraction between $dI_{(N)}/d\omega d\Omega$ and $dI_{(1)}/d\omega d\Omega$ for one-dimensional bunch & Eq.(\ref{cohN}), Section \ref{sec4}\\
$K_\nu$ & Modified Bessel function & Section \ref{sec2}\\
$L$ & Length of a bunch & Section \ref{sec41}\\
$\mathcal{L}$ & Luminosity & Section \ref{sec71}\\
$\dot M$ & Wind mass-loss rate & Section \ref{sec722}\\
$\mathcal{M}$ & Multiplicity of electron-positron pair in pulsar magnetosphere & Section \ref{sec33}\\
$N$ & Electron number of a bunch & Section \ref{sec41}\\ 
$N_e(\gamma)d\gamma$ & Energy distribution of electrons in a bunch & Eq.(\ref{spece}), Section \ref{sec32} \\
$N_{e,0}$ & Normalization of the energy distribution of electrons in a bunch & Eq.(\ref{spece}), Section \ref{sec32} \\
$N_{e,{\rm eff}}$ & the effective electron (net charge) number in the bunch & Section \ref{sec63} \\
$N_t$ & Trajectory number of a trajectory family & Section \ref{sec42}\\ 
$N_B$ & Bunches number in a certain trajectory & Section \ref{sec412}\\
$P$ & Period of a neutron star & Section \ref{sec63} \\
$P_B$ & Magnetic pressure & Eq.(\ref{pb}), Section \ref{sec722} \\
$P_s$ & Ram pressure & Eq.(\ref{ps}), Section \ref{sec722} \\
$\mathcal{R}$ & Distance between field point and retarded position of electron & Section \ref{sec2}; Appendix \ref{seca}\\
$R$ & Radius of a neutron star & Section \ref{sec63} \\
$R_{\rm LC}$ & Radius of light cylinder & Section \ref{sec721} \\
$R_{\max}$ & Distance at which the field line crosses the equator & Eq.(\ref{dline}), Appendix \ref{sece} \\
$T$ & Mean time interval between adjacent bunches & Eq.(\ref{power}), Section \ref{sec2} \\
$T_b$ & Period of a beat wave & Eq.(\ref{beatperiod}), Section \ref{sec43} \\
$T_p$ & Pulse duration of curvature radiation & Eq.(\ref{duration}), Section \ref{sec2} \\
$T_{\rm obs}$ & Observed duration of a pulse of pulsars or FRBs & Section \ref{sec1} \\
$T_B$ & Brightness temperature & Section \ref{sec1} \\
$V$ & Volume of a bunch & Eq.(\ref{volume}), Section \ref{sec63} \\
\end{tabular}

\begin{tabular}{lll}
$\alpha$ & Angle between the magnetic axis and the magnetic field & Section \ref{sec63}; Appendix \ref{sece} \\
$\beta$ & Dimensionless velocity of an electron & Section \ref{sec2}; Appendix \ref{seca}\\
 & Angle between the radial direction and the magnetic field & Section \ref{sec63}; Appendix \ref{sece}\\
$\gamma$ & Lorentz factor of an electron & Section \ref{sec1}\\
$\gamma_{1,2}$ & The minimum (and maximum) Lorentz factor of electrons & Eq.(\ref{spece}), Section \ref{sec32} \\
$\eta$ & A parameter describing the cross section of a bunch in the comb model & Section \ref{sec721} \\
$\theta$ & Angle between the line of sight and the trajectory plane & Section \ref{sec2}\\
 & Poloidal angle of the dipole field & Section \ref{sec63}; Appendix \ref{sece}\\ 
$\theta_c(\omega)$ & Frequency-dependent spread angle of curvature radiation & Eq.(\ref{angle}), Section \ref{sec2} \\
$\mu$ & Cosine of poloidal angle, e.g., $\mu=\cos\theta$ & Appendix \ref{sece} \\
$\mu_c$ & Normalized fluctuating Goldreich-Julian density & Eq.(\ref{muc}), Section \ref{sec63} \\
$\nu_a$ & Synchrotron self-absorption frequency & Eq.(\ref{nebula}), Section \ref{sec71} \\
$\nu_c$ & Critical frequency of curvature radiation, $\nu_c=\omega_c/2\pi$ & Eq.(\ref{omegac}), Section \ref{sec2} \\
$\nu_l$ & Critical frequency for bunch length, $\nu_l=\omega_l/2\pi$ & Eq.(\ref{omegal}), Section \ref{sec411} \\
$\nu_{\rm obs}$ & Observed frequency & Section \ref{sec1}\\
$\nu_{\rm peak}$ & Peak frequency of the spectrum of curvature radiation & Eq.(\ref{fmax}), Section \ref{sec63} \\
$\nu_\varphi$ & Critical frequency for bunch half-opening angle, $\nu_\varphi=\omega_\varphi/2\pi$ & Eq.(\ref{omegaphi}), Section \ref{sec422} \\
$\xi_c$ & Compression factor in the comb model & Section \ref{sec721} \\
$\rho$ & Curvature radius of an electron trajectory & Section \ref{sec2} \\
$\tau_\nu$ & Optical depth at frequency $\nu$ & Section \ref{sec71} \\
$\varphi$ & Minimum angle between the line of sight and the bunch velocity & Section \ref{sec411}\\
 & Bunch half-opening angle of a trajectory family & Section \ref{sec42}\\
 & --- Bunch half-opening angle of Family III for Family A & Section \ref{sec51}\\
 & --- Maximum bunch half-opening angle of Family B & Eq.(\ref{openangleb}), Section \ref{sec52}\\ 
$\varphi_{1,2,3}$ & Bunch half-opening angle of Family I, II and III & Section \ref{sec5}\\
$\varphi_d$ & Drifting angle due to curvature drift effect & Section \ref{sec8}\\
$\varphi_{\times,+}$ & A pair of the orthogonal bunch opening angles & Section \ref{sec5}\\
$\phi$ & Toroidal angle of the dipole field & Section \ref{sec63} \\
$\psi$ & Angle between the magnetic field direction and the line of sight & Section \ref{sec63} \\
$\omega_{bm}$ & Maximum inter-bunch coherent angle frequency & Eq.(\ref{omegabm}), Section \ref{sec412} \\
$\omega_c$ & Critical angle frequency of curvature radiation & Eq.(\ref{omegac}), Section \ref{sec2} \\
$\omega_{c1,c2}$ & Critical angle frequency of curvature radiation when $\gamma=\gamma_{1,2}$ & Section \ref{sec32}\\
$\omega_l$ & Critical angle frequency for bunch length & Eq.(\ref{omegal}), Section \ref{sec411} \\
$\omega_m$ & Upper limit of the coherent angle frequency & Eq.(\ref{omm}), Section \ref{sec411} \\
$\omega_\varphi$ & Critical angle frequency for bunch half-opening angle & Eq.(\ref{omegaphi}), Section \ref{sec422} \\
$\Gamma$ & Gamma function & Section \ref{sec2}\\
$\Delta\bm{r}_j$ & The relative displacement between the first electron and the $j$th electron & Section \ref{sec41}\\
$\Delta S$ & Cross section of a bunch & Section \ref{sec63} \\
$\Omega$ & Solid angle of radiation & Section \ref{sec2}\\
 & Angle frequency of a neutron star & Section \ref{sec63}\\
\hline
\hline
\end{tabular}

\end{document}